\documentclass[]{pasj02} 
\usepackage[switch,mathlines]{lineno} 

\jyear{2026}
\Received{}
\Accepted{}

\begin{document} 

\title{Subaru/HSC {\it NB395} view of NGC~5466: metallicity, mass function, and the nature of its tidal stream}
\author{
 Itsuki \textsc{Ogami},\altaffilmark{1,2,3}\altemailmark\orcid{0000-0001-8239-4549} \email{itsuki.ogami@nao.ac.jp} 
 Miho N. \textsc{Ishigaki},\altaffilmark{1,4,5,6}\orcid{0000-0003-4656-0241}
 Pete B. \textsc{Kuzma},\altaffilmark{1,7}\orcid{0000-0003-1980-8838}
 Takanobu \textsc{Kirihara},\altaffilmark{8}\orcid{0000-0001-6503-8315}
 Nozomu \textsc{Tominaga},\altaffilmark{1,4,9}\orcid{0000-0001-8537-3153}
 Masashi \textsc{Chiba},\altaffilmark{10}\orcid{0000-0002-9053-860X}
 Yutaka \textsc{Komiyama},\altaffilmark{3,1}\orcid{0000-0002-3852-6329}
 Mohammad \textsc{Mardini},\altaffilmark{11}\orcid{0000-0001-9178-3992}  and 
 Hiroko \textsc{Okada}\altaffilmark{12,1}\orcid{0009-0009-6151-8157}
}
\altaffiltext{1}{National Astronomical Observatory of Japan, 2-21-1 Osawa, Mitaka, Tokyo 181-8588, Japan}
\altaffiltext{2}{The Institute of Statistical Mathematics, 10-3 Midoricho, Tachikawa, Tokyo 190-8562, Japan}
\altaffiltext{3}{Department of Advanced Sciences, Faculty of Science and Engineering, Hosei University, 3-7-2 Kajino-cho, Koganei, Tokyo 184-8584, Japan}
\altaffiltext{4}{Astronomical Science Program, The Graduate University for Advanced Studies (SOKENDAI), 2-21-1 Osawa, Mitaka, Tokyo 181-8588,
Japan}
\altaffiltext{5}{Subaru Telescope, National Astronomical Observatory of Japan, 650 North A'ohoku Place, Hilo, HI 96720, USA}
\altaffiltext{6}{Kavli Institute for the Physics and Mathematics of the Universe (WPI),The University of Tokyo
Institutes for Advanced Study, The University of Tokyo, Kashiwa, Chiba 277-8583, Japan}
\altaffiltext{7}{Institute for Astronomy, University of Edinburgh, Royal Observatory, Blackford Hill, Edinburgh, EH9 3HJ, UK}
\altaffiltext{8}{Kitami Institute of Technology, 165, Koen-cho, Kitami, Hokkaido 090-8507, Japan}
\altaffiltext{9}{Department of Physics, Faculty of Science and Engineering, Konan University, 8-9-1 Okamoto, Kobe, Hyogo 658-8501, Japan}
\altaffiltext{10}{Astronomical Institute, Tohoku University, 6-3 Aoba, Aramaki, Aoba-ku, Sendai, Miyagi 980-8578, Japan}
\altaffiltext{11}{Department of Physics, Zarqa University, Zarqa 13110, Jordan}
\altaffiltext{12}{Nishi-Harima Astronomical Observatory, Center for Astronomy, University of Hyogo, 407-2 Nishigaichi, Sayo-cho, Sayo, Hyogo
679-5313, Japan}

\maketitle

\begin{abstract}
We present a deep photometric study of the globular cluster NGC~5466 and its tidal stream using Subaru/Hyper Suprime-Cam (HSC) imaging with the metallicity-sensitive narrowband filter {\it NB395}. We develop an improved member-selection technique based on a k-nearest neighbor algorithm applied to the color–color–magnitude diagram (CCMD), enabling reliable candidate identification down to $i_{2,0} < 23.5$. Photometric metallicities derived from {\it NB395} colors agree with previous measurements, supporting the robustness of our calibration. While modest residual contamination and possible offsets - potentially driven by variations in light-element abundances - may remain beyond 10 arcmin, the metallicity distribution of high-probability inner members matches the known mean metallicity of NGC~5466, demonstrating the effectiveness of our method. The spatial distribution of {\it NB395}-selected stars clearly delineates the tidal stream. Beyond the tidal radius, the azimuthally averaged radial surface density profile follows a power law with slope $\alpha = -4.53_{-0.14}^{+0.13}$. We also detect a power-law component perpendicular to the stream, suggestive of multiple apogalactic passages. A density gap is identified at a projected distance of $\sim200$ pc from the cluster center, consistent with eTidal N-body predictions and possibly associated with a recent pericentric passage or Galactic disk interaction. Analysis of the main-sequence mass function reveals a strong negative radial gradient in the slope within the tidal radius, whereas the slope along the outer stream is relatively flat, consistent with preferential tidal stripping of low-mass stars. These results highlight the power of HSC/{\it NB395} photometry for identifying metal-poor populations and deriving photometric metallicities, underscoring its value for future wide-field surveys.
\end{abstract}


\section{Introduction}
Accretion of small stellar systems into a larger, host galaxy is one of the major processes of galaxy formation (e.g., \cite{1991ApJ...379...52W,2008ApJ...689..936J}). An accreting system loses its constituent stars as a result of interaction with the host galaxy's gravitational field, leading to the formation of tidal tails or streams, so the variation of stellar number density along tidal streams is sensitive to both the process of tidal disruption and the Galactic dynamical structure. 

Among the various accreting systems, globular clusters (GCs) provide particularly clean and well-defined laboratories for studying these processes. GCs consist of old, nearly single stellar populations with small velocity dispersion ($<$ a few km s$^{-1}$), making them ideal systems for studying and validating stellar dynamical processes. GCs gradually eject their stars through the combined effects of internal two-body relaxation and the external tidal field of the Milky Way (MW), producing extended stellar envelopes and tidal streams that can persist over many orbital periods. These extended structures not only trace the orbit of the progenitor cluster within the Galactic potential (e.g., \cite{2010MNRAS.401..105K,2012MNRAS.424L..16L,2015ApJ...811..123F,2022MNRAS.513..853Y}), but also encode detailed information about the internal dynamical evolution of the cluster itself (e.g.,\cite{2001AJ....122.3231G,2004ASPC..327..333K,2016MNRAS.463.2383W,2019ApJ...887L..12B}). In particular, the preferential escape of low-mass stars driven by mass segregation leads to characteristic signatures in both the surface density distribution and the mass function (MF) along and around tidal streams (e.g., \cite{2003MNRAS.340..227B,2013MNRAS.433.1378L}).

NGC~5466 is among the most prominent and well-studied examples of a GC hosting an extended stellar stream. Using Sloan Digital Sky Survey data, \citet{2006ApJ...637L..29B} and \citet{2006ApJ...639L..17G} independently revealed a remarkably long and thin stream extending over tens of degrees on the sky, making NGC~5466 one of the earliest and clearest cases of a GC tidal stream. Subsequent studies refined the stream morphology and orbital properties using deeper photometry and Gaia astrometry (e.g. \cite{2007MNRAS.380..749F,2012MNRAS.424L..16L,2021MNRAS.507.1923J}), establishing NGC~5466 as a benchmark system for studying GC disruption in the Galactic halo. However, previous observational studies of NGC~5466 have been limited to relatively bright stellar tracers (red giant branch and horizontal branch stars; RGB and HB stars), and have therefore not been able to probe the detailed morphology of the stream, such as the presence of gaps.

From a dynamical perspective, NGC~5466 is particularly intriguing. It is a low-mass, metal-poor cluster with a low central density and a relatively long half-mass relaxation time, suggesting that it is in an early stage of dynamical evolution (e.g., \cite{2013ApJ...776...60B,2015ApJ...814..144B}). Such systems are expected to exhibit strong mass segregation and a pronounced radial variation in the MF slope, both within the cluster and in the surrounding extratidal regions. Numerical simulations predict that the MF slope becomes progressively flatter with increasing radius as low-mass stars are preferentially removed and populate the tidal tails (e.g. \cite{1997MNRAS.287..915V,2014MNRAS.445.1048W}).

A particularly useful diagnostic of this process is the radial gradient of the MF slope. Using N-body simulations, \citet{2014MNRAS.445.1048W} demonstrated that the radial gradient of the MF slope correlates strongly with a cluster's dynamical age and remaining mass fraction, providing a physically motivated link between observable MF gradients and the cumulative effects of two-body relaxation and tidal stripping. More recently, \citet{2017MNRAS.471.3668S} extended this framework by showing that the MF slope and its radial variation can be used to distinguish between clusters in different dynamical regimes, including tidally filling and underfilling systems, and to constrain the efficiency of mass loss into tidal tails.

Despite the prominence of the NGC~5466 stream, observational constraints on its stellar mass distribution and morphology remain limited. Most previous studies have focused on the spatial extent of the stream based on bright stars (e.g., \cite{2006ApJ...637L..29B,2021MNRAS.507.1923J}), while a comprehensive study of the MF and morphology using low-mass main-sequence (MS) stars as tracers beyond the tidal radius has not yet been performed. Measuring the MF slope and morphological variation using MS stars at large radii is observationally challenging, requiring deep photometry with both sufficient depth to reach low stellar masses and wide spatial coverage to sample regions well beyond the nominal tidal radius, where contamination from field stars becomes severe.

Wide-field imaging with Subaru/Hyper Suprime-Cam (HSC; \cite{2012SPIE.8446E..0ZM,2018PASJ...70S...1M,2018PASJ...70S...3F,2018PASJ...70...66K,2018PASJ...70S...2K}) provides a unique opportunity to overcome these limitations. In particular, the {\it NB395} narrow-band filter, sensitive to Ca\,\emissiontype{II}\,H\&K absorption features, enables efficient discrimination of metal-poor cluster members from foreground contaminants when combined with broad-band photometry. In previous studies with the Ca\,\emissiontype{II}\,H\&K-sensitive narrow-band (e.g., Pristine survey; \cite{2017MNRAS.471.2587S,2024A&A...692A.115M}; Hubble Space Telescope (HST)/{\it F395N} observation; \cite{2022ApJ...925....6F}; MAGIC survey; \cite{2025ApJ...993...77B}), this capability allows us to extract clean samples of Galactic dwarf galaxies and to estimate the photometric metallicities. In addition, recent works with synthesized {\it CaHK} colors revealed the extended structures of Galactic GCs combined with the Gaia dataset \citep{2025MNRAS.537.2752K,2025AJ....170..157K}. Therefore, HSC/{\it NB395} also allows us to extract a clean sample of NGC~5466 member stars from the cluster core to its extended tidal features, and to trace both the surface density profile and the MF slope continuously across the tidal boundary.

In this paper, we present a detailed analysis of the NGC 5466 stellar stream based on Subaru/HSC {\it NB395} imaging. We focus on (i) constructing the surface density profile beyond the classical tidal radius, and (ii) quantifying the radial variation of the MS mass profile slope and its gradient. By directly linking these observational measurements to theoretical expectations from dynamical models, we aim to clarify how internal relaxation and tidal stripping shape the present-day structure and stellar content of one of the Galactic most iconic GC streams. In addition, since this is the first paper based on Subaru/HSC {\it NB395} imagings, we aim to demonstrate the effectiveness of {\it NB395} by presenting results on photometric metallicity estimates and membership determination.

The paper is organized as follows. Section \ref{sec:data} describes the observations, data reduction, and member selection. Section \ref{sec:methods} presents the surface density profile of NGC~5466 from the cluster core to the extratidal regions and the radial behavior of the MS mass profile slope. In section \ref{sec:results}, we present the results of photometric metallicity estimation based on the {\it NB395} filter, the spatial distribution and radial profiles of the member stars, and an investigation of radial variations in MF based on low-mass MS stars. Section \ref{sec:discussion} discusses the implications of our results in the context of GC dynamical evolution and tidal disruption models. Our conclusions are summarized in Section \ref{sec:conclusions}.

\section{Data and reduction}\label{sec:data}

\subsection{Observations, reduction and photometry}\label{subsec:photometry}

In this study, we analyze imaging data obtained with Subaru/Hyper Suprime-Cam (HSC). The observations were conducted in 2022 and 2023 using the {\it g}, and ${\it i_2}$ filters (PI: M. Ishigaki, ID: S22A-113, S23A-040). Because several exposures were taken under bad weather conditions, some frames suffer from poor seeing and bad transparency. To ensure uniform data quality, we perform the reduction and calibration only on images that satisfy seeing $< 1.3$ and transparency $> 0.1$.

\begin{figure*}
 \begin{center}
  \includegraphics[width=0.95\linewidth, trim=0 0 0 0, clip]
  {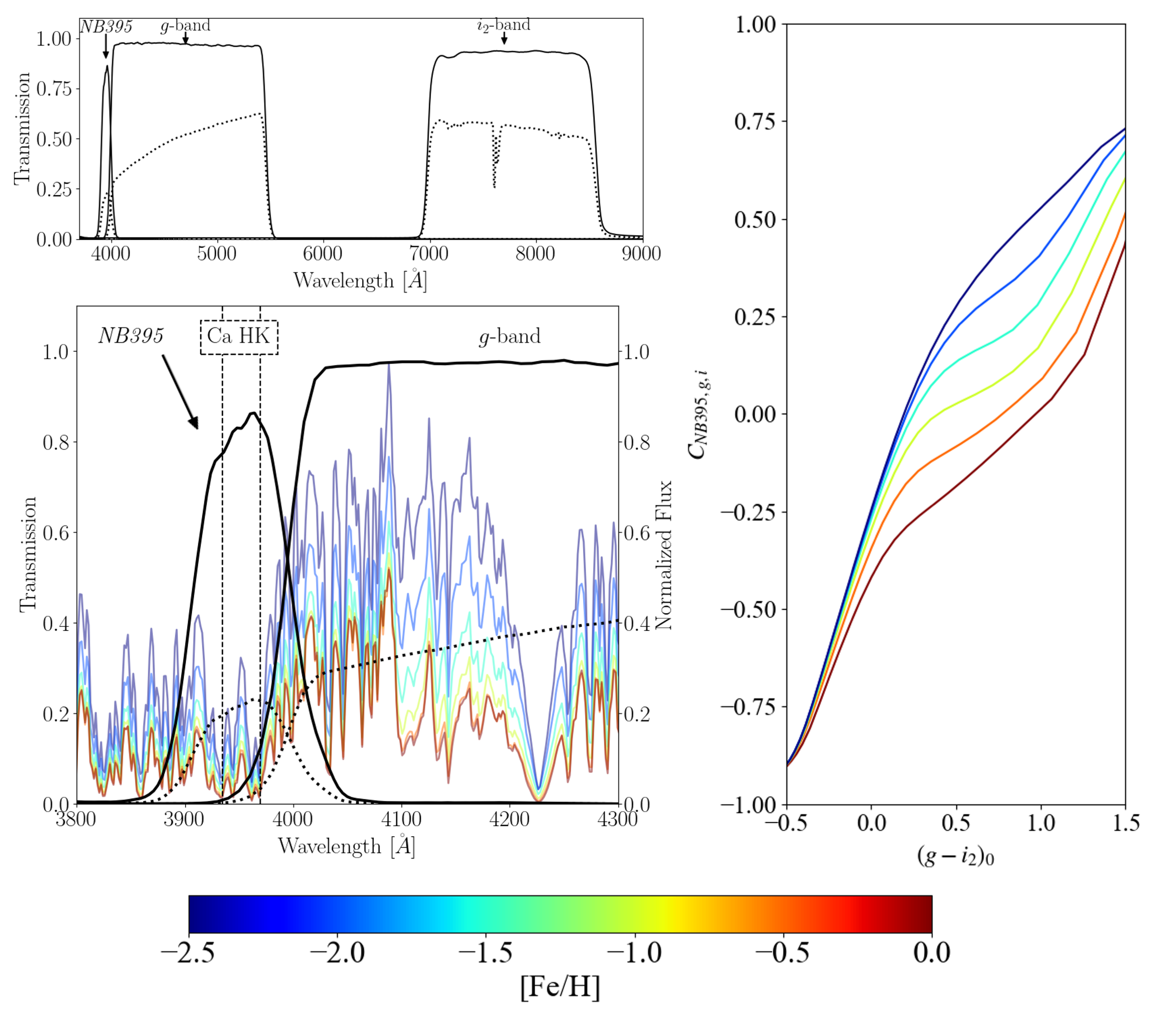}
 \end{center}
 \caption{(Upper left) Response curves of the Subaru/HSC filters used in this study. The solid lines show the intrinsic filter response curves, while the dotted lines indicate the total system throughput. (Lower left) Zoomed-in view of the upper-left panel over $3,800 - 4,300$~$\AA$, with BOSZ model spectra (\cite{2017AJ....153..234B,2024A&A...688A.197M}) overplotted. The model spectra correspond to $\log g=+5.0$, $T_{\rm eff}=3500$~K, and $[\alpha/{\rm Fe}]=+0.25$, for metallicities $[{\rm Fe/H}]=-2.5,~-2.0,~-1.5,~-1.0,~-0.5,$ and $0.0$. The spectra are color-coded by metallicity according to the color bar shown at the bottom. Two vertical black dashed lines indicate the Ca\,\emissiontype{II}\,H\&K absorption lines. (Right) A color-color diagram using {\it NB395}-based color indices defined in equation (\ref{eq:NBcolor}). The color-coded curves represent synthetic model loci computed by convolving BOSZ model spectra with $\log g=+5.0$, $[\alpha/{\rm Fe}]=+0.25$, and $[{\rm Fe/H}]=-2.5,~-2.0,~-1.5,~-1.0,~-0.5,$ and $0.0$ with the HSC filter transmission curves, instrumental sensitivity, and atmospheric absorption.
  {Alt text: Subaru/HSC filter response curves for {\it NB395}, $g$, and $i_2$, shown as intrinsic transmissions (solid) and total system throughput (dotted). A zoomed-in view from $3,800$-$4,300$ $\AA$ overlays BOSZ spectra ($\log g=5.0$, $T_{\rm eff}=3,500$ K, $[\alpha/{\rm Fe}]=0.25$) color-coded by metallicity ($[{\rm Fe/H}]=-2.5$ to $0.0$). The right panel shows an NB395-based color–color diagram with synthetic model loci for the same metallicities.}}\label{fig:NB395_filter}
\end{figure*}

We also obtained images using the narrowband filter {\it NB395}. {\it NB395} is equivalent to the {\it CaHK} filter in Canada-France-Hawaii Telescope (CFHT)/MegaCam, {\it F395N} in HST, {\it N395} in 4 m Blanco Telescope/DECam, which is useful for estimating photometric metallicities. Figure \ref{fig:NB395_filter} summarizes the filter characteristics of {\it NB395} and demonstrates its feasibility for estimation of the photometric metallicity. The upper-left panel of figure \ref{fig:NB395_filter} shows the transmission curve of {\it NB395} (most blue side), together with those of the HSC $g$- and $i_2$-bands. The solid lines represent the original filter transmission curves, while the dashed lines indicate the effective throughputs after accounting for the CCD quantum efficiency, the transmittance of the dewar window, the transmittance of the prime focus unit of HSC, the reflectivity of the primary mirror, and atmospheric absorption. The lower-left panel presents a zoomed-in view of the upper-left panel over the wavelength range from 3,800 to 4,300~$\AA$. In the lower-left panel, we also overplot BOSZ model spectra (\cite{2017AJ....153..234B,2024A&A...688A.197M}) with $\log g=+5.0$, $T_{\rm eff}=3500$~K, and $[\alpha/{\rm Fe}]=+0.25$ for metallicities $[{\rm Fe/H}]=-2.5,~-2.0,~-1.5,-1.0,~-0.5,$ and $0.0$. These spectra are color-coded according to metallicity following the scale shown in the color bar at the bottom of figure \ref{fig:NB395_filter}. {\it NB395} has a bell-shaped transmission curve centered on 3,954 $\AA$ with a full width at half maximum of 89 $\AA$ and samples Ca\,\emissiontype{II}\,H\&K features. In the lower left panel, Ca\,\emissiontype{II}\,H\&K absorption lines are plotted as two vertical black dashed lines, for reference.

The right panel of figure \ref{fig:NB395_filter} displays a color-color diagram using {\it NB395}. In this diagram, we define the color index:
\begin{align}
C_{\it NB395,g,i_2} := 1.5({\it g-i_2})_0-({\it NB395-g})_0~.
\label{eq:NBcolor}
\end{align}
\noindent
This color primarily traces stellar metallicity: the $({\it NB395-g})$ color term reflects the strength of the Ca\,\emissiontype{II}\,H\&K absorption, while the $1.5({\it g-i_2})$ color serves as a corrective term that arranges metal-poor to metal-rich populations from top to bottom in the diagram. The color-coded curves represent synthetic model loci computed by convolving BOSZ model spectra with $\log g=+5.0$, $[\alpha/{\rm Fe}]=+0.25$, and $[{\rm Fe/H}]=-2.5,~-2.0,~-1.5,~-1.0,~-0.5,$ and $0.0$ with the HSC filter transmission curves, instrumental sensitivity, and atmospheric absorption. As in the lower-left panel, the curves in the right panel are color-coded by metallicities. Figure \ref{fig:NB395_filter} clearly demonstrates that the HSC/{\it NB395} filter enables us to infer stellar metallicities from broadband–narrowband colors. Although the {\it NB395} imaging is performed on only 2022, the data are deep enough to identify the faint metal-poor MS stars of NGC~5466.

\begin{figure}
 \begin{center}
  \includegraphics[width=0.95\linewidth, trim=0 0 0 0, clip]
  {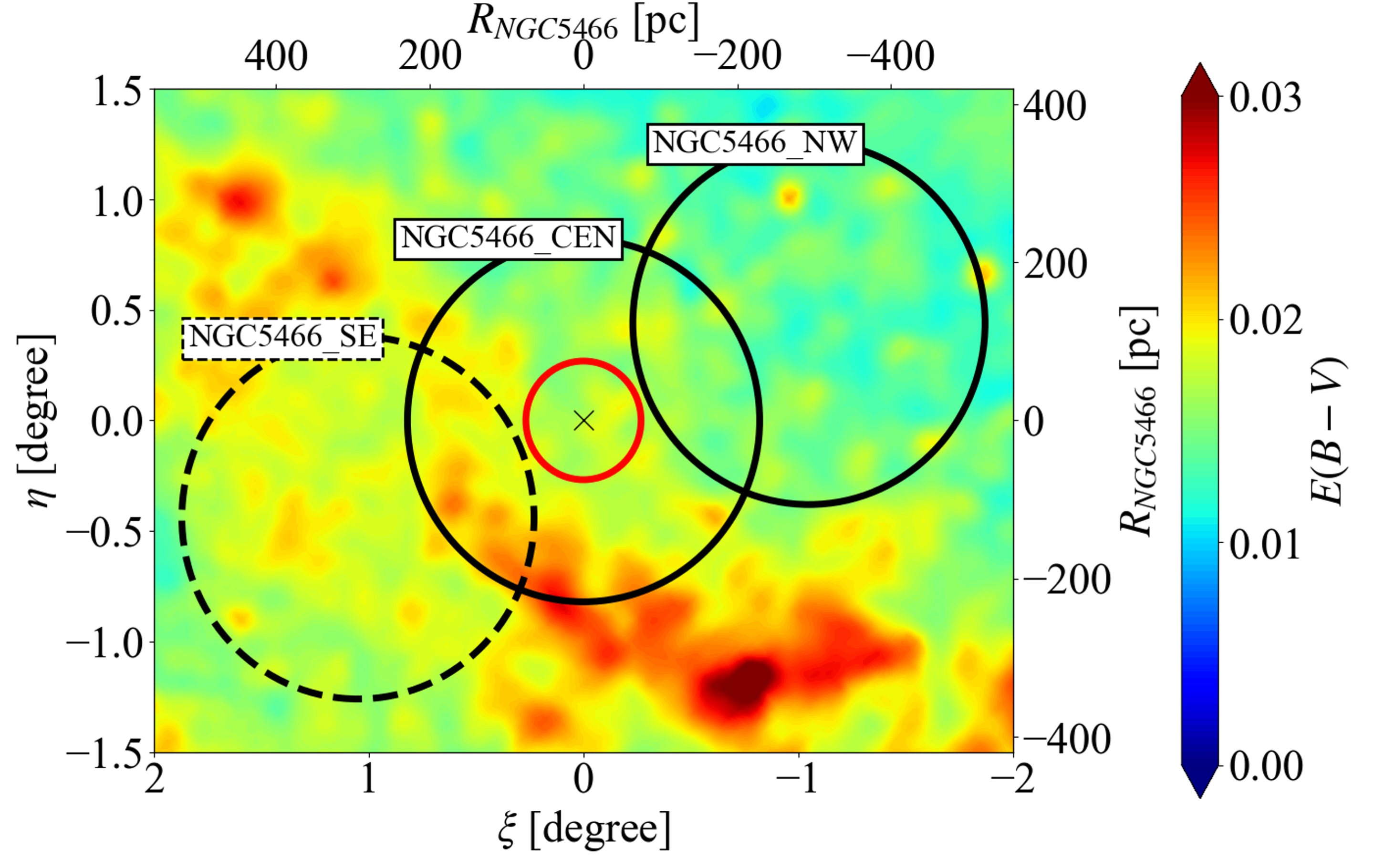}
 \end{center}
 \caption{Illustrating the coverage of our HSC imaging. The black circles show our observational footprints, where the solid circle corresponds to the region observed with {\it g}, ${{\it i}_{{\it 2},0}}$, {\it NB395}, and the dashed circle indicates the region observed with only {\it g}-band. The cross mark denotes the center of the NGC~5466 main body, and the red circle indicates the tidal radius estimated in this study (16.1 arcmin). The background color map is the Galactic dust extinction map.
  {Alt text: Map of the Subaru/HSC observational footprint showing regions observed in multiple bands and in {\it g}-band only. The plot marks the center of NGC~5466 and highlights the estimated tidal radius of 16.1 arcmin.}}\label{fig:ObsMap}
\end{figure}

Figure \ref{fig:ObsMap} displays the spatial distribution of the HSC pointings in the tangential plane $(\xi, \eta)$, centered on the main body of NGC~5466. In figure \ref{fig:ObsMap}, the background color map corresponds to the Galactic dust map (\cite{1998ApJ...500..525S,2012ApJ...756..158S}) taken from \texttt{python/dustmaps} \citep{2018JOSS....3..695G}, and this extinction map shows that there is extinction variations in our observational fields. Details of the extinction correction are described later. Black circles show the observational pointings. For one of the fields (NGC5466\_SE; dashed circle in figure \ref{fig:ObsMap}), only {\it g}-band data were obtained due to observational constraints. Therefore, in this work, we restrict our analysis to the two fields with complete coverage - NGC5466\_CEN and NGC5466\_NW (solid circles in figure \ref{fig:ObsMap}), but it should be noted that the NGC5466\_SE field is also included in the stacked image in order to improve the signal-to-noise ratio. Table \ref{tab:ObsInfo} summarizes the datasets used in the analysis and provides details of the corresponding observations.

\begin{table*}
  \tbl{The details of our observations in HSC.}{
  \begin{tabular}{cccccccc}
  \hline
    Field & $\alpha$ & $\delta$ & Band & Date & Total Exposure Time & Seeing FWHM\\
  \hline
   NGC5466\_CEN & \timeform{14h05m27s28} & $+$\timeform{28D32'03''.90} & {\it g} & 2022 May 1, 2023 March 29 & 575 sec & $0\farcs88 \pm 0\farcs17$\\
                & & & {\it $i_2$} & 2022 May 1 & 1,614 sec & $1\farcs12 \pm 0\farcs14$\\
                & & & {\it NB395} & 2022 April 30 & 16,530 sec & $0\farcs98 \pm 0\farcs40$\\
   NGC5466\_NW  & \timeform{14h00m39s28} & $+$\timeform{28D58'08''.70} & {\it g} & 2022 May 1, 2023 March 29 & 605 sec & $0\farcs92 \pm 0\farcs19$\\
                & & & {\it $i_2$} & 2023 March 20 & 720 sec & $0\farcs70 \pm 0\farcs05$ \\
                & & & {\it NB395} & 2022 April 30 & 4,400 sec & $1\farcs21 \pm 0\farcs33$\\
   NGC5466\_SE  & \timeform{14h10m12s89} & $+$\timeform{28D32'05''.21} & {\it g} & 2023 March 29 & 480 sec & $0\farcs83 \pm 0\farcs13$\\
  \hline
  \end{tabular}}\label{tab:ObsInfo}
  \begin{tabnote}
  \end{tabnote}
\end{table*}

The raw data are reduced and calibrated using the HSC pipeline (hscPipe; \cite{2018PASJ...70S...5B}) version 8.4, a branch of the pipeline for Vera C. Rubin Observatory data (\cite{2010SPIE.7740E..15A,2017ASPC..512..279J,2019ApJ...873..111I}). The pipeline performs standard CCD processing—including de-biasing, dark subtraction, flat-fielding, and sky subtraction—and subsequently carries out astrometric and photometric calibration for each visit using Pan-STARRS1 (PS1; \cite{2012ApJ...756..158S,2012ApJ...750...99T,2013ApJS..205...20M,2020ApJS..251....7F}) reference data. After the individual visits are calibrated, all exposures that satisfy our quality criteria are coadded, and source detection and forced photometry are performed on the resulting stacked images.

Point-source selection is carried out based on the determinant radius, following the procedure described in Ogami et al. in prep. In this method, we compare the PSF radius measured from the PSF model ($R_{\rm det,psf}$) with the radius obtained by fitting a two-dimensional Gaussian to each object ($R_{\rm det}$). We classify an object as a point source if their $R_{\rm det}-R_{\rm det,psf}$ values are within $3\sigma$ of the distribution determined from the artificial stars described below, in the ${\it g}$- or ${\it i_2}$-band.

For the selected sources, we apply the extinction correction for each source, based on the Galactic dust extinction (\cite{1998ApJ...500..525S,2012ApJ...756..158S}). The extinction coefficients are derived following the methodology described in \citet{2024ApJ...971..107O}. Briefly, we multiply the interstellar extinction law with $R_V=+3.1$ (\cite{1999PASP..111...63F}) to the spectral energy distribution of a dwarf star, which is our primary target, characterized by $\log{\rm Z/Z_{\odot}}=-1$ and $\log{g}=+4.5$. Then, this multiplied spectrum is integrated with the response curves of each filter and the instrumental sensitivity functions. Based on these coefficients, we correct the magnitudes using the following relation:
\begin{align}
{{\it g}_0} &= {\it g} - 3.076\times E(B-V) \notag \\    
{\it i}_{{\it 2},0} &= {\it i_{2}} - 1.540\times E(B-V)\\
{\it NB395}_0 &= {\it NB395} - 3.819\times E(B-V) \notag
\end{align}
\noindent
where the subscript ``$_0$'' means the extinction corrected magnitude.

\begin{figure*}
 \begin{center}
  \includegraphics[width=\linewidth, trim=50 0 50 0, clip]
  {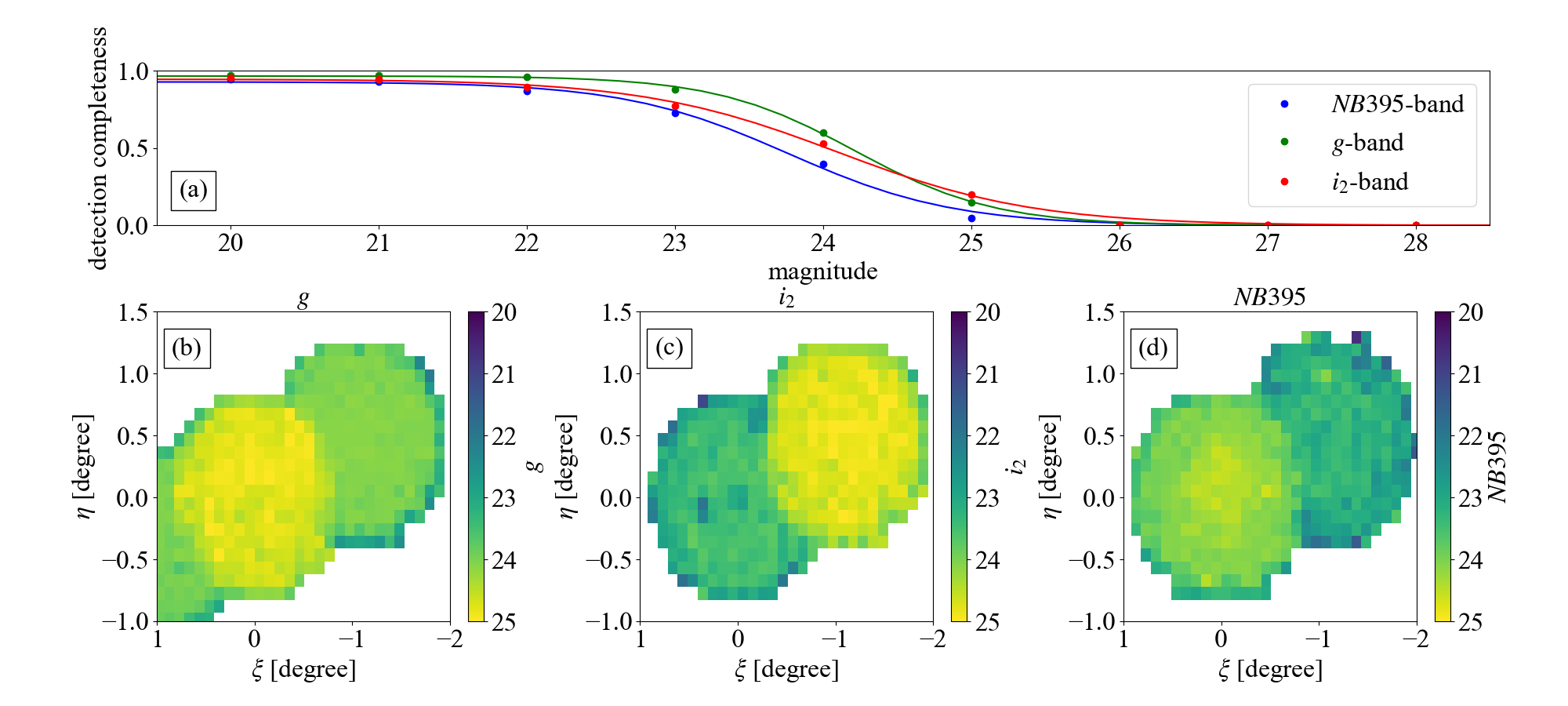}
 \end{center}
 \caption{Results of artificial star tests. Panel (a): the completeness functions for each band over the full analysis field. Panel (b-d): the spatial distribution of the 50\% completeness magnitude for each band.
  {Alt text: Figure showing the results of the artificial star tests. One panel shows the result in all our fields, the other three panels are the spatial information of 50\% detection completeness.}}\label{fig:DetComp}
\end{figure*}

To assess the limiting depth of the observations, we perform artificial star tests (ASTs) to estimate the detection completeness. The procedures follow the method in \citet{2025MNRAS.536..530O}. Briefly, for each band, we inject artificial point sources of given magnitudes into the stacked images using \texttt{injectStar.py} (\cite{2025MNRAS.536..530O}). These injected images are then processed with hscPipe, following the same detection and photometry steps used in our main analysis. Artificial stars are embedded every 300 pixels, with a total of $\sim 56,000$ (g-band) and $\sim 40,000$ (${\it i_2}$-band/{\it NB395}) injected sources per test.

The resulting detection completeness curves are presented in figure \ref{fig:DetComp}. The panel (a) shows the completeness functions for each band over the full analysis field, and the panels (b-d) display the spatial distribution of the 50\% completeness magnitude. In figure \ref{fig:DetComp}(a), red, blue, and green curves represent the results of fitting the estimated data points, shown as colored dots, with equation (6) of \citet{2016ApJ...833..167M}. These fitted curves indicate that our field-averaged 50\% completeness limits are 24.21, 24.10, and 23.77 mag for the {\it g} (green), {\it $i_2$} (red), and {\it NB395} (blue) bands, respectively. These limits are broadly consistent with the $5\sigma$ limiting magnitudes. These facts indicate that our dataset is deeper by approximately 2 mag compared with previous Pristine studies \citep{2017MNRAS.471.2587S} and is comparable to that of the DELVE (the DECam Local Volume Exploration Survey; \cite{2021ApJS..256....2D,2022ApJS..261...38D}) broadband observations used in the MAGIC survey (\cite{2025ApJ...993...77B}) for broadband data. Panels (b-d) of figure \ref{fig:DetComp} show that the detection completeness varies from field to field, so in the following analysis, we correct for these variations using this spatial completeness information. 

\subsection{Photometric calibration using Gaia XP data}\label{subsec:calibration}

\begin{figure*}
 \begin{center}
  \includegraphics[width=\linewidth, trim=30 0 30 0, clip]{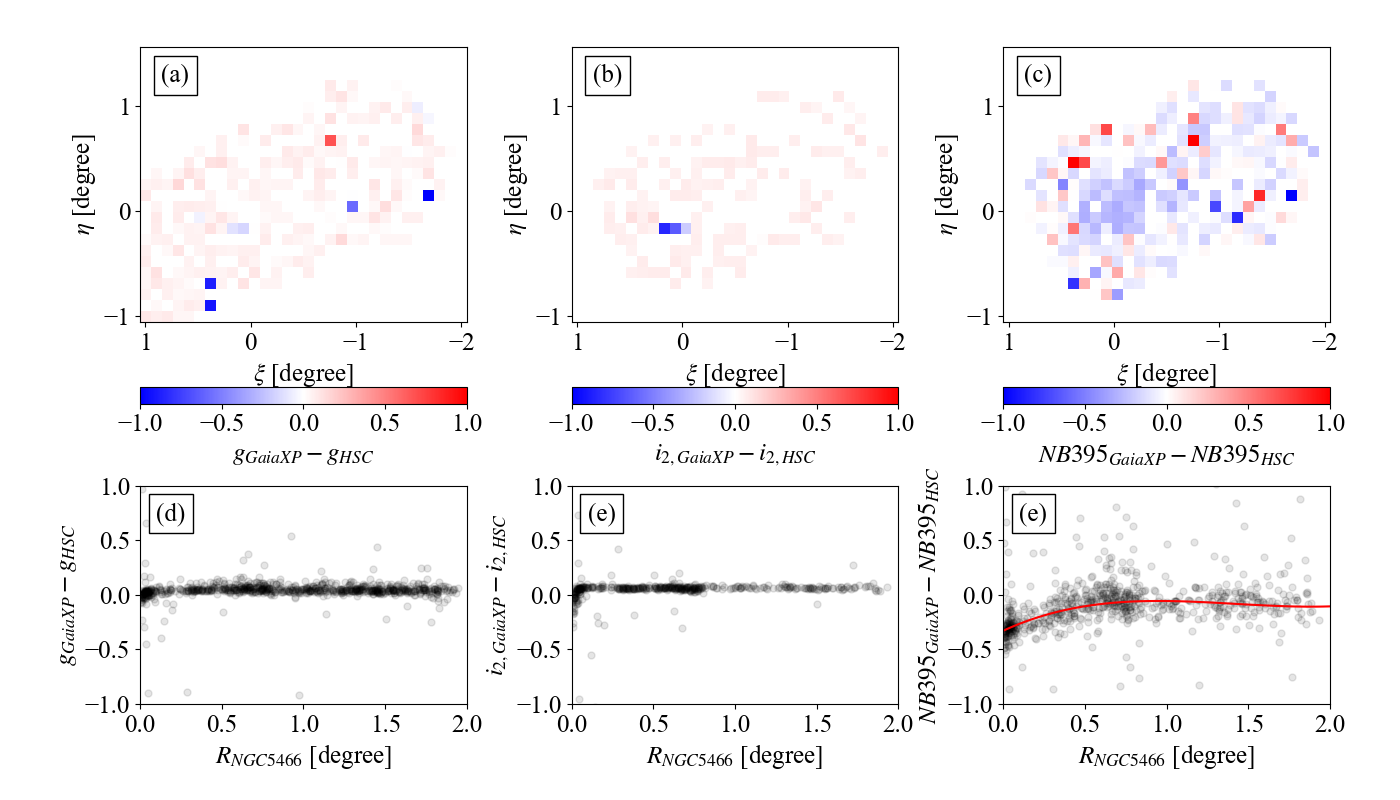}
 \end{center}
 \caption{Comparison between the synthesized Gaia XP magnitudes and our HSC measurements. Panels (a-c): spatial distributions of the median magnitude difference within 0.1 $\times$ 0.1 degree$^{2}$ bins. Panel (d-e): magnitude differences as a function of projected distance from the cluster center. In the panel (e), the red solid curve is a third-order polynomial function to correct our {\it NB395} magnitude.
  {Alt text: Figure comparing the Gaia XP magnitudes with our HSC measurements. The top three panels are spatial information of magnitude differences, and the bottom three panels are the radial information of magnitude differences.} }\label{fig:GaiaXP}
\end{figure*}

To assess the reliability of the photometry used in our analysis, we compare our measurements with existing external datasets. Because several of the HSC exposures were taken under suboptimal weather conditions and the transmission curve of {\it NB395} is located near the edge of the instrumental sensitivities of Subaru/HSC, we use Gaia DR3 BP/RP spectra (hereafter, XP spectra; \cite{2023A&A...674A...2D,2023A&A...674A...3M,2016A&A...595A...1G,2021A&A...649A...1G,2023A&A...674A...1G}) within our survey footprint to perform an independent photometric cross-check.

For each source with both HSC detections and Gaia XP spectra (the number of cross-matched objects is 1,743), Gaia XP-based magnitudes are synthesized by convolving the XP spectrum with the HSC instrumental sensitivity, atmospheric transmission, and the filter response curves, and integrating the resulting flux. Figure \ref{fig:GaiaXP} shows the comparison between these synthesized Gaia XP magnitudes and our HSC measurements. The top panel presents the spatial distribution of the median magnitude difference within 0.1 $\times$ 0.1 degree$^{2}$ bins, while the bottom panel shows the magnitude differences as a function of projected distance from the cluster center.

For the {\it g}- and ${\it i_2}$-bands, most of Gaia XP-based magnitudes are broadly consistent with the HSC magnitudes within $\sim 0.01$ mag, in panels (a), (b), (d), and (e) of figure \ref{fig:GaiaXP}. In panels (a) and (b), small offsets are visible in several regions. These correspond to areas where artifacts have caused a lack of data, so these regions are masked in the subsequent analysis. In contrast, the {\it NB395} data exhibit a systematic offset in the central region ($R < 0.5$ degree). To determine whether this offset originates from the HSC photometry or from the Gaia XP-based synthetic magnitudes, we compare datasets with theoretical predictions from BaSTi isochrones \citep{2018ApJ...856..125H,2021ApJ...908..102P,2022MNRAS.509.5197S,2024MNRAS.527.2065P} with 12.5 Gyr, $[{\rm Fe/H}] = -2.0$, $[{\rm \alpha/Fe}]=+0.4$. In the BaSTi database, isochrones are available for many photometric systems, including the Subaru/HSC photometric system. However, HSC/{\it NB395} isochrones are not publicly available, so we transform the CFHT/MegaCam {\it CaHK} magnitudes into HSC/{\it NB395}. The details of this photometric transformation are described in the Appendix \ref{appendix:PhtometricTransformation}. We convert the CFHT/MegaCam {\it CaHK} magnitudes to the Subaru/HSC {\it NB395} system based on equation (\ref{eq:NH-CM}) which introduces magnitude uncertainties at the level of $\sim 0.01$ mag.

\begin{figure}[!ht]
 \begin{center}
  \includegraphics[width=\linewidth, trim=0 0 0 0, clip]{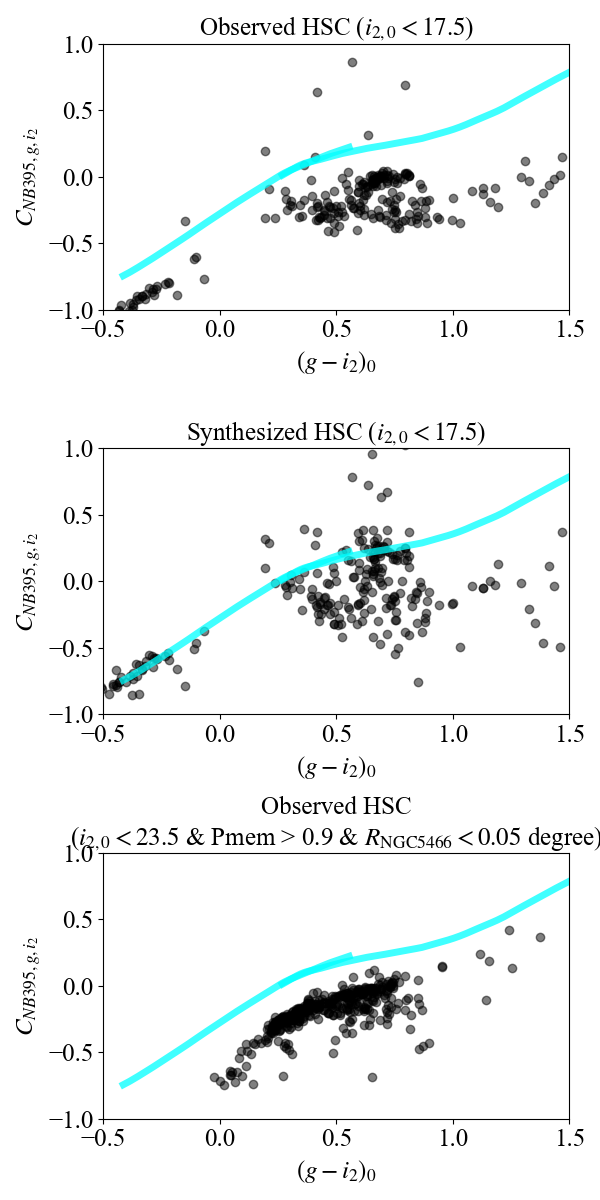}
 \end{center}
 \caption{Top panel: Color–color diagram of HSC observational data. The sample is restricted to sources with $i_{{\it 2},0}<17.5$ mag within 0.5 degree from the cluster center. The cyan solid curves correspond to BaSTI isochrones with $[{\rm Fe/H}]=-2.00$, Age$=12.5$ Gyr, and $[{\rm \alpha/Fe}]=+0.4$ Middle panel: The same as in the top panel, but for the observed HSC/{\it g}- and ${\it i_2}$-bands and the synthesized {\it NB395} magnitude from the Gaia XP spectra. Bottom panel: Color–color diagram of sources with membership probabilities greater than 0.9, as derived in Section \ref{subsec:selection}. The sample is restricted to objects with $i_{{\it 2},0}<23.5$ mag within 0.05 from the cluster center.
  {Alt text: Color–color diagrams illustrating the selection and properties of cluster stars; the top panel shows bright sources near the cluster center with BaSTI isochrones overplotted; the middle panel shows the same sample using HSC photometry combined with {\it NB395} magnitudes synthesized from Gaia XP spectra; and the bottom panel shows high-probability cluster members concentrated in the central region.} }\label{fig:HSC-XP}
\end{figure}

Figure \ref{fig:HSC-XP} shows the resulting color–color diagram for HSC observed stars and Gaia XP-synthesized stars. The top panel of figure \ref{fig:HSC-XP} shows the color-color diagram of HSC-observed objects cross-matched with Gaia XP within 0.5 degree of the center of NGC~5466 (with $i_{{\it 2},0}<17.5$). The middle panel presents the color-color diagram for the same sample, using {\it NB395} magnitudes predicted by the Gaia XP spectra. The bottom panel focuses on the color–color diagram of fainter objects ($i_{{\it 2},0} < 23.5$) located in the region of interest ($< 0.05$ degree), for which the membership probabilities derived in Section \ref{subsec:selection} exceed 0.9. In each panel, observational data are shown as black dots, and BaSTi isochrones are shown as cyan solid lines. Since more than half of the objects under comparison are bright stars ${\it i}_{{\it 2},0} < 17.5$ located near the cluster center ($<0.5$degree), the RGB and HB populations of NGC~5466 are expected to dominate in these datasets. Therefore, RGB and HB isochrones are plotted with the NGC~5466 mean metallicity of $[{\rm Fe/H}]\sim-2.00$ (\cite{1996AJ....112.1487H,2010arXiv1012.3224H}) in figure \ref{fig:HSC-XP}.

In top and bottom panels of figure \ref{fig:HSC-XP}, the observed HSC/{\it NB395} data deviate systematically from the $[{\rm Fe/H}] \sim -2$ model locus, whereas the Gaia XP-synthesized {\it NB395} magnitudes align well with the sequence with $[{\rm Fe/H}] = -2$. This comparison demonstrates that our {\it NB395} HSC photometry exhibits a systematic offset in the crowded central region of the cluster. It should be noted that some sources that deviate from the isochrones are unlikely to be associated with NGC~5466 and are therefore likely contaminants, in the top and middle panels.

Possible reasons for the magnitude offset in the HSC/{\it NB395} photometry are related to sky subtraction and data quality in our HSC data. In the hscPipe reduction, over-subtraction of the sky background around large objects ($> \sim$ 1 arcmin) has been reported (\cite{2018PASJ...70S...8A}). Although this issue has been partially mitigated by modifying the mesh grid size for the local sky subtraction and the background sky model (\cite{2019PASJ...71..114A,2022PASJ...74..247A}), it is possible that a similar over-subtraction problem re-emerged in our {\it NB395} images. In addition, our dataset includes exposures obtained under relatively poor seeing conditions. Therefore, to robustly assess the {\it NB395} photometric accuracy in crowded regions, it will be necessary to accumulate additional observations and perform comparisons across datasets in future work.

To remove the photometric offset in the {\it NB395} data, we apply a correction based on the Gaia XP-synthesized magnitudes. For the radial profile shown in panel (e) of figure \ref{fig:GaiaXP}, a third-order polynomial function is fitted to the magnitude differences between the Gaia XP synthetic magnitudes and our observed HSC magnitudes. The resulting correction function is overplotted as a red solid line in panel (e) of figure \ref{fig:GaiaXP}. We then apply this function to the {\it NB395} photometry as a function of radius. The corrected magnitudes are hereafter denoted as {\it NB395}$_c$.



\begin{figure*}[!t]
 \begin{center}
  \includegraphics[width=\linewidth]{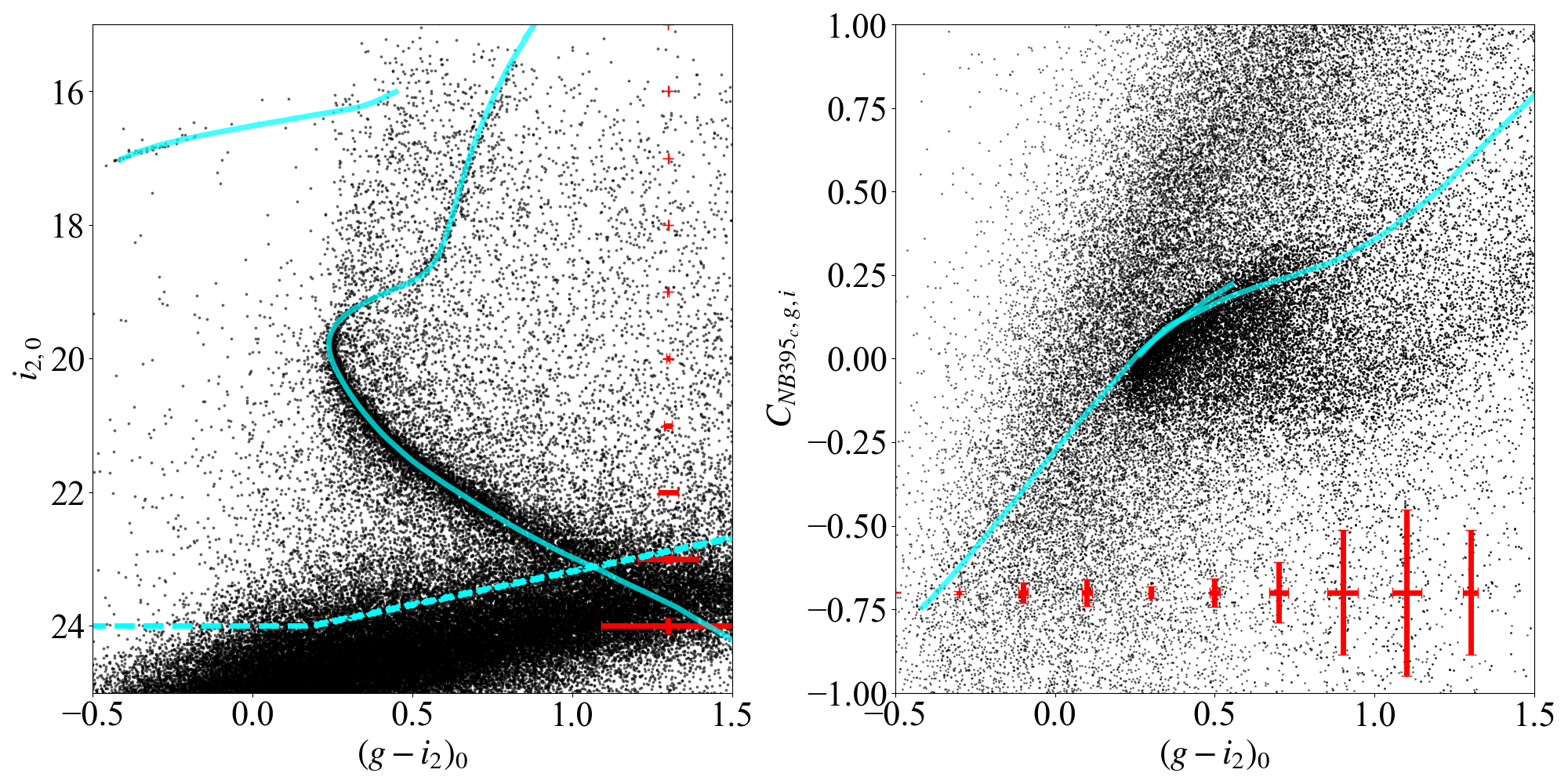}
 \end{center}
 \caption{Left: The color-magnitude diagram of point sources in the NGC~5466 field. Cyan solid lines are an isochrone curve with 12.5 Gyr, $[{\rm Fe/H}] = -2.0$, $[{\rm \alpha/Fe}]=+0.4$. The cyan dashed line corresponds to the 50 \% detection completeness, and red crosses indicate the photometric uncertainties. Right: The color-color diagram of point sources in the NGC~5466 field. Cyan solid lines and red crosses are the same as in the left panel.
  {Alt text: The color–magnitude diagram showing red giant branch (RGB), subgiant branch (SGB), and main-sequence (MS) stars as a dense sequence extending from the upper right to the lower right. The color-color diagram showing giant and MS stars as a dense clump.}}\label{fig:CMD_CCD}
\end{figure*}

\section{Methods}\label{sec:methods}
\subsection{Color-magnitude and color-color diagrams}\label{subsec:diagrams}

Figure \ref{fig:CMD_CCD} is the extinction-corrected CMD and color-color diagram of 84,257 cataloged point sources in our survey fields. In this CMD, photometric error in each band, which is the mean value of every 1 mag bin, is calculated, and the errors in the $i_{2,0}$ magnitude and color (calculated for $-0.5 < (g - i_2)_0 < 1.5$) are plotted as red crosses in this figure. The 50 \% completeness limit, which is the mean value over the survey field, is plotted as a cyan dashed line. Cyan solid lines indicate a BaSTI isochrone curve with 12.5 Gyr, $[{\rm Fe/H}] = -2.0$, $[{\rm \alpha/Fe}]=+0.4$, assuming the distance to NGC~5466 (16 kpc).

In this CMD, several characteristic stellar populations are identified. The stars associated with NGC~5466 occupy a thin sequence extending from the RGB at $(({\it g} - {\it i_2})_0,~{\it i}_{{\it 2},0}) \sim (0.8,~15.5)$, through the main-sequence turnoff (MSTO) at $(({\it g} - {\it i_2})_0,~{\it i}_{{\it 2},0}) \sim (0.3,~19.5)$, down to the MS at $(({\it g} - {\it i_2})_0,~{\it i}_{{\it 2},0}) \sim (1.0,~23)$. In addition, the HB stars of NGC~5466 appear at $(({\it g} - {\it i_2})_0,~{\it i}_{{\it 2},0}) \sim (-0.3,~17)$. Stars that are widely distributed across the CMD at $({\it g} - {\it i_2})_0>0$ correspond primarily to Galactic disk and halo stars, whereas objects dominating the CMD at $i_{2,0}>24$ are unresolved background galaxies. Below the 50\% completeness limit, photometric uncertainties increase significantly, making it difficult to distinguish the characteristic features of NGC~5466 and background galaxies.

The right panel of figure \ref{fig:CMD_CCD} presents the color–color diagram for NGC~5466. The red crosses indicate representative photometric uncertainties for sources with $-1.0<C_{\it NB395_c,g,i_2}<1.0$, sampled over $({\it g-i})=-0.5$ to $1.5$ at regular intervals. The cyan curve corresponds to a $12.5$ Gyr, $[{\rm Fe/H}]=-2.0$ BaSTI isochrone, which is transformed to the HSC photometric system.

In the right panel of figure \ref{fig:CMD_CCD}, the sequence extending from $(({\it g-i_2})_0,~C_{{\it NB395_c,g,i_2}}) \sim (0.25,~-0.75)$, to $(({\it g-i_2})_0,~C_{{\it NB395_c,g,i_2}}) \sim (0.70,~0.25)$ represents the metal-poor member stars of NGC~5466. More metal-rich Milky Way disk stars lie below this sequence, whereas more metal-poor halo stars appear above it. The sequence extending from $(({\it g-i_2})_0,~C_{{\it NB395_c,g,i_2}}) \sim (0.00,~0.00)$, to $(({\it g-i_2})_0,~C_{{\it NB395_c,g,i_2}}) \sim (0.75,~0.75)$, which overlaps with Galactic metal-poor halo stars, shows the unresolved background galaxies, and the number of these background objects is decreased when we cut faint objects ($i_{2,0} > 24.0$). In the following analysis, we use both the CMD and the color–color diagram to identify and select NGC~5466 member stars.

\subsection{Selection of NGC~5466 stars}\label{subsec:selection}
To securely identify member stars of NGC~5466, we apply a k-nearest-neighbour (kNN) method (\cite{1967ITIT...13...21C}) in color–color–magnitude diagram (CCMD). The kNN algorithm is a widely used clustering and classification technique, and its variants have been successfully adopted to select GC members on the basis of colors, magnitudes, proper motions, and parallaxes (e.g., \cite{2020IAUS..351..516S,2025MNRAS.537.2752K,2025AJ....170..157K}). In this study, we refine and extend the methodology of \citet{2025MNRAS.537.2752K}, hereafter K25a, by using the photometric parameters $C_{\it NB395_c, g, i_2}$, $({\it g-i_2})_0$, and ${\it i_{2}}$-band magnitude to estimate membership probabilities for stars associated with NGC~5466 without Gaia astrometry. The procedure is outlined below.

\begin{figure}
 \begin{center}
  \includegraphics[width=\linewidth, trim=0 0 0 0, clip]{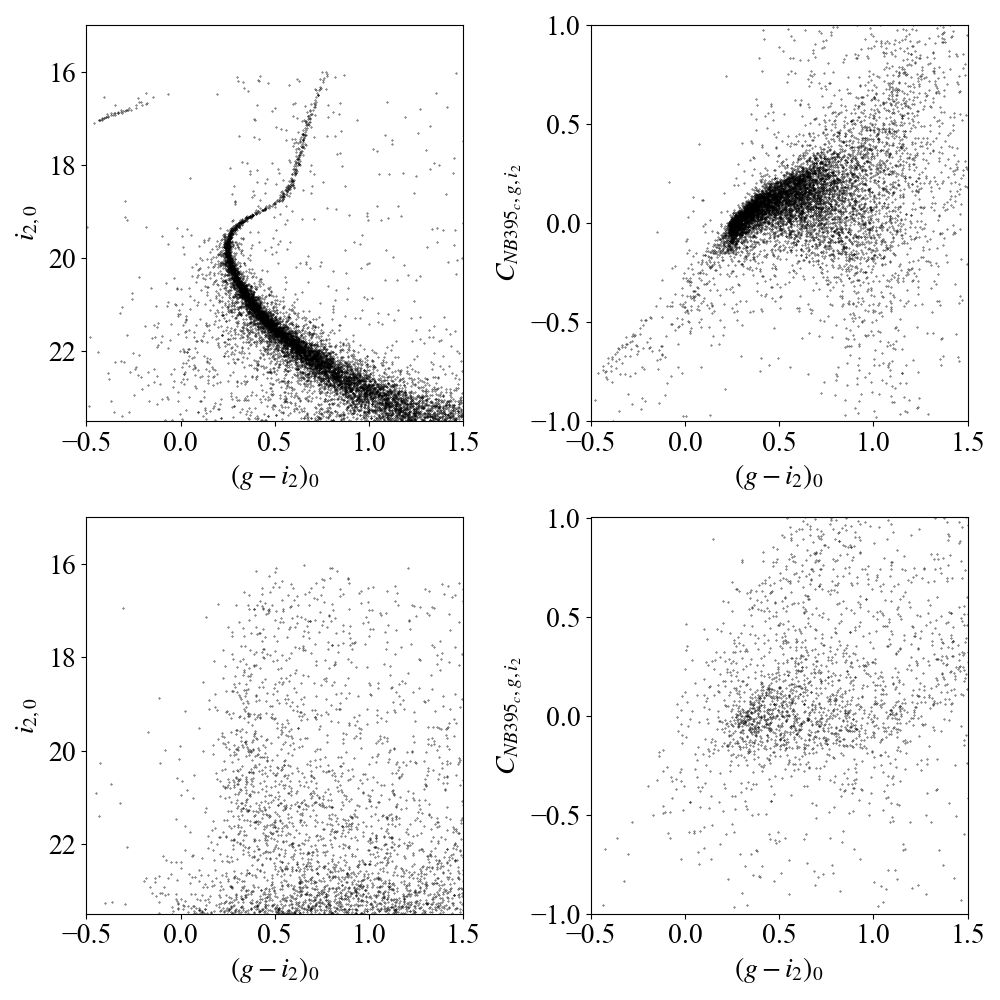}
 \end{center}
 \caption{CMDs and color–color diagrams for the reference samples. The top rows correspond to the cluster sample, while the bottom rows correspond to the field sample.
  {Alt text: CMDs and color–color diagrams of the reference samples, with the top panels showing the cluster sample and the bottom panels showing the field sample.} }\label{fig:sample}
\end{figure}

We first construct two reference samples, hereafter the ``cluster'' ($_{\rm cl}$) and ``field'' ($_{\rm fld}$) sets. The ``cluster'' reference sample is intended to represent the intrinsic properties of NGC~5466 and is defined using stars located within the nominal tidal radius of the cluster (48 arcmin; \cite{2014ApJ...793..110M}). Conversely, the ``field'' reference sample is constructed from all stars located beyond a radial distance of 2 degree from the cluster center.

Figure \ref{fig:sample} shows the comparison of the distributions of the two reference sets with $16 < i_{2,0} < 23.5$ in each CCMD, clearly illustrating that the ``cluster'' population (top panels) is well separated from the ``field'' (bottom panels). No plausible cluster members occupy the region with $({\it g-i_2})_0>2$. Therefore, objects falling in this color range not expected to contain NGC~5466 stars are excluded from the analysis. We restrict the parameter space to $-0.5<({\it g-i_2})_0<2$ and $16<i_{{\it 2},0}<23.5$, and apply the probability calculations only within this domain.

Using the two reference catalogs, we then apply the kNN algorithm in the CCMD. For any given star in the full sample, we estimate the local density relative to each reference set by computing the distances to the $k=10$ nearest neighbours drawn separately from the ``cluster'' and ``field'' reference samples. Then, it is important to account for the relative standard deviation in magnitude (i.e., the spread in the ${\it i_{{\it 2},0}}$ magnitude distribution within the CCMD) is typically larger than those in colors, since an unscaled Euclidean metric would be dominated by $\Delta_{{\it i}_{{\it 2},0}}$ due to its large scale and would fail to capture the intrinsic structure. To properly weight these differences, we compute the mean ratio of colors to magnitude deviations in the ``cluster'' reference sample. This ratio is found to be $\Delta_{\it C_{NB395_c,g,i_2}} : \Delta_{{\it (g-i_2)}_0} : \Delta_{{\it i}_{{\it 2},0}} = 0.9 : 1.1 : 6.2$ in the range of $-1.0 < C_{\it NB395_c, g, i_2} < 1.0$, $-0.5 < (g - i_2)_0 < 2$, and $16 < i_{2,0} < 23.5$. We then adopt this ratio as the metric scaling parameter in the CCMD local-density estimator. The resulting probability expression for each star therefore takes the following form:
\begin{align}
& \ln P_{\rm CCMD} = \nonumber \\
& \ln \!\left(
\frac{10}{\tfrac{4\pi}{3}
\Bigl(
 \bigl( \tfrac{\Delta_{\it C_{NB395_c,g,i_2}}}{0.9} \bigr)^{2}
 + \bigl( \tfrac{\Delta_{{\it (g-i_2)}_0}}{1.1} \bigr)^{2}
 + \bigl( \tfrac{\Delta_{{\it i}_{{\it 2},0}}}{6.2} \bigr)^{2}
\Bigr)^{3/2}}
\right) \nonumber \\
&\quad - \ln (N)~.
\label{eq:log_prob}
\end{align}
\noindent
In this expression,  $\Delta_{\it C_{NB395_c,g,i_2}}$, $\Delta_{\it (g-i_2)_0}$, and $\Delta_{{\it i}_{{\it 2},0}}$ denote the differences in color and magnitude between a target star and its 10th nearest neighbour, while $N$ represents the total number of stars in each reference sample. Using each reference sample and this equation, we derive the probabilities $P_{\rm CCMD}^{\rm cl}$ and $P_{\rm CCMD}^{\rm fld}$, which represent the probabilities that each star is a ``cluster'' member or a ``field'' star, respectively.

Using the membership probabilities defined above, we determine the relative fraction of cluster stars in the observed field stars, $f_{\rm cl}$, by adopting the following likelihood function:
\begin{align}
\mathcal{L} = f_{\rm cl}P_{\rm CCMD}^{\rm cl} + (1-f_{\rm cl})P_{\rm CCMD}^{\rm fld}~.
\label{eq:frac_likelihood}
\end{align}
\noindent
Once $f_{\rm cl}$ is determined, we assign to each star a membership probability, $P_{\rm mem}$, ranging from 0 to 1. A value of $P_{\rm mem} = 1$ indicates that the star is most likely a member of NGC~5466, whereas $P_{\rm mem} = 0$ implies that its position in CCMD is fully consistent with that of the field population. The membership probability is computed following:
\begin{align}
P_{\rm mem} = \frac{f_{\rm cl}P_{\rm CCMD}^{\rm cl}}{\mathcal{L}}~.
\label{eq:prob}
\end{align}

Using these expressions, we compute the membership probability for every star in the sample. For equation (\ref{eq:frac_likelihood}), we evaluate fraction $f_{\rm cl}$ using Markov Chain Monte Carlo (MCMC) sampling with \texttt{python/emcee} package (\cite{2013PASP..125..306F}). We adopt non-informative priors and employ 100 walkers with 110,000 steps, discarding the first 100,000 steps as burn-in. The resulting posterior distribution yields $f_{\rm cl} = 0.0470_{-0.005}^{+0.005}$. This random uncertainty, and all subsequent uncertainties from our MCMC fitting, correspond to the 68\% Bayesian credible interval of the posterior distribution. The full posterior distributions obtained from the MCMC samples are presented in Appendix \ref{appendix:MCMC_Results}.

\section{Results}\label{sec:results}

In this section, we present the derived properties of NGC~5466 based on the membership probabilities established in section \ref{subsec:selection}. In section \ref{subsec:metallicity}, we derive photometric metallicities using the metallicity-sensitive {\it NB395} filter and assess the level of potential contamination remaining in our sample. In section \ref{subsec:SpatialDistribution}, we examine the spatial distribution of member stars and their extent. Finally, in section \ref{subsec:MF}, we present MF derived from low-mass MS stars, extending the analysis to regions beyond the tidal radius.

\subsection{Metallicity distribution}\label{subsec:metallicity}

The {\it NB395} filter covers absorption features that are sensitive to stellar metallicity, enabling estimates of the photometric metallicities of member stars. To evaluate the accuracy of metallicities inferred from {\it NB395}, CMD-based metallicities are also independently derived for comparison. Photometric metallicities based on {\it NB395} are estimated by comparing the observed magnitudes with predictions from BaSTi isochrones.

\begin{figure}[h]
 \begin{center}
  \includegraphics[width=\linewidth, trim=0 0 0 0, clip]{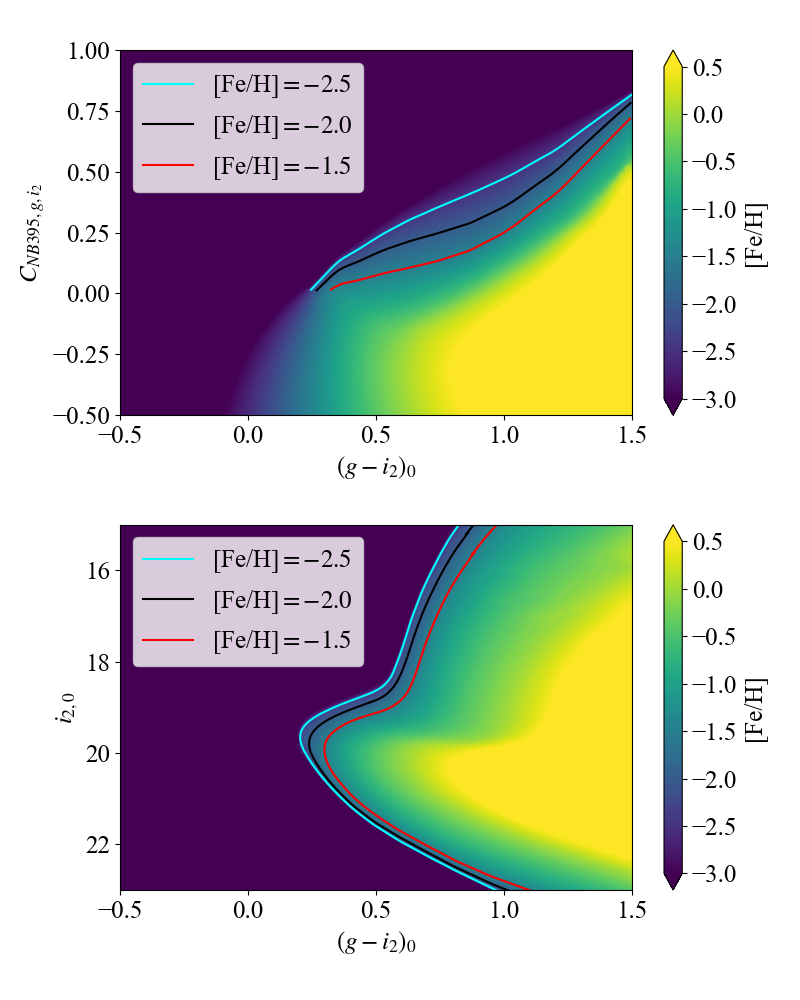}
 \end{center}
 \caption{Interpolated metallicity map on the color-color diagram (top panel) and CMD (bottom panel) for NGC~5466. Cyan, black, red lines show the isochrones with $[{\rm Fe/H}] = -2.5,~-2.0,~-1.5$, Age$=12.5$ Gyr and $[{\rm \alpha/Fe}]=+0.4$. 
  {Alt text: Interpolated metallicity maps for NGC 5466 shown on a color–magnitude diagram (top) and a color–color diagram (bottom), with overplotted isochrones at $[{\rm Fe/H}] = -2.5$ (cyan), $-2.0$ (black), and $-1.5$ (red) for an age of $12.5$ Gyr and $[{\rm \alpha/Fe}] = +0.4$.} }\label{fig:MD_RBF}
\end{figure}

To estimate the photometric metallicity for each star, we construct 31 isochrones with $-3.0 \leq [{\rm Fe/H}] \leq 0.0$ in 0.1 dex intervals, assuming that the age is $12.5$ Gyr (e.g., \cite{2013ApJ...775..134V}) and an alpha-enhanced system ($[{\rm \alpha/Fe]} = +0.4$). It should be noted that, for a given metallicity, the $C_{\it NB395_c, g, i_2}$ color index exhibits systematic offsets as a function of surface gravity, $\log {g}$. Since the majority of our targets are MS stars with high surface gravity, photometric metallicities are estimated using only MS isochrones.

Using this isochrone set, we construct the metallicity model on the color-color diagram by using the radial basis function (RBF) interpolation of the \texttt{python/scipy} package. The top panel of figure \ref{fig:MD_RBF} shows the interpolated metallicity model. The metallicity of each star at a given position $(({\it g - i_2})_0, C_{\it NB395_c, g, i_2})$ on the color-color diagram is estimated by interpolating $[{\rm Fe/H}]$ at that position in the metallicity model. By this method, we estimate the photometric metallicity of all point sources.

\begin{figure}[!ht]
 \begin{center}
  \includegraphics[width=0.95\linewidth, trim=0 0 10 0, clip]{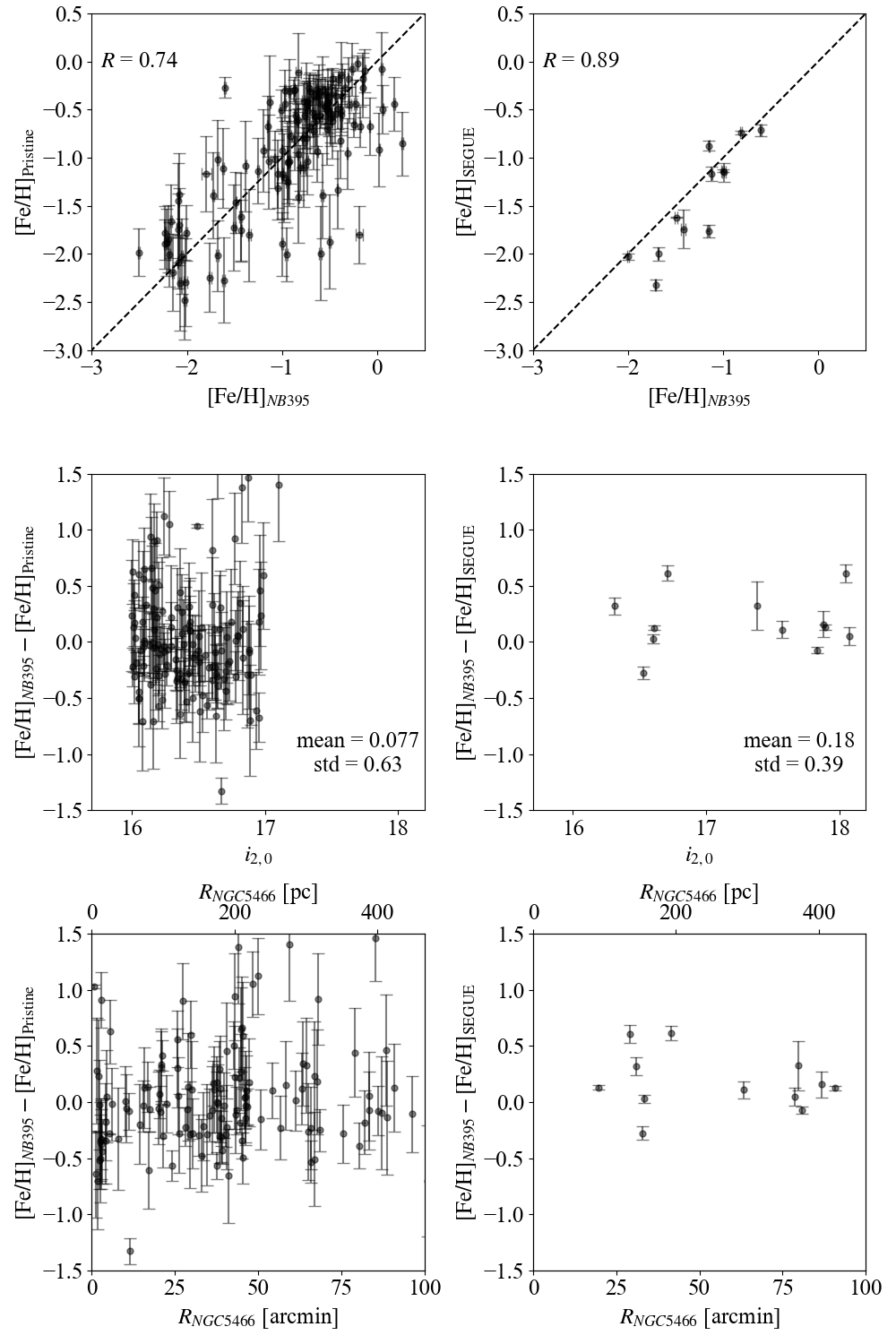}
 \end{center}
 \caption{Comparison of metallicities between HSC/{\it NB395} and Pristine/{\it CaHK} (left column), and between HSC/{\it NB395} and SEGUE spectroscopic metallicity (right column). (Top) Direct comparison of metallicities between our {\it NB395}-based metallicity and the reference values. In the upper-left corner, the correlation coefficient, $R$, for each dataset is shown. (Middle) Differences between our {\it NB395}-based metallicity and the reference values as a function of the HSC ${\it i}_{{\it 2},0}$-band magnitude. Mean and standard deviation values of the metallicity differences are described in the lower-left corner. (Bottom) Metallicity differences as a function of projected distance from the cluster center. 
  {Alt text: Comparisons of {\it NB395}-based metallicities with Pristine and SEGUE reference values. Top panels show metallicity differences versus ${{\it i}_{{\it 2},0}}$ magnitude, including mean and standard deviation. Bottom panels show metallicity differences versus projected distance from the cluster center.} }\label{fig:MetallicityComparison}
\end{figure}

The estimation of photometric metallicities from the CMD follows the same methodology as described in \cite{2025MNRAS.536..530O}. Briefly, we construct a metallicity model on the CMD by preparing a grid of theoretical isochrones and applying RBF interpolation across the grid. The metallicity of each star is then inferred by comparing its CMD position with this interpolated model shown in the bottom panel of figure \ref{fig:MD_RBF}. In this study, we adopt isochrones with an age of 12.5 Gyr and $[{\rm \alpha/Fe}]=+0.4$, spanning a metallicity range of $-3.0<[{\rm Fe/H}]<+0.5$ in steps of 0.1 dex. These isochrones are used to construct the final metallicity model applied to our dataset.

\begin{figure*}
 \begin{center}
  \includegraphics[width=0.9\linewidth, trim=0 0 10 0]{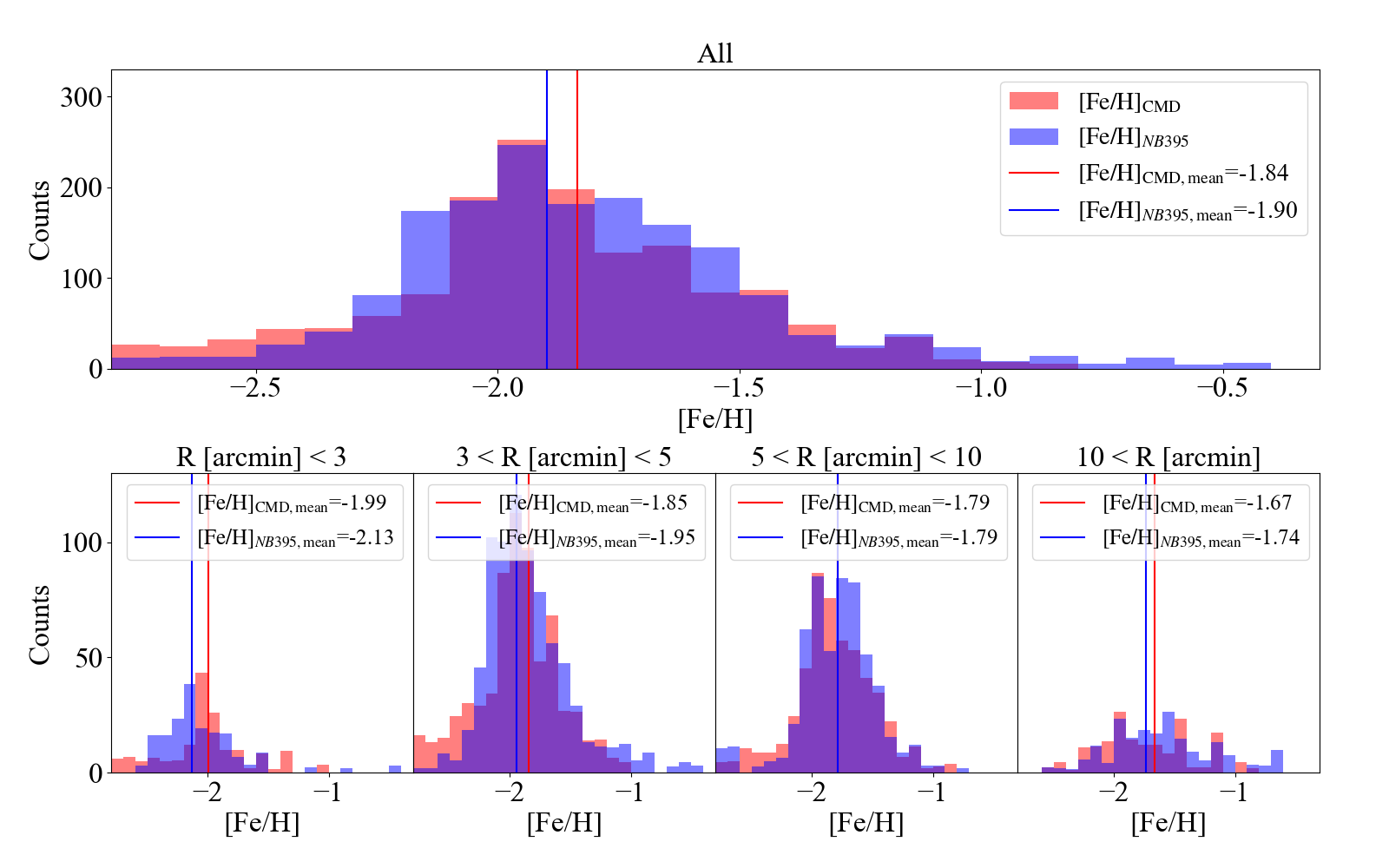}
 \end{center}
 \caption{Metallicity distributions of stars with membership probability $P_{\rm mem} > 0.99$. The upper panel is the metallicity distribution in the all field. The lower panels are the distributions at different radial distance from the cluster center. Red histograms correspond to the photometric metallicities derived from the CMD, and the blue ones represent the metallicities obtained from the color-color diagram. The vertical lines indicate the mean metallicity of each distribution.
  {Alt text: Figure presents the metallicity distributions of NGC~5466. The upper panel shows the distribution for the full field, while the lower panels show distributions in separate radial bins. Red histograms indicate photometric metallicities derived from the CMD, and blue histograms show metallicities from the color-color diagram. Vertical lines mark the mean metallicity in each panel.} }\label{fig:MD}
\end{figure*}

\begin{figure*}
 \begin{center}
  \includegraphics[width=0.9\linewidth, trim=0 0 10 0, clip]{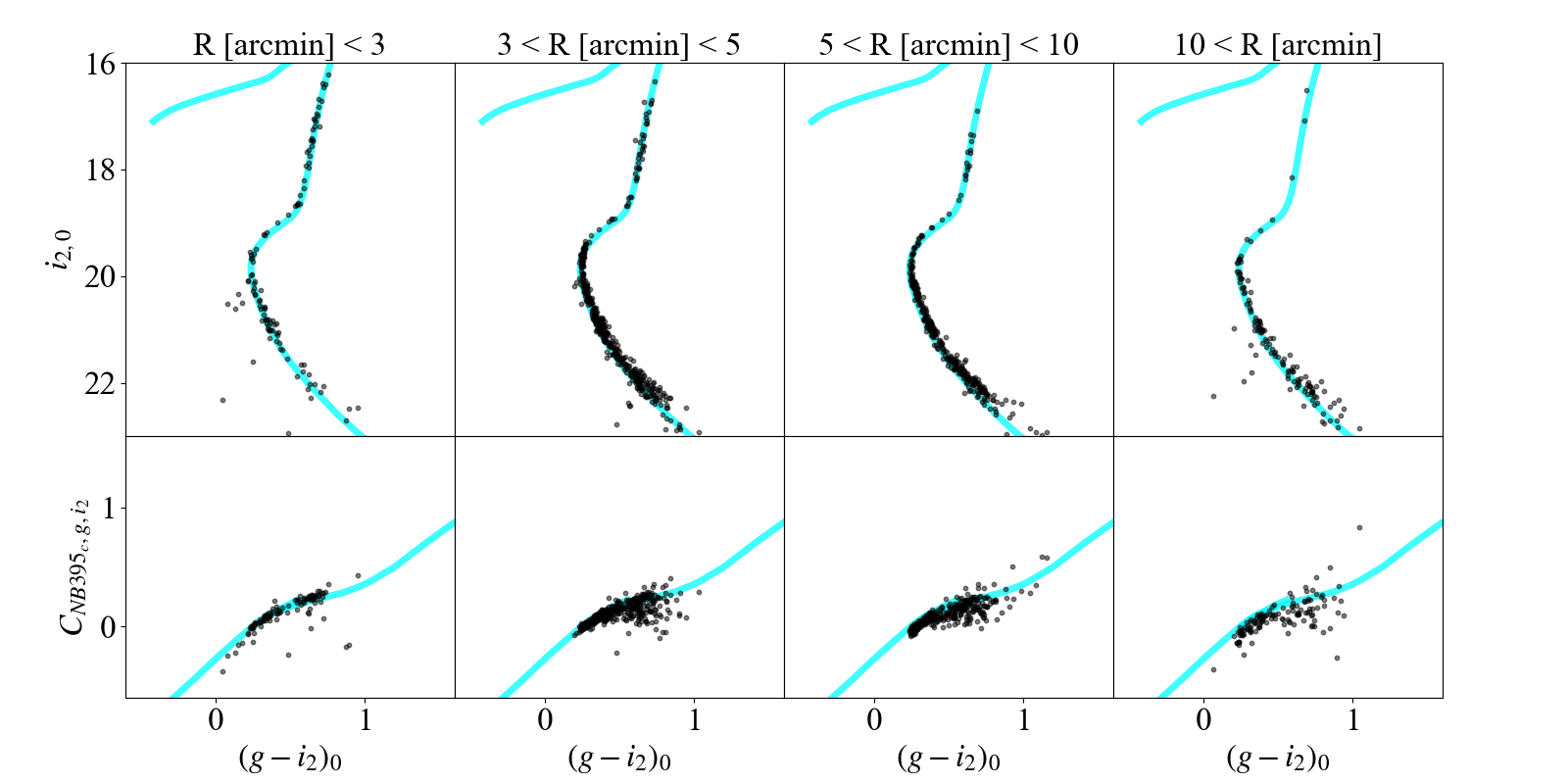}
 \end{center}
 \caption{CMDs (top panels) and color–color diagrams (bottom panels) in different radial bins from the cluster center. Black points denote sources with membership probabilities of $P_{\rm mem} > 0.99$, and the cyan solid lines show the NGC~5466-like isochrone.
  {Alt text: Figure shows CMDs (top panels) and color–color diagrams (bottom panels) in several radial bins from the cluster center. Black points indicate stars with membership probability $P_{\rm mem}>0.99$. Cyan solid lines overlaid on each panel represent an NGC~5466-like isochrone used as a reference.} }\label{fig:CMD_CCD_RadialProfile}
\end{figure*}

To validate the reliability of our metallicity estimation based on {\it NB395} ($[{\rm Fe/H}]_{\it NB395}$), we first compare our results with previous observational data. As reference datasets, we adopt the photometric metallicities inferred from the Pristine–Gaia synthetic catalog ($[{\rm Fe/H}]_{\rm Pristine}$; \cite{2024A&A...692A.115M}) based on the CFHT/{\it CaHK} filter, as well as the spectroscopic metallicities from the SEGUE observation ($[{\rm Fe/H}]_{\rm SEGUE}$; \cite{2022MNRAS.513..853Y}). Our catalog is cross-matched with these reference datasets under the condition that the position of each star is $< 1.0$ arcsec. The comparison is presented in figure \ref{fig:MetallicityComparison}. The left column in figure \ref{fig:MetallicityComparison} shows the comparison with Pristine, while the right column corresponds to the comparison with SEGUE. The top panels show the direct comparisons of metallicity measurements between the different datasets. In the upper-left corner of the top panels, the correlation coefficient $R$ is represented, and our {\it NB395}-based metallicites are agreement with other reference datasets. The middle panels show the differences between our {\it NB395}-based metallicity and the reference values as a function of the ${\it i}_{{\it 2},0}$ magnitude, while the bottom panels show the metallicity differences as a function of projected distance from the cluster center. Only stars with $0<(g-{\it i}_{{\it 2}})_0<2$ and metallicity uncertainties smaller than 0.5 dex in both our measurements and the reference datasets are plotted. For reference, the mean and standard deviation of the metallicity differences are indicated in the lower-left corner of each middle panel.

Figure \ref{fig:MetallicityComparison} demonstrates that our {\it NB395}-based metallicities are broadly consistent with both the Pristine and SEGUE measurements across wide metallicity, magnitude, and radial ranges. In the middle-right and bottom-right panels of figure \ref{fig:MetallicityComparison}, small offsets relative to the SEGUE metallicities are apparent. This discrepancies are likely attributable to the fact that all our samples are predominantly RGB stars (${\it i}_{{\it 2},0}<18$), whereas {\it NB395} metallicities, $[{\rm Fe/H}]_{\it NB395}$, are estimated under the assumption that the targets are MS stars, which constitute the primary science targets of this study. Indeed, the Pristine metallicities - derived under an MS star assumption - show no significant offset relative to our results, although the standard deviations are larger due to the large uncertainties of Pristine metallicities. Therefore, figure \ref{fig:MetallicityComparison} indicates that, provided the assumed stellar parameters are appropriate, metallicities can be accurately estimated using HSC/{\it NB395} photometry.

Figure \ref{fig:MD} presents metallicity distributions of stars with membership probability $P_{\rm mem}>0.99$, which are defined as secure member stars. These secure member stars minimize contamination with low membership probabilities and ensure that their cumulative contribution does not alter the shape of the metallicity distribution. The top panel shows the metallicity distribution obtained using all available data, while the bottom panels display the distributions in different radial intervals from the cluster center. The red histograms correspond to the photometric metallicities derived from the CMD, and the blue histograms represent the metallicities obtained from the color-color diagram. The vertical lines indicate the mean metallicity of each distribution. All histograms are corrected by the detection completeness. 

Overall, the metallicity distributions derived from {\it NB395} photometry and from the CMD exhibit broadly similar shapes. In the bottom panel, the peaks of the metallicity distributions derived from the CMD and the color-color diagram shift toward more metal-rich values than the established metallicity of NGC~5466 ($[{\rm Fe/H}] \sim -2.0$) at larger radii, despite the assumption in our membership determination that NGC~5466 member stars share a common metallicity. 

One possible origin of such a metallicity gradient is the effect of multiple stellar populations (MPs), which appear to be ubiquitous in Galactic GCs. In most GCs, MPs do not appear as an intrinsic spread in [Fe/H], but primarily as systematic variations in light elements (C-N, O-Na, Mg-Al) and helium abundance (e.g., \cite{2017MNRAS.464.3636M,2019MNRAS.487.3815M}). In particular, enhanced nitrogen (and depleted carbon) alters the strength of molecular bands such as CN, NH, and CH in stellar atmospheres, producing systematic differences in the spectral energy distribution at ultraviolet and blue wavelengths. A previous synthetic spectral study showed that a typical second-generation (2G) C-N and O-Na abundance patterns can significantly modify the flux distribution at wavelengths shorter than $\sim$400 nm, providing a key physical basis for the prominent photometric signatures of MPs in the ultraviolet/blue regime (\cite{2011A&A...534A...9S}). Moreover, the spatial distributions of the 2G (N-rich/C-poor) and first-generation (1G; N-poor/C-rich) populations often differ, and tidal stripping preferentially samples stars from the cluster outskirts; therefore, the relative contribution of 1G and 2G stars may change in the tidal stream compared to the cluster body (\cite{2011A&A...525A.114L,2012ApJ...744...58M}). In this case, differences in CNO (especially N) and He between 1G and 2G stars can modify the observed UV/blue flux through CN/CH/NH absorption, potentially shifting the zero-point of {\it NB395}-based photometric metallicity proxies (\cite{2011A&A...525A.114L,2017MNRAS.464.3636M}). Consequently, the apparently more metal-rich [Fe/H] inferred in the outer regions or in the stream may not directly reflect a true iron abundance difference, but instead represent an apparent gradient caused by population mixing effects being projected onto the photometric metallicity estimates.

As an alternative explanation for an apparent metallicity gradient, contamination from non-members should also be considered. While MPs have been widely detected in globular clusters, they are generally characterized by anti-correlations in light elements rather than by genuine [Fe/H] bimodality. A small number of massive globular clusters have been reported to show bimodal or broadened metallicity distributions; however, such behavior has not been established for clusters in the same mass range as NGC 5466. Therefore, because the surface density of true members decreases with increasing radius, the relative fraction of residual metal-rich contaminants may increase in the outskirts and along the stream, potentially producing an apparent metallicity gradient in the observed sample.

Figure \ref{fig:CMD_CCD_RadialProfile} presents the CMDs and color–color diagrams of stars with $P_{\rm mem} > 0.99$ in each radial bin, corresponding to the same bins shown in the bottom panels of figure \ref{fig:MD}. In each panel, the cyan lines indicate isochrones matched to the properties of NGC 5466 (12.5 Gyr, $[{\rm Fe/H}] = -2.00$, $[{\rm \alpha/Fe}]=+0.4$), and the majority of stars are distributed along these isochrones. In the outer regions, a slight broadening of the sequence is visible, particularly in the color–color diagrams, despite most stars remaining closely aligned with the isochrones. This offset might be the photometric offset from multiple stellar populations of light-abundance elements. In the metallicity distribution constructed from the full field (see top panel of figure \ref{fig:MD}), the mean metallicity appears more metal-rich than previously reported values for NGC~5466 ($[{\rm Fe/H}] \sim -2.0$; \cite{1996AJ....112.1487H,2015MNRAS.448...42L,2024RAA....24f5014G}), whereas the distribution measured near the cluster center yields a mean metallicity fully consistent with previous studies. Therefore, our metallicity estimation based on {\it NB395} works well, and future large-scale spectroscopic follow-up observations, such as the Subaru/Prime Focus Spectrograph (PFS or `\=Onohi`ula in its Hawaiian name; \cite{2014PASJ...66R...1T}) will be essential for clarifying the interpretation of this result.

We note that projection effects may dilute the intrinsic metallicity gradient, since stars located at larger three-dimensional radii can be projected onto the inner regions. Therefore, the true gradient may be slightly steeper than observed.

\subsection{Spatial distribution}\label{subsec:SpatialDistribution}

\begin{figure*}
 \begin{center}
  \includegraphics[width=\linewidth, trim=0 0 10 0, clip]{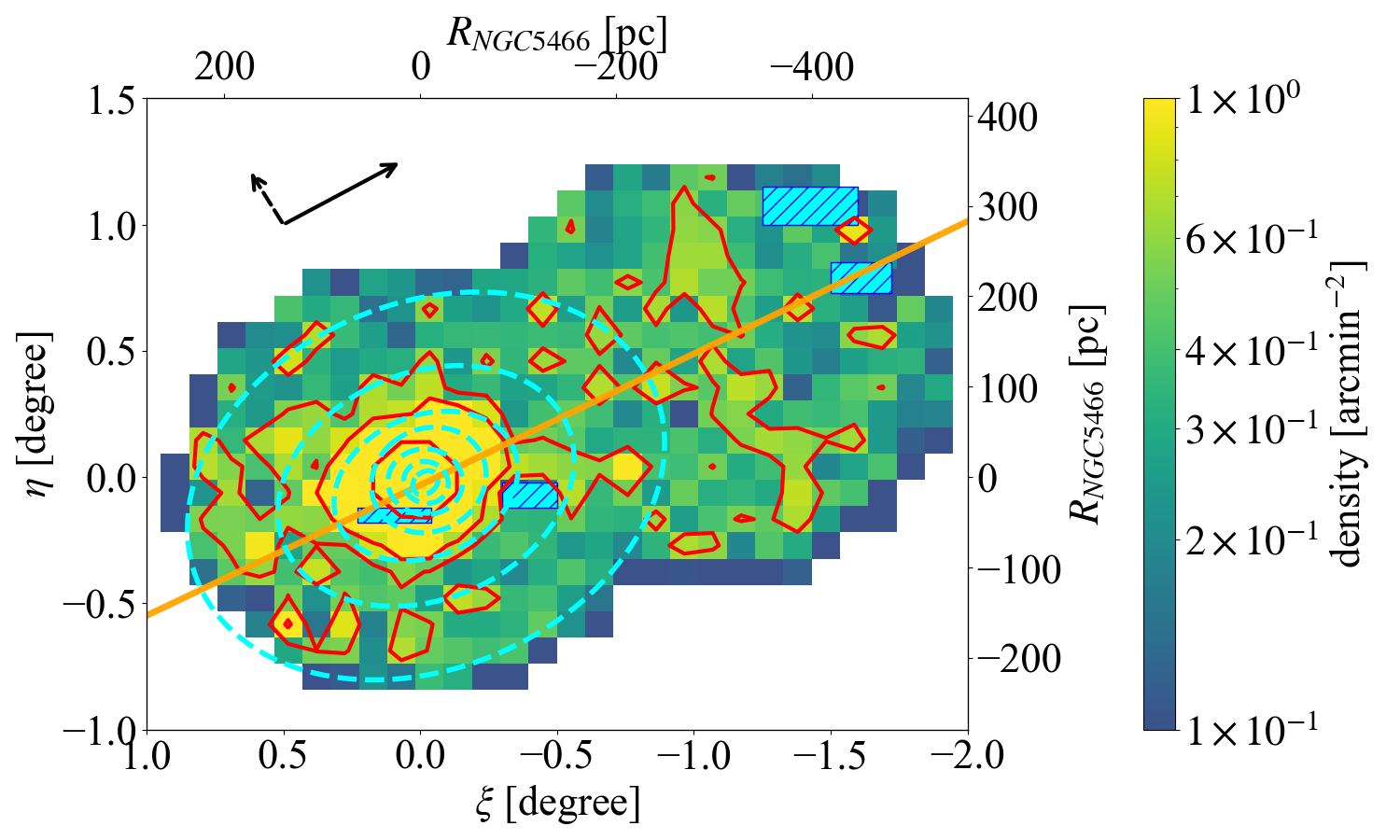}
 \end{center}
 \caption{Surface density distribution of NGC~5466 member stars. The red contours correspond to surface density levels of 0.5, 1, and 10 arcmin$^{-2}$. The orange line represents the stream ridge line obtained from \texttt{galstreams}. The cyan dashed ellipses indicate the spatial extent and ellipticity of the member stars within each isophotal radius. The regions masked by cyan and blue rectangles correspond to areas affected by chip gaps or artifacts, where data are missing. The black arrows in the upper-left corner indicate the directions of the major-axis (aligned with the stream; solid arrow) and the minor-axis (perpendicular to the stream; dashed arrow), as estimated from the \texttt{python/isophote}.
  {Alt text: Surface density map of NGC~5466 member stars showing density contours, the stream ridge line, isophotal ellipses, and masked regions affected by chip gaps or artifacts.} }\label{fig:SurfceDensityMap}
\end{figure*}

\begin{figure}
 \begin{center}
 \vspace{-5mm}
 \includegraphics[width=0.9\linewidth, trim=0 -20 0 -10, clip]{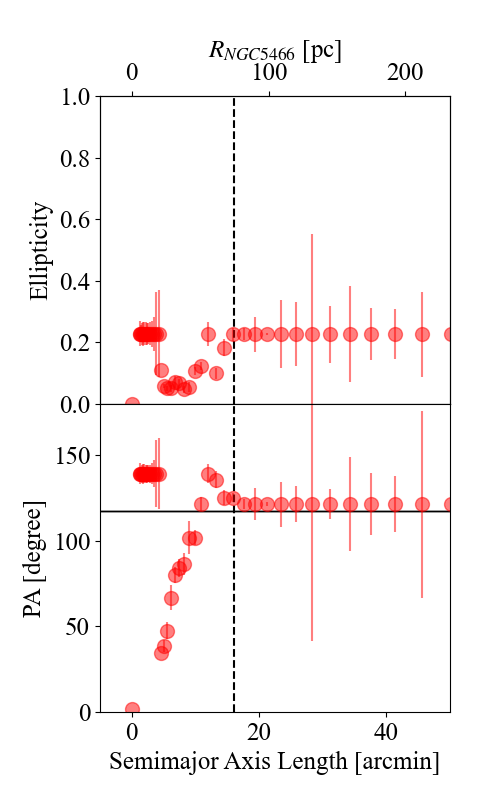}
 \end{center}
 \vspace{-5mm}
 \caption{Results of the isophotal analysis performed using the \texttt{isophote} package. (Top) Ellipticity as a function of cluster-centric distance measured along the semi-major axis. (Bottom) Position angle as a function of cluster-centric distance measured along the semi-major axis. The horizontal black solid line indicates the position angle of \texttt{galstreams} track. In each panel, the black dashed line indicates the tidal radius of NGC~5466.
  {Alt text: Two-panel plot of isophotal analysis results for NGC 5466, showing ellipticity (top) and position angle (bottom) as functions of cluster-centric distance along the semi-major axis, with the tidal radius indicated by a black dashed line.} }\label{fig:isophote}
  \vspace{-5mm}
\end{figure}


Using the derived membership probability, we construct the spatial density map shown in figure \ref{fig:SurfceDensityMap}. In this figure, the red contours correspond to density levels 0.5, 1, and 10 arcmin$^{-2}$. The orange line indicates the previously reported ridge line of the NGC~5466 stellar stream (\cite{2021MNRAS.507.1923J}) obtained from \texttt{galstreams} (\cite{2023MNRAS.520.5225M}). Regions masked in blue squares correspond to chip gaps or areas where bad pixels or detection failures prevented reliable photometry. In figure \ref{fig:SurfceDensityMap}, the spatial extents of the member stars at each radius from the center of NGC 5466 are illustrated by cyan dashed ellipses. These ellipses reflect the ellipticity and position angle (from North to East) of the member-star distribution as estimated from \texttt{python/photutils.isophote} package (\cite{2022zndo....596036B}). In the isophotal analysis, \texttt{isophote} computes the ellipticity, position angle, and density center of the two-dimensional map at each semi-major axis radius. In figure \ref{fig:isophote}, we present the ellipticity (top panel) and position angle (bottom panel) as functions of cluster-centric distance measured along the semi-major axis. In each panel, the black dashed line indicates the tidal radius of NGC 5466 derived using the method described below. In the bottom panel of figure \ref{fig:isophote}, the horizontal black solid line indicates the average position angle of \texttt{galstreams} track in our observational region.

In the upper-left corner of the figure \ref{fig:SurfceDensityMap}, the black arrows indicate the directions of the major-axis (aligned with the stream; solid arrow) and the minor-axis (perpendicular to the stream; dashed arrow), as estimated from the \texttt{python/isophote}. As shown in figure \ref{fig:isophote}, the position angle varies with cluster-centric distance, but it becomes constant beyond the tidal radius. We therefore define the position angle (120 degree) measured at radii beyond the tidal radius as the major-axis of the NGC~5466 stream. The minor-axis direction is defined to be perpendicular to the major-axis.

In figure \ref{fig:SurfceDensityMap}, NGC~5466 exhibits an elliptical morphology, with tidal distortions extending to both sides of the cluster. From the cyan ellipses in figure \ref{fig:SurfceDensityMap} and the trend of the bottom panel in figure \ref{fig:isophote}, the position angle beyond the tidal radius is aligned with the direction of the previously reported stellar stream. In figure \ref{fig:isophote}, no significant variation in the position angle is observed outside the tidal radius, but figure \ref{fig:SurfceDensityMap} reveals that, beyond the tidal radius, member stars are distributed around $(\xi,~\eta)\sim(-0.5,~0.0)$ in a direction offset from the \texttt{galstreams} ridge line. This structure is likely an S-shaped feature similar to those reported for Palomar 5. In the isophotal analysis, the lack of a detectable radial variation in the position angle beyond the tidal radius is likely due to low-level noise from remaining contaminations and these contaminants prevent a reliable determination of the position angle. A more detailed assessment will require the identification of a larger sample of member stars through future spectroscopic follow-up observations with Subaru/PFS.

\begin{figure*}
 \begin{center}
  \includegraphics[width=\linewidth, trim=0 0 0 0, clip]{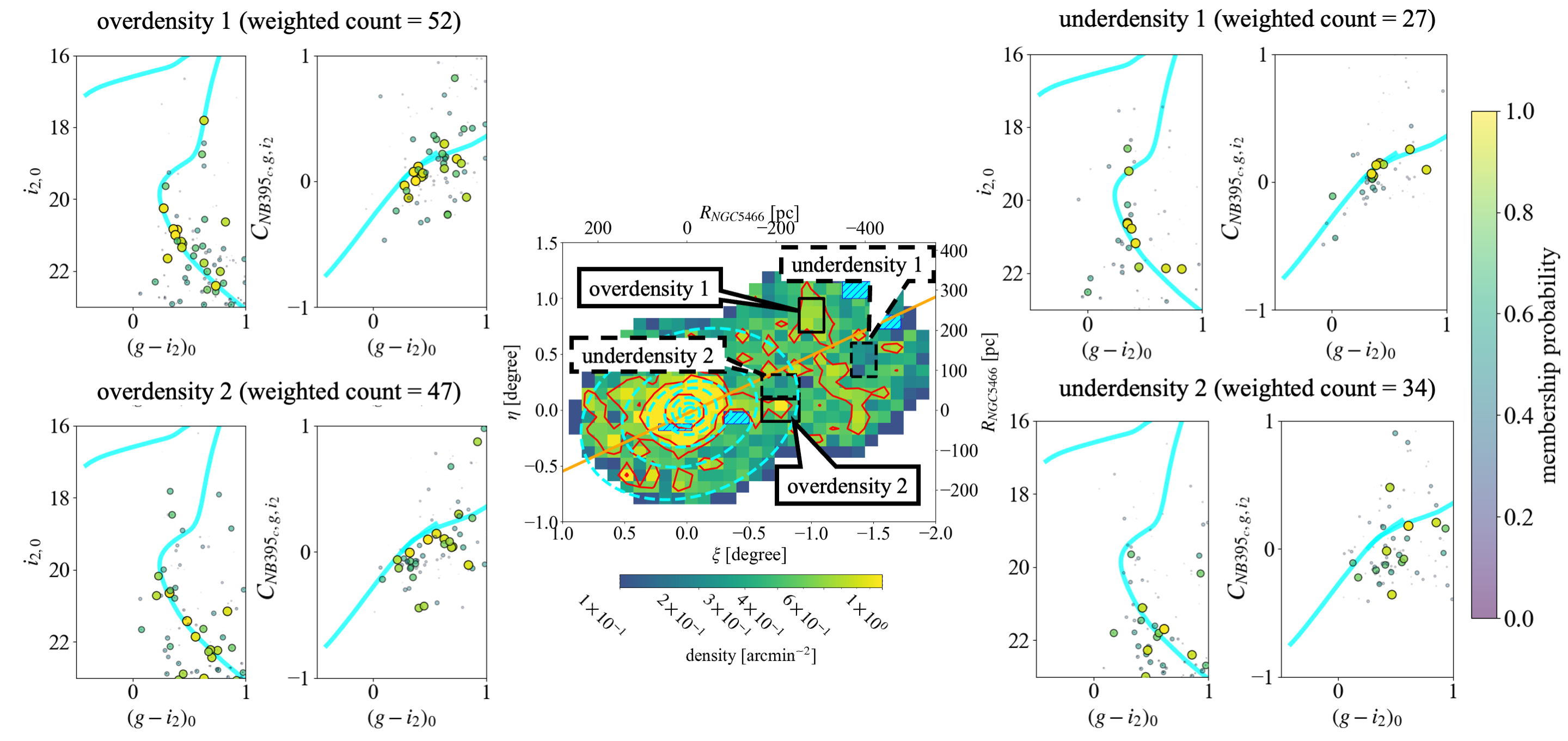}
 \end{center}
 \caption{Surface density distribution of NGC~5466 member stars, and color-magnitude/color-color diagrams for overdensity and underdensity regions. The central panel is the same as figure \ref{fig:SurfceDensityMap}. The black solid rectangles indicate the overdense regions of interest, while the black dashed rectangles correspond to the underdense regions. The left two rows present the CMDs and color–color diagrams for the selected overdense regions, whereas the right two rows show the CMDs and color–color diagrams for the selected underdense regions. In each CMD and color–color diagram, the symbol size and color are scaled according to the membership probability of each object, with higher-probability objects shown as larger symbols. The cyan solid lines represent isochrones corresponding to the properties of NGC 5466. Above each diagram, weighted counts by membership probabilities $P_{\rm mem}$ are shown for the corresponding region.
  {Alt text: Surface density map of NGC 5466 member stars with overdense (black solid rectangles) and underdense (black dashed rectangles) regions highlighted, accompanied by CMDs and color–color diagrams for each region. Symbol size and color encode membership probability, and cyan lines show NGC 5466 isochrones.} }\label{fig:Gap}
\end{figure*}

In figure \ref{fig:SurfceDensityMap}, an overdense region is also visible to the northwest of the cluster center at approximately $(\xi,~\eta)\sim(-1,~1)$. Moreover, a local decrease in the stellar density is confirmed in the inner region at $(\xi,~\eta)\sim(-0.75,~0.25)$. This region coincides with the location of the previously identified stream ridge line, shown as the orange line. No artifacts are detected in this underdense area, suggesting that this may represent an intrinsic gap in the stellar stream. 

To examine the overdense and underdense regions in more detail, we inspect CMDs and color–color diagrams. Figure \ref{fig:Gap} shows CMDs and color-color diagrams for overdense and underdense regions. The central panel of figure \ref{fig:Gap} is identical to figure \ref{fig:SurfceDensityMap}. The black solid rectangles indicate the selected overdense regions, while the black dashed rectangles mark the selected underdense regions. The left two rows present the CMDs and color–color diagrams of the overdense regions of interest, whereas the right two rows show those of the underdense regions. In each CMD and color–color diagram, the symbol size and color are scaled according to the membership probability of each object, with higher-probability members displayed as larger symbols. The cyan solid lines correspond to isochrones matching the properties of NGC 5466 (12.5 Gyr, $[{\rm Fe/H}] = -2.0$, $[{\rm \alpha/Fe}]=+0.4$). Above each diagram, weighted counts by membership probabilities $P_{\rm mem}$ are shown for the corresponding region. As shown in figure \ref{fig:Gap}, the overdense regions exhibit a clear concentration of member stars along the isochrone, whereas the underdense regions contain relatively few stars consistent with the isochrone. This contrast indicates that the observed gap is likely to be a genuine physical feature rather than a result of statistical fluctuations. We discuss the nature and possible origin of this gap in Section \ref{subsection:stream}.

To further characterize the properties of the stream, we construct three types of radial density profiles. (i) the full azimuthally integrated profile (``full data''), (ii) major-axis profile along the stellar stream (``major-axis data''), and (iii) minor-axis profile perpendicular to the stream (``minor-axis data''). The major-axis and minor-axis data are used to assess how strongly the radial profile is affected by the stellar stream. We define the major-axis and minor-axis datasets as objects located within $\pm 0.25$ degrees of the respective axes. The major-axis and minor-axis are defined as the direction of the position angle derived from the isophotal analysis described in section \ref{subsec:SpatialDistribution}, and their directions are shown in figure \ref{fig:SurfceDensityMap} as black arrows.

Previous studies employed various functional forms to describe the radial profiles of GCs and their tidal debris. In this work, we adopt a three-component fitting approach, extending earlier methodologies. Briefly, we fit the observed radial density profile using the following functional expression:

\begin{align}
    \rho(r) = 
    \begin{cases}
    \Sigma_0 \left( \frac{1}{1+(r/r_c)^2} - \frac{1}{1+(r_t/r_c)^2} \right) + \Sigma_{\rm cont}, & r < r_t \\
    k \times r^{\alpha} + \Sigma_{\rm cont}, & r > r_t
    \end{cases}
    \label{eq:radial_density_profile}
\end{align}

In this model, we assume that the stellar density inside the tidal radius $r_t$ follows a classical King profile (\cite{1962AJ.....67..471K}), whereas the outer regions are dominated by a power-law profile representing the stellar stream (e.g., \cite{2007MNRAS.380..749F,2016MNRAS.461.3639K,2012MNRAS.419...14C,2018MNRAS.474..683C}). In both components, we include a constant contamination term $\Sigma_{\rm cont}$, which accounts for the surface density of remaining non-member foreground and background sources. The parameters are $\Sigma_0$, $k$, $r_c$, $r_t$, $\alpha$, and $\Sigma_{\rm cont}$, where $\Sigma_0$ denotes the central surface density, $k$ is a scale factor linking the inner King and outer power-law components, $\alpha$ is the power-law index, and $\Sigma_{\rm cont}$ represents the uniform contaminant surface density.

\begin{figure}
 \begin{center}
  \includegraphics[width=0.95\linewidth, trim=0 0 10 0, clip]{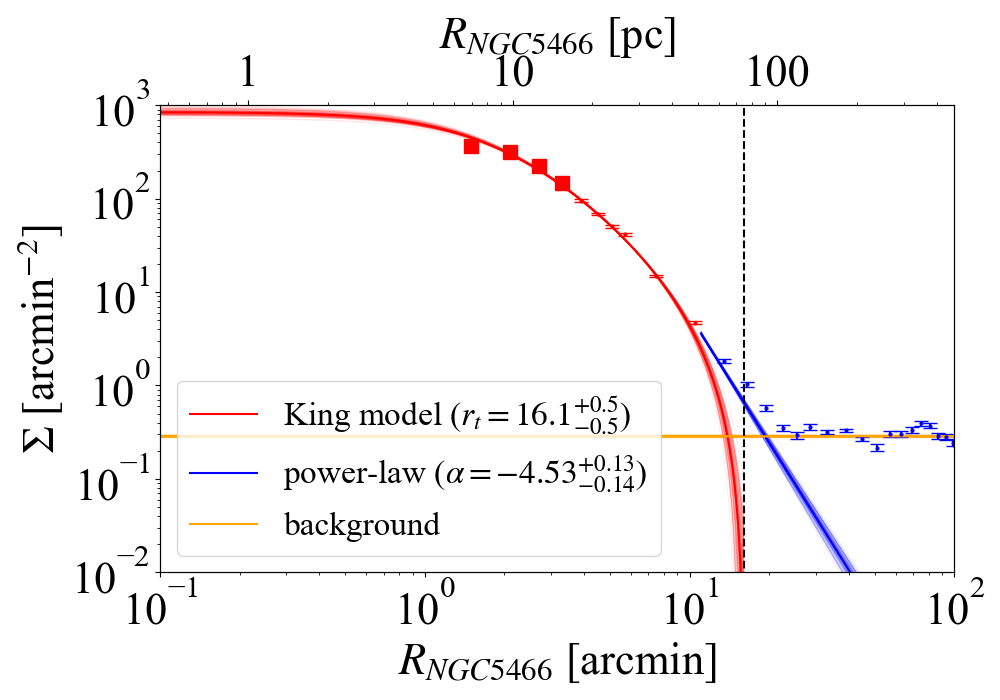}
 \end{center}
 \caption{The radial stellar density profile of full data. The error-bar plot shows the HSC data, while the square symbols represent the Pristine–Gaia catalog from K25b. The red points indicate the data used for fitting to the King model plus uniform contamination model, and the blue points correspond to the data fitted to the power-law plus uniform contamination model. The red, blue, and orange lines show the best-fitting King, power-law, and uniform-contamination models, respectively. These curves are drawn by 100 randomly sampling from the posterior distributions. The black vertical line corresponds to the estimated tidal radius ($r_t = 16.1$ arcmin).
  {Alt text: Radial stellar density profile of NGC~5466 showing HSC and Pristine–Gaia data with fitted King, power-law, and uniform-contamination models. The plot includes posterior-sampled model curves and marks the estimated tidal radius at 16.1 arcmin.} }\label{fig:RadialProfile}
\end{figure}

\begin{figure}
 \begin{center}
  \includegraphics[width=0.95\linewidth, trim=0 0 10 0, clip]{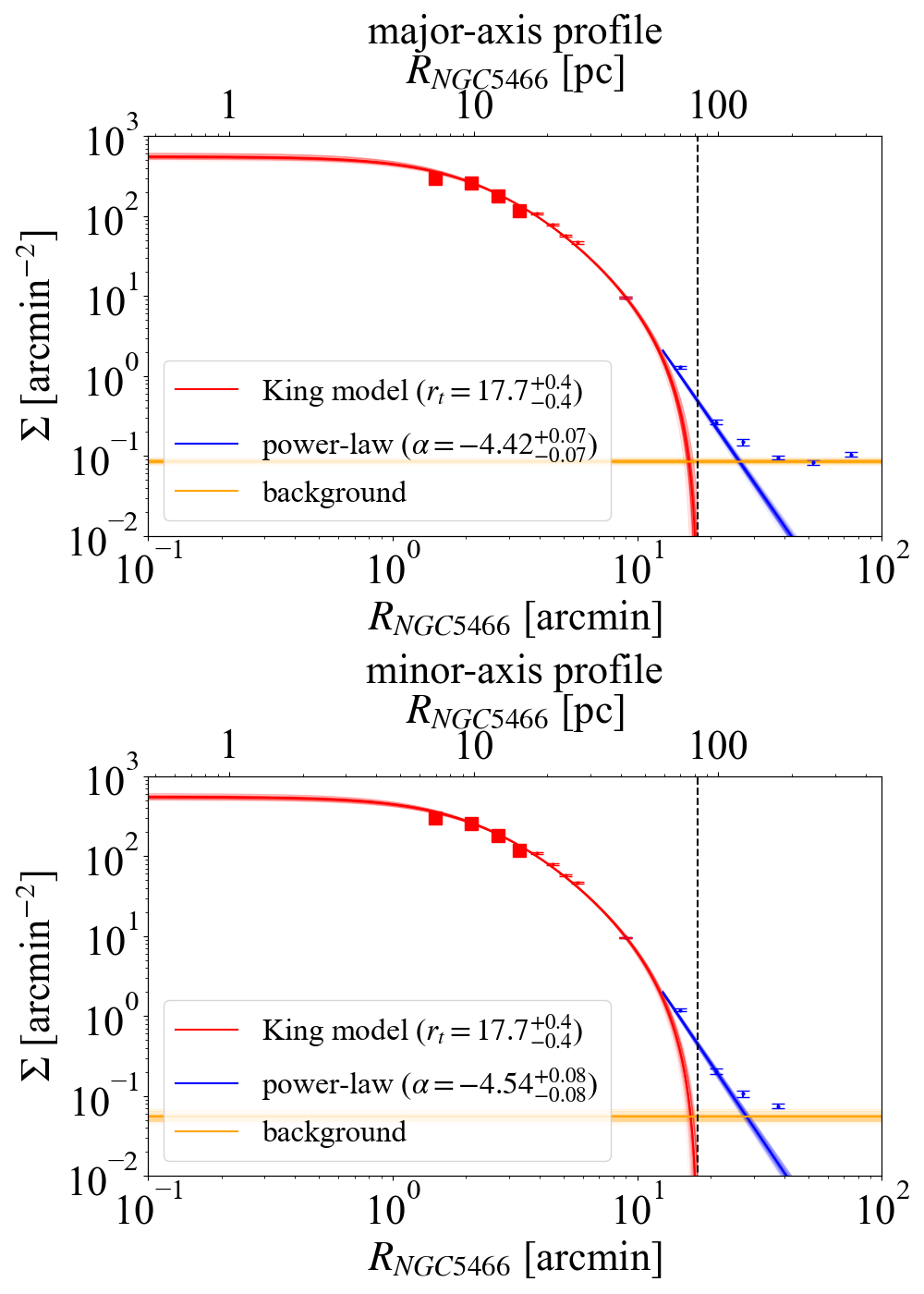}
 \end{center}
 \caption{The same as in figure \ref{fig:RadialProfile} but for the major-axis (top panel) and minor-axis (bottom panel) profiles of NGC 5466.
  {Alt text: Radial stellar density profiles of major-axis data (top panel) and minor-axis data (bottom panel).} }\label{fig:RadialProfile_Major_Minor}
\end{figure}

Figure \ref{fig:RadialProfile} presents the radial density profile of full data. The points show the observed profile, computed by weighting each star by its membership probability and detection completeness. Because the number of stars decreases toward the outer regions and the Poisson uncertainties become large, we adopt different bin sizes: the radial range from 0 to 6 arcmin in steps of 0.6 arcmin, from 6 to 30 arcmin in steps of 6 arcmin, from 30 to 60 arcmin in steps of 15 arcmin, and from 60 to 120 arcmin in steps of 30 arcmin. Blue and red points indicate, respectively, the regions mainly fitted by the King component and those fitted by the power-law component of the model.

The red square symbols in figure \ref{fig:RadialProfile} show the Gaia-based radial profile constructed using the selection criteria of \citet{2025AJ....170..157K}, hereafter K25b. This dataset is incorporated into our dataset to mitigate incompleteness in the crowded central region. We construct radial profiles for both the HSC and K25b catalogs. Inside a radius of 3.9 arcmin - where the HSC profile begins to decline due to incompleteness - we adopt the K25b profile, while the HSC profile is used at larger radii. To ensure a seamless transition between the two profiles, we compute the ratio of their densities at 3.9 arcmin and apply this factor so that the profiles match at that radius.

Figure \ref{fig:RadialProfile_Major_Minor} shows the radial density profile of major-axis data (top panel) and minor-axis data (bottom panel). These profiles are constructed using the same procedure as for the radial profile of full data. In addition, the symbols are the same as those used for the radial profile of the full data shown in figure \ref{fig:RadialProfile}.

We fit all radial density profiles to the model described in equation (\ref{eq:radial_density_profile}). Following \citet{2024ApJ...971..107O}, we adopt a log-likelihood framework and use \texttt{emcee} for MCMC sampling. We employ 300 walkers and run 11,000 steps, discarding the first 10,000 as burn-in. Because incompleteness in the crowded core may bias the estimation of the core radius $r_c$, we impose a Gaussian prior on $r_c$ with mean $\mu = -1.43$ arcmin and standard deviation $\sigma=0.1$ arcmin, based on previous measurements (\cite{1996AJ....112.1487H,2010arXiv1012.3224H}). All other parameters are assigned non-informative priors. The adopted priors and the resulting posterior constraints are summarized in Table \ref{tab:MCMCparameters}. The all marginalized posterior distributions are provided in Appendix \ref{appendix:MCMC_Results}.

Figures \ref{fig:RadialProfile} and \ref{fig:RadialProfile_Major_Minor} display the best-fit model as colored lines. The red, blue, and orange curves correspond to the King component, power-law component, and constant background, respectively. To illustrate parameter uncertainties, we plot 100 model realizations randomly drawn from the posterior distributions. Owing to the reduced background level achieved with our deep HSC photometry and the {\it NB395}-based selection, all radial profiles clearly demonstrate that NGC~5466 possesses an extended component beyond that predicted by a simple King profile. Moreover, in every profile, this extended component is well described by a power-law tail. From the MCMC analysis of full data, we find a power-law index of $\alpha=-4.53_{-0.14}^{+0.13}$, consistent with the previous reported value ($-4.05_{-0.27}^{+0.43}$; \cite{2012MNRAS.419...14C}).

Beyond a radius of $\sim 30$ pc, the background level indicated by the orange curve becomes dominant, making the reliable selection of stream stars increasingly challenging. This highlights the importance of narrow-band photometric selection for future spectroscopic follow-up observations. The tidal radius is constrained to be $r_t = 16.1_{-0.5}^{+0.5}$ arcmin, which corresponds to $75_{-2}^{+2}$ pc assuming a cluster distance of $16.12\pm0.16$ kpc. This estimated value is broadly consistent with previous measurements of the tidal radius (79 pc; \cite{2014ApJ...793..110M}).

From figures \ref{fig:RadialProfile} and \ref{fig:RadialProfile_Major_Minor}, the power-law slope of major-axis data is shallower than that of the other two cases. This result is consistent with the presence of the stream. In addition, we detect a power-law component in the minor-axis profile as well. This indicates that a fraction of the tidally stripped stars in NGC~5466 are distributed away from the major-axis direction. Such stars may be associated with the formation of multiple tidal tails, as predicted if NGC~5466 has experienced multiple passages through apogalacticon (\cite{2015MNRAS.446.3100H}). However, the minor-axis extent of these features lies beyond the observational footprint of this study, and further observations will be required to test this scenario.

\begin{table*}
 \caption{Summary of the results of our MCMC sampling and a brief description of the parameters.}
 \label{tab:MCMCparameters}
 \begin{tabular*}{2\columnwidth}{@{\hspace*{40pt}}c@{\hspace*{40pt}}c@{\hspace*{40pt}}c@{\hspace*{40pt}}c@{\hspace*{40pt}}}
  \hline
  Parameter & Prior distribution & Estimated value & sample\\
  \hline
  & & $1063_{-50}^{+54}$ & all\\
  $\Sigma_{\rm king}$ [arcmin$^{-2}$] & $\mathrm{Uniform}[0,10000000]$ & $720_{-26}^{+27}$ & major-axis\\
  & & $720_{-26}^{+27}$ & minor-axis\\  
  & & $1.75_{-0.07}^{+0.07}$ & all\\
  $r_c$ [arcmin] & $\mathrm{Gaussian}(\mu=1.43,\sigma=0.1)$ & $2.22_{-0.06}^{+0.06}$ & major-axis\\
  & & $2.22_{-0.06}^{+0.06}$ & minor-axis\\
  & & $16.1_{-0.5}^{+0.5}$ & all\\
  $r_t$ [arcmin] & $\mathrm{Uniform}[0,1000]$ & $17.7_{-0.4}^{+0.4}$ & major-axis\\
  & & $17.7_{-0.4}^{+0.4}$ & minor-axis\\
  & & $190,000_{-50,000}^{+80,000}$ & all\\
  $\Sigma_{\rm power}$ [arcmin$^{-2}$] & $\mathrm{Uniform}[0,10000000]$ & $160,000_{-20,000}^{+30,000}$ & major-axis\\
  & & $210,000_{-30,000}^{+40,000}$ & minor-axis\\
  & & $-4.53_{-0.14}^{+0.13}$ & all\\
  $\alpha$ & $\mathrm{Uniform}[-10,1]$ & $-4.42_{-0.07}^{+0.07}$ & major-axis\\
  & & $-4.54_{-0.08}^{+0.08}$ & minor-axis\\
  & & $0.29_{-0.01}^{+0.01}$ & all\\
  $\Sigma_{\rm const}$ [arcmin$^{-2}$]& $\mathrm{Uniform}[0,10000000]$ & $0.085_{-0.003}^{+0.003}$ & major-axis\\
  & & $0.055_{-0.005}^{+0.005}$ & minor-axis\\
  \hline
 \end{tabular*}
\end{table*}

\subsection{Mass function}\label{subsec:MF}

\begin{figure*}
 \begin{center}
  \includegraphics[width=\linewidth, trim=0 0 10 0, clip]{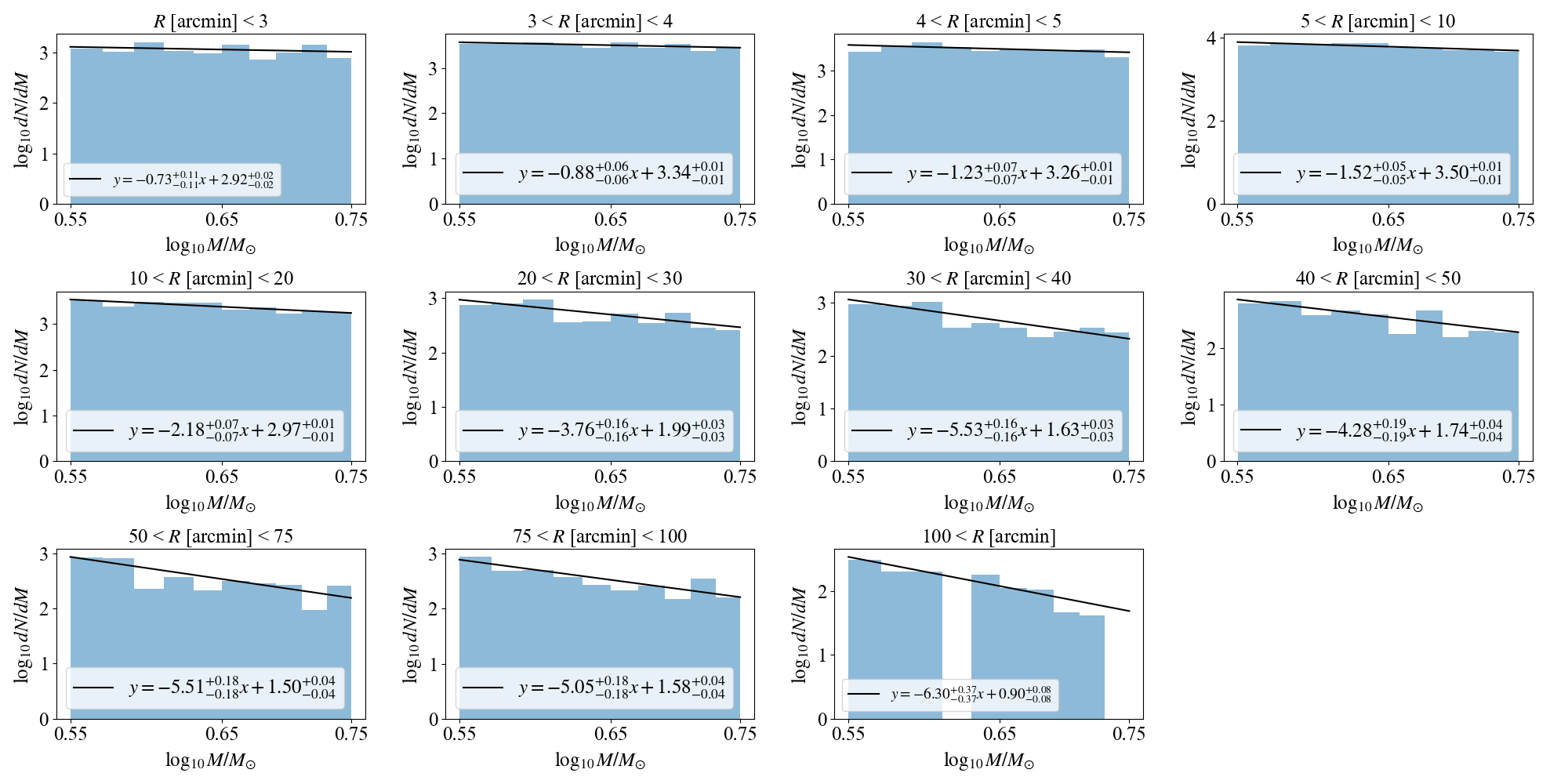}
 \end{center}
 \caption{MFs at each radial bin constructed using full data. For each panel, a black solid line is the best-fit model.
  {Alt text: Figure showing the stellar MFs in each radial bin derived from full data. Each panel displays the observed MF with the best-fit model overplotted as a black solid line. The upper-left panel is a most-inner one.}}\label{fig:MF_AllData}
\end{figure*}

\begin{figure*}
 \begin{center}
  \includegraphics[width=\linewidth, trim=0 0 10 0, clip]{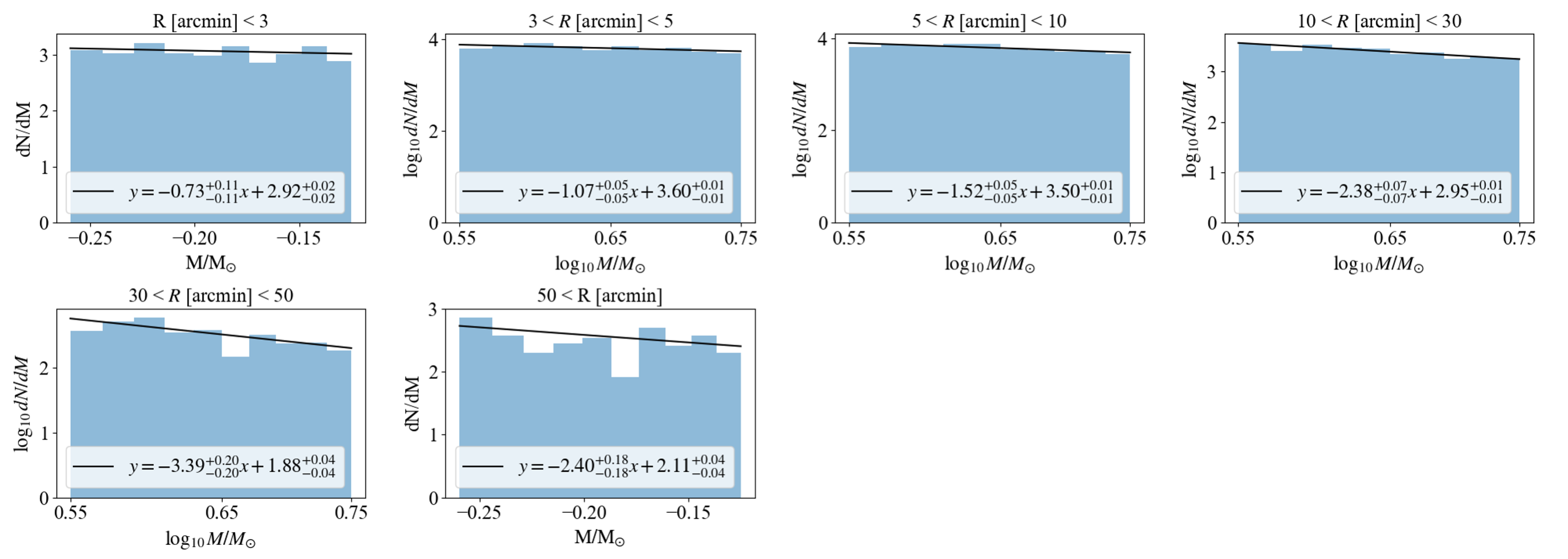}
 \end{center}
 \caption{The same as in figure \ref{fig:MF_AllData}, but for major-axis sample.
  {Alt text: Figure showing the stellar MFs in each radial bin derived from major-axis data.}}\label{fig:MF_MajorAxis}
\end{figure*}

\begin{figure*}
 \begin{center}
  \includegraphics[width=\linewidth, trim=0 0 10 0, clip]{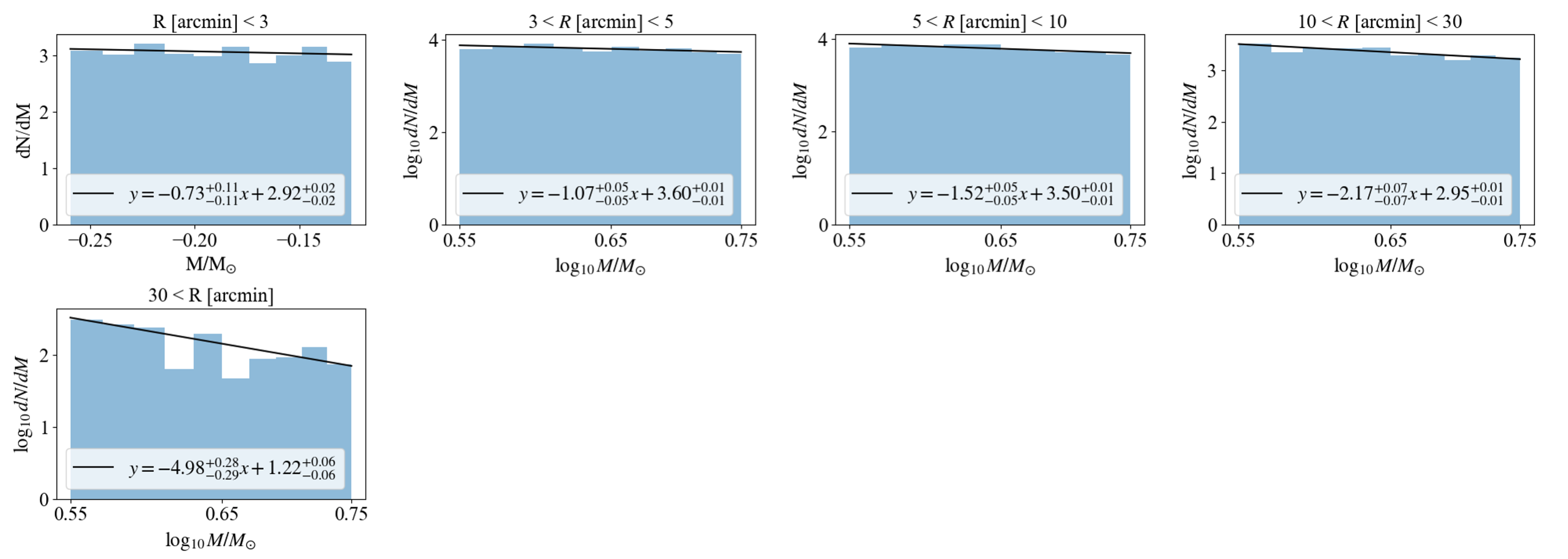}
 \end{center}
 \caption{The same as in figure \ref{fig:MF_AllData}, but for minor-axis sample.
  {Alt text: Figure showing the stellar MFs in each radial bin derived from minor-axis data.}}\label{fig:MF_MinorAxis}
\end{figure*}

MF is derived using a method based on \citet{2015ApJ...814..144B}. The transformation of the apparent $i_{{\it 2},0}$ magnitudes into solar masses is conducted by adopting the mass-to-light law of MS stars from the BaSTi isochrone of metallicity $[{\rm Fe/H]} = -2.00$ and $[{\rm \alpha/Fe}] = +0.4$, assuming a distance to NGC~5466 (16.12 kpc; \cite{2021MNRAS.505.5957B}). 

To select MS stars and exclude unresolved binary stars, we identify sources lying within 2.5 times the mean photometric uncertainty from the BaSTI isochrone and treat these as bona-fide MS stars for constructing the MF. The MF is computed over the range $0.55 < M/M_{\odot} < 0.75$ ($20.22\lesssim i_{2,0}\lesssim22.64$), divided into 10 bins. During this procedure, we apply a completeness correction based on the detection completeness derived in section \ref{subsec:photometry}. As shown in figure \ref{fig:DetComp}, the completeness of our dataset reaches 50 \% at approximately $((g-i_2)_0,~i_{2,0}) \sim(1.0,~23.5)$. Also, unresolved background galaxies increase rapidly at $i_{2,0}\sim23.5$ mag and may lead to an overestimation of the MF on the low-mass side. For this reason, we adopt a narrower mass range and limit our analysis to $M/M_{\odot}>0.55$, which is more restrictive than the range used in \citet{2015ApJ...814..144B}.

We construct the MF for three different datasets: (i) the full azimuthally integrated bona-fide MS stars (``full data''), (ii) MS stars located along the direction of the stellar stream (``major-axis data''), and (iii) MS stars located perpendicular to the stream (``minor-axis data''). These data are defined in the same manner as in subsection \ref{subsec:SpatialDistribution}, where the radial density profile is constructed. For each dataset, MFs are computed in radial bins from the cluster center. For full data, we adopt bins of $R=[3,4,5,10,20,30,40,50,75,100,<100]$ arcmin. For the major-axis and minor-axis datasets, we use wider bins, $R=[3,5,10,30,50,<50]$ arcmin, to ensure a sufficient number of statistics and reduce Poisson noise.

Figure \ref{fig:MF_AllData} shows the MFs at each radial bin constructed using full data, and figures \ref{fig:MF_MajorAxis} and \ref{fig:MF_MinorAxis} are the MFs at each radial bin using the major-axis and minor-axis data, respectively. For each derived MF, we perform an MCMC fitting assuming a power-law form of the MF, $dN/dM \propto M^{\alpha}$. The fitting function is given by
\begin{align}
\log_{10}{\left( \frac{dN}{dM} \right)} = \alpha \log_{10}{\left(M/M_{\odot} \right)} + \beta.
\label{eq:MF_model}
\end{align}

Adopting this expression, and assuming that the observational data follow a Gaussian distribution with their uncertainties as the standard deviations, we evaluate the following logarithmic likelihood:
\begin{align}
\log{\mathcal{L}} = \prod_i\frac{1}{\sqrt{2\pi}\sigma_i}\exp{\left[ -\frac{1}{2} \left\{ \frac{y_i - (\alpha x_i + \beta)}{\sigma_i} \right\} ^2 \right]},
\label{eq:MF_likelihood}
\end{align}
\noindent
where $(x_i,y_i)$ represents the $i$-th data point in $(\log_{10}{M/M_{\odot}}_i,\log_{10}{dN/dM}_i)$ and $\sigma_i$ denotes its Poisson uncertainty propagated into logarithmic space, $\sigma_i = 1/[ (\ln{10})\sqrt{dN/dM_i} ]$. Although the logarithmic likelihood is constructed under the assumption that the observational data follow a Gaussian distribution, the number statistics are sufficiently large that the Poisson errors can be treated as the standard deviations of an equivalent Gaussian distribution. In this likelihood, $\alpha$ and $\beta$ correspond to the parameters of Equation (\ref{eq:MF_model}) and are the quantities to be inferred in the MCMC fitting. We adopt uninformative uniform priors for both parameters, and the posterior distribution is obtained by combining these priors with the likelihood. The MCMC sampling is performed using the \texttt{python/emcee} module, employing 100 walkers, 110,000 steps, and burn-in of 100,000 steps.

The resulting best-fit model is shown as a black curve in figures \ref{fig:MF_AllData}, \ref{fig:MF_MajorAxis} and \ref{fig:MF_MinorAxis}, and the corresponding parameters are discribed in the legend of each panel. Posterior distributions of estimated parameters are shown in Appendix \ref{appendix:MCMC_Results}. As seen in figure \ref{fig:MF_AllData}-\ref{fig:MF_MinorAxis}, the estimated statistical uncertainties range from 0.05 to 0.4, and the best-fit models provide a satisfactory reproduction of the observed MFs across all radial bins.

In figure \ref{fig:MF_AllData}-\ref{fig:MF_MinorAxis}, MFs for all three datasets are relatively shallow (i.e., top-heavy MF) in the central regions, whereas they become progressively steeper (bottom-heavy MF) toward larger radii. This trend likely reflects the effects of mass segregation, whereby massive stars migrate toward the cluster center while lower-mass stars move outward, as well as the impact of tidal stripping, which can further disperse low-mass stars to larger radii. Surprisingly, we confirm a top-heavy MF in the inner region despite the presence of projection effects. When considered as a function of projected radius, stars located at larger three-dimensional distances can be superposed onto the central region, thereby introducing additional low-mass stars and potentially diluting the intrinsic top-heavy trend of the central mass function. A more detailed discussion of these processes is provided in Section \ref{subsec:MF_Discussion}.

\section{Discussion}\label{sec:discussion}
\subsection{Spatial gaps along the stellar stream}\label{subsection:stream}

In figure \ref{fig:SurfceDensityMap}, we identify a discontinuity in the spatial structure along the stellar stream. Several physical mechanisms have been proposed to explain the origin of gap-like features in GC streams. Encounters with dark matter subhaloes perturb the continuous structure of a stream, producing localized gaps (e.g., \cite{2012ApJ...748...20C,2013ApJ...768..171C,2015MNRAS.454.3542E}). In addition, epicyclic motions of stars escaping from the main body of the cluster are known to induce periodic variations in the stream density (e.g., \cite{2008MNRAS.387.1248K,2010MNRAS.401..105K,2012MNRAS.423.2845L,2014ApJ...788..181N}). Enhanced mass-loss episodes during pericentric passages or passages of the Galactic disk also generate density inhomogeneities along the stream (e.g., \cite{2003AJ....126.2385O,2015MNRAS.450..575A}). Therefore, gaps in stellar streams serve as important tracers of both the host-galaxy potential and past accretion and interaction events.

\begin{figure*}
 \begin{center}
  \includegraphics[width=\linewidth, trim=0 0 10 0, clip]{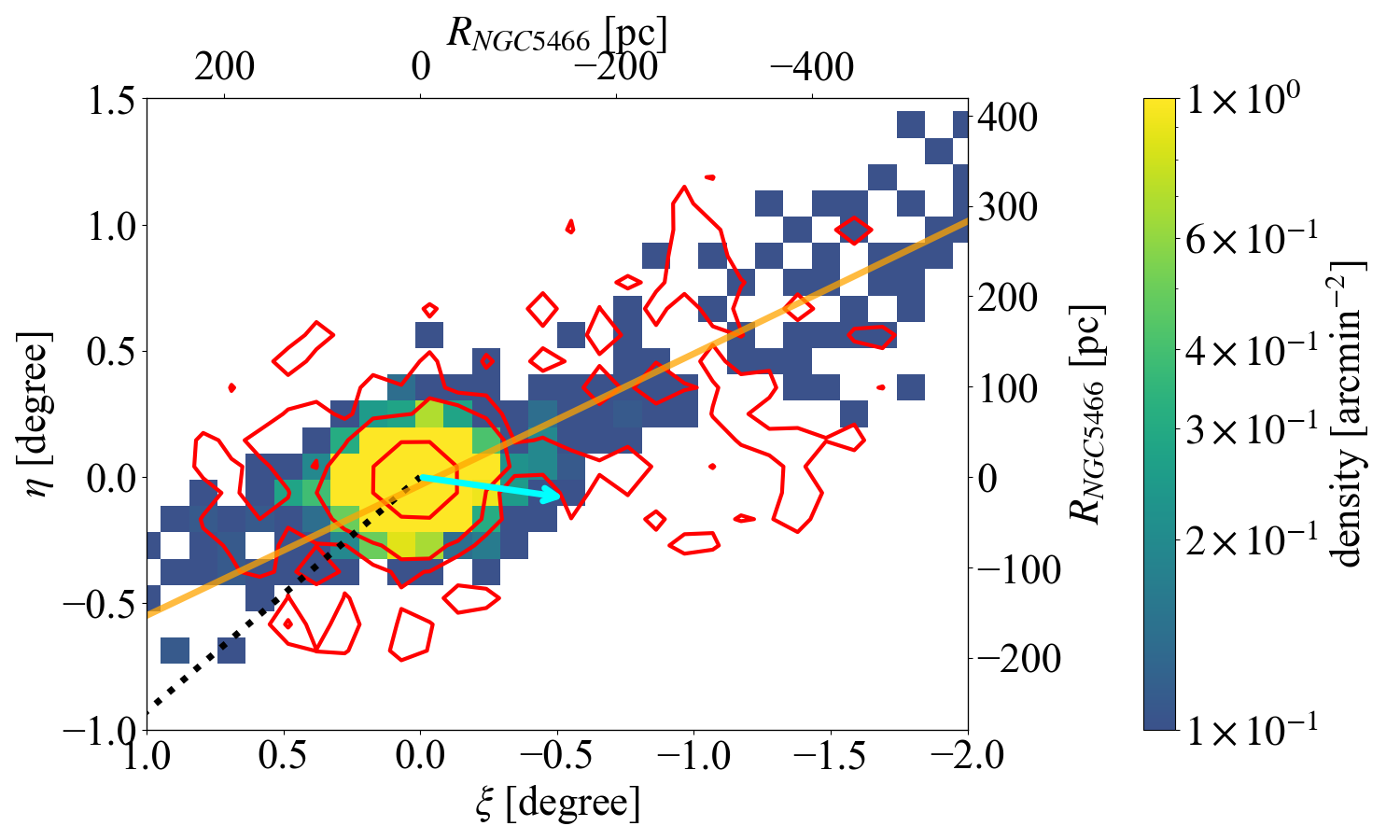}
 \end{center}
 \caption{Surface density distribution of simulated NGC 5466 from eTidals. The red contours correspond to surface density levels of 0.5, 1, and 10 arcmin$^{-2}$ from our HSC data. The black dotted line corresponds to the direction of the Galactic center, and the cyan arrow indicates the proper motion of NGC~5466. The orange solid line is the stream track from the \texttt{galstreams} package.
 {Alt text: Surface density map of simulated NGC 5466 member stars from eTidals and contour maps of our HSC data.} }\label{fig:SurfaceDensityMap_eTidals}
\end{figure*}

To investigate the origin of the detected gap, we compare the observed surface density distribution with {\it N}-body models from the e-TidalGCs project (hereafter eTidals; \cite{2023A&A...673A..44F}). The eTidals framework models the disruption of present-day Galactic GCs using their observed structural and kinematic parameters. In brief, each cluster orbit is integrated backward for 5 Gyr, after which a Plummer-model representation of the cluster, realized with $N=100,000$ particles, is evolved forward to the present time. Several Galactic gravitational potentials are explored. In this study, following K25b, we adopt the publicly available PII model, which includes thin and thick disks and a spherically symmetric dark matter halo, but no bulge component. Figure \ref{fig:SurfaceDensityMap_eTidals} shows the eTidals surface-density map with contours of the observed surface density overlaid. In this map, the direction of the Galactic center, the proper motion of NGC~5466, and the \texttt{galstreams} track are shown as a black dotted line, a cyan arrow, and an orange solid line for reference. The proper motion of NGC~5466 is adopted $\bar{\mu_{\alpha}} = -5.342$~[mas yr$^{-1}$], $\bar{\mu_{\delta}} = -0.822$~[mas yr$^{-1}$] (\cite{2021MNRAS.505.5978V}), and we assume that the relations $\mu_{\alpha,*} \approx \mu_{\xi}$ and $\mu_{\delta} \approx \mu_{\eta}$ within our observed field.

From figure \ref{fig:SurfaceDensityMap_eTidals}, the location of the gap detected in our data is slightly offset from the gap predicted by the eTidals model, but it is broadly consistent within the 0.2 degree uncertainty. Although the exact location of the gap is not reproduced by the eTidal models, the overall agreement of this gap may indicate that the stream gap in NGC~5466 could be explained without invoking an external encounter with a dark matter subhalo. Instead, it may arise from the cluster's orbital evolution during accretion. NGC~5466 follows a highly eccentric orbit ($e = 0.74$; \cite{2021MNRAS.507.1923J}) and has experienced a recent pericentric passage, with a perigalactic distance of $\sim 5 - 6$ kpc (\cite{2007MNRAS.380..749F,2021MNRAS.507.1923J}), and likely a recent disk crossing (\cite{2004ASPC..327..284O}). Strong tidal stripping during pericentric passages and disk crossings can therefore plausibly generate the observed density variations, giving rise to the gap-like structure.

\begin{figure}
 \begin{center}
  \includegraphics[width=\linewidth, trim=0 0 0 0, clip]{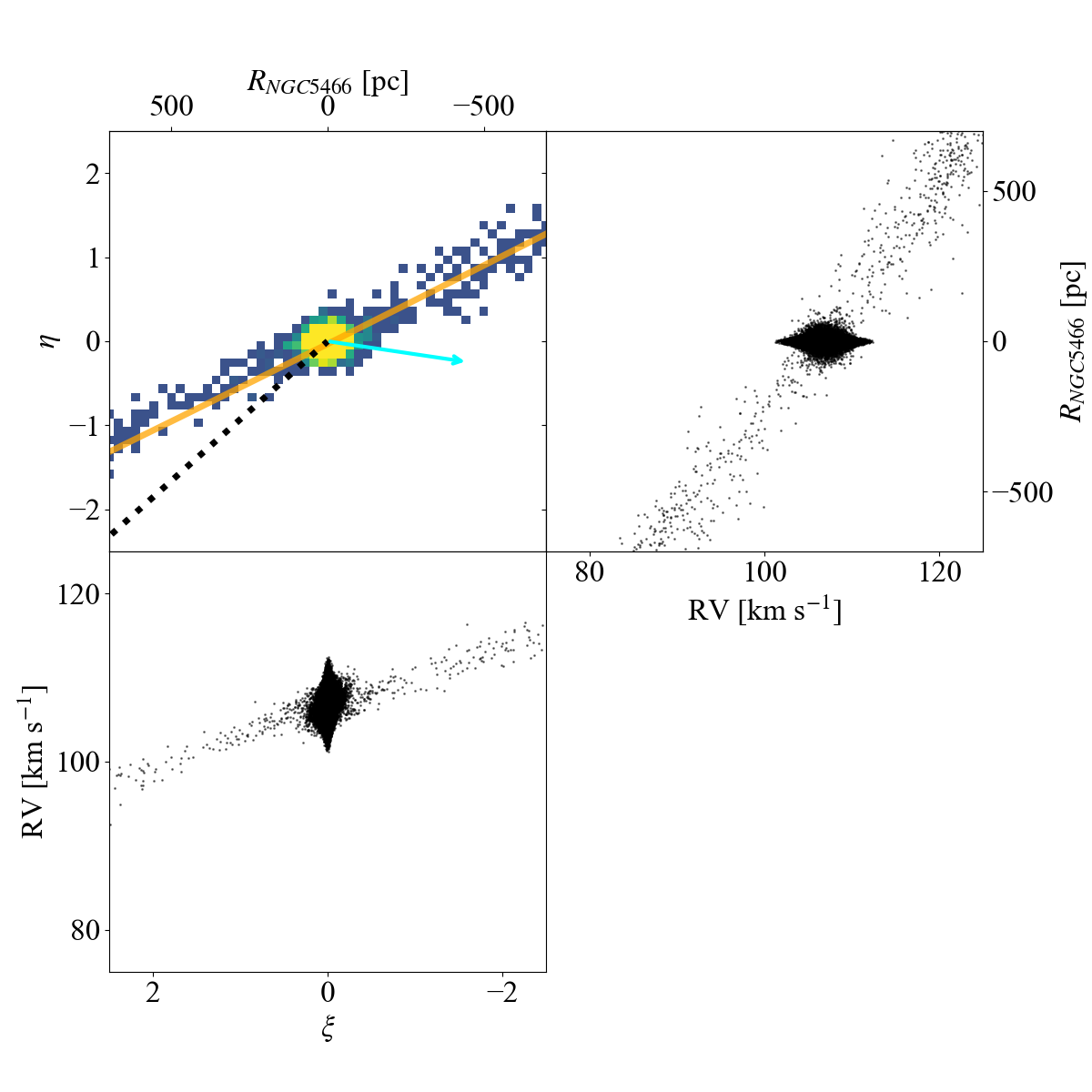}
 \end{center}
 \caption{The surface density map and the predicted radial velocities (RVs) of NGC~5466 provided by eTidal. The upper-left panel is the same as figure \ref{fig:SurfaceDensityMap_eTidals}. The lower-left panel shows the eTidal-predicted RVs as a function of $\xi$. The lower-right panel presents the simulated RV results as a function of $\eta$.
 {Alt text: Multi-panel figure showing the eTidal surface density map and predicted radial velocities for NGC~5466. The upper-left panel repeats figure ~\ref{fig:SurfaceDensityMap_eTidals}. The lower-left panel plots predicted RV as a function of $\xi$, and the lower-right panel shows simulated RV as a function of $\eta$.} }\label{fig:RV_eTidals}
\end{figure}

Figure \ref{fig:RV_eTidals} presents the surface-density map of NGC~5466 and the corresponding radial velocity (RV) predictions provided by the \texttt{eTidal} model. The upper-left panel is identical to figure \ref{fig:SurfaceDensityMap_eTidals}, while the lower-left and lower-right panels show the simulated RV distributions as functions of $\xi$ and $\eta$, respectively. If the gap identified in this study is primarily produced by recent Galactic disk crossings or pericentric passages, we expect a relatively continuous RV field along the stream, as illustrated in figure \ref{fig:RV_eTidals}. In contrast, if the gap is induced by an encounter with a dark-matter subhalo or a giant molecular cloud, a localized velocity perturbation (i.e., a ``velocity kick''; e.g., \cite{2015MNRAS.450.1136E}) is expected near $(\xi,~\eta)\approx(-0.75,~0.25)$.

A more definitive identification of the gap origin will require radial velocity information from future spectroscopic observations of member stars. In particular, Subaru/PFS can efficiently obtain spectra for stars as faint as $g\sim22$ mag. The sample presented in this work thus provides a valuable spectroscopic target set for PFS and represents an important test case for future studies of the physical nature of tidal streams.

\subsection{Radial variation of the slope of the stellar mass function and dynamical process of NGC~5466}\label{subsec:MF_Discussion}

\begin{figure}
 \begin{center}
  \includegraphics[width=0.95\linewidth, trim=0 0 10 0, clip]{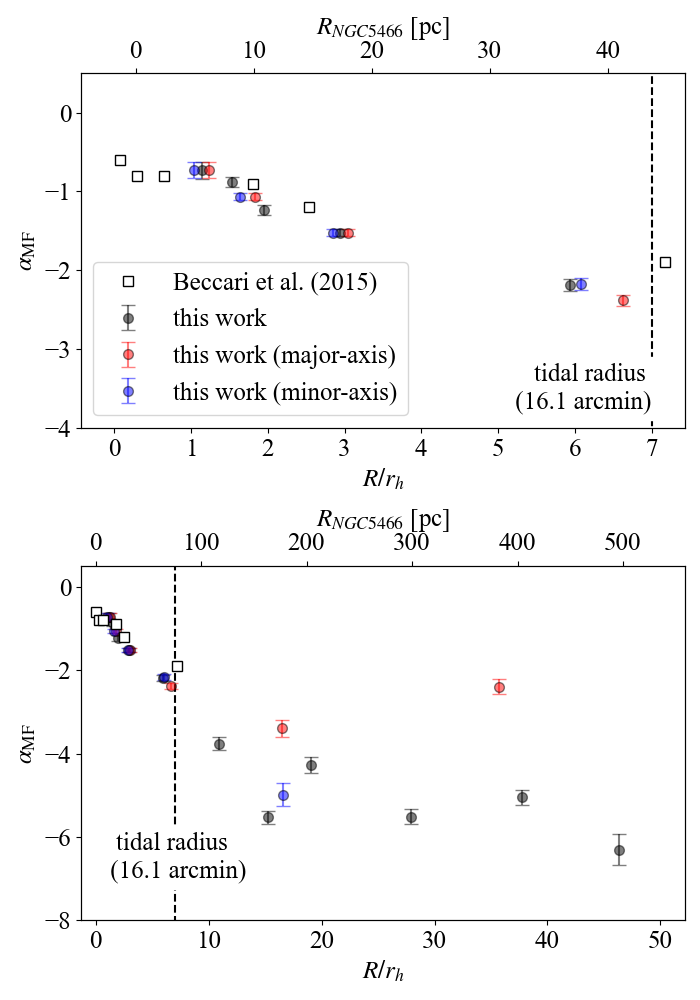}
 \end{center}
 \caption{Radial distribution of the MF slope. The horizontal axis is normalized by the half-mass radius, $r_h$. The top panel focuses on the distribution within the tidal radius, while the bottom panel shows the entire observed field. Black, red, and blue circles represent our results obtained from the ``full data'', ``major-axis data'', and ``minor-axis data'', respectively. For clarity of visualization, the red and blue points are offset by $\pm 0.1$ along the x-axis. The white squares indicate the measurements reported by \citet{2015ApJ...814..144B}, and the black vertical dashed line marks the tidal radius estimated in Section \ref{subsec:SpatialDistribution}.
  {Alt text: Radial profile of the mass function slope normalized by the half-mass radius, showing measurements from all stars, major-axis stars, and minor-axis stars, alongside literature values. The plot highlights differences inside and outside the tidal radius, indicated by a vertical dashed line.}}\label{fig:MF_AlphaProfile}
\end{figure}

In figures \ref{fig:MF_AllData}-\ref{fig:MF_MinorAxis}, we detect a radial gradient in the slope of the MFs. Such a radial variation is expected to trace the internal dynamical evolution of GC. To investigate this in more detail, figure \ref{fig:MF_AlphaProfile} shows the MF slope $\alpha_{\rm MF}$ measured at different radii as a function of distance from the cluster center. The horizontal axis represents both the distance normalized by the half-mass radius $r_h$ (lower x-axis) and the physical distance (upper x-axis). The top panel of figure \ref{fig:MF_AlphaProfile} focuses on the inner region ($0 < R/r_h < 4$), while the bottom panel displays the measurements over the entire observed field. In figure \ref{fig:MF_AlphaProfile}, black, red, and blue circles correspond to the results derived from the full data, major-axis, and minor-axis samples, respectively; for clarity, small offsets are applied along the horizontal axis. White squares indicate the results from the deep HST and Large Binocular Telescope (LBT) observations of \citet{2015ApJ...814..144B}. In the bottom panel, the vertical black dashed line marks the tidal radius of NGC~5466 as estimated in this study. In the top panel, our measurements are in good agreement with the radial trend reported by \citet{2015ApJ...814..144B}, indicating that our HSC-based analysis and calibration reliably recover MS stars down to $\sim 0.5~M/M_{\odot}$.

The radial variation of the MF slope, $\delta \alpha$, evolves in response to internal relaxation processes and therefore provides a key diagnostic of the degree of mass segregation in GC. $\delta \alpha$ corresponds to the slope of the linear fit applied to the data points shown in figure \ref{fig:MF_AlphaProfile}. Using the measurements within the tidal radius (black points in the top panel of figure \ref{fig:MF_AlphaProfile}), we estimate $\delta \alpha$ by fitting a linear relation to the full data (black circles) as a function of radius. The fit is performed using an MCMC approach, adopting the same likelihood function as in equation (\ref{eq:MF_likelihood}) for a first-order model and assuming uninformative priors. We run the MCMC with 100 walkers, 110,000 steps, and a burn-in phase of 100,000 steps; the resulting posterior distributions are presented in the Appendix \ref{appendix:MCMC_Results}. From this analysis, we obtain $\delta \alpha = -0.91 \pm 0.06$. This value is larger than the estimate of $\delta \alpha = -0.6$ reported by \citet{2015ApJ...814..144B}, which can be attributed to differences in the stellar mass ranges considered. While \citet{2015ApJ...814..144B} measured $\delta \alpha$ over $0.3 < M/M_{\odot} < 0.8$, our estimate is based on a narrower and higher-mass interval, $0.55 < M/M_{\odot} < 0.75$. \citet{2016MNRAS.463.2383W} showed that, although the qualitative evolution of $\delta \alpha$ is similar across different mass ranges, values derived from higher-mass intervals ($0.5 < M/M_{\odot} < 0.8$) evolve toward systematically lower slopes than those obtained from broader mass ranges. From figures 4 and 5 of \citet{2016MNRAS.463.2383W}, after the dynamical time, the difference in $\delta \alpha$ between these two mass ranges reaches $\sim 0.5$, consistent with the discrepancy between our result and that of \citet{2015ApJ...814..144B}. This suggests that NGC~5466 has undergone significant internal dynamical evolution, with stars in the higher-mass range experiencing more rapid and more advanced mass segregation than lower-mass stars.

In the bottom panel of figure \ref{fig:MF_AlphaProfile}, the MF slopes derived from all three samples (black, red, and blue circles) systematically decrease with increasing radius, corresponding to a steepening of the MF in the outer regions. This behavior is consistent with mass segregation in NGC~5466, whereby more massive stars preferentially sink toward the cluster center while lower-mass stars migrate outward, resulting in a flatter MF in the core and a steeper MF at larger radii. Within the tidal radius, all samples exhibit similar radial trends in $\alpha$. Beyond the tidal radius, however, the major-axis (stream-focused) sample (red circles) shows systematically larger MF slopes than the full data and minor-axis samples, which include off-stream regions. This indicates a relative deficiency of low-mass stars within the stream at the same cluster-distance. This result suggests that low-mass stars were first displaced outward by internal dynamical processes such as mass segregation and were then further removed from the stream through tidal stripping.

Finally, in figure \ref{fig:MF_AlphaProfile}, the MF slope of the full data sample beyond the tidal radius appears to show quasi-periodic radial fluctuations. One possible explanation for such behavior is the epicyclic motion of tidally stripped stars (e.g., \cite{2008MNRAS.387.1248K,2010MNRAS.401..105K,2012MNRAS.423.2845L,2014ApJ...788..181N}). Stars escaping from the GC generally follow orbits similar to those of the parent cluster, but small velocity offsets at the time of escape can induce epicyclic oscillations. These oscillations are predicted to generate alternating overdense and underdense regions along the stream. In our data, the observed periodic radial variations in the MF slope may reflect locations where low-mass stars preferentially accumulate due to such epicyclic motions. However, similar epicyclic gaps reported in streams such as Palomar 5 and GD-1 typically occur on spatial scales of several hundred parsecs to a few kiloparsecs, which are larger than the $\sim 100$ pc-scale structures suggested by our measurements (e.g., \cite{2017MNRAS.470...60E,2020ApJ...891..161I}). Moreover, the predicted location of the first epicyclic overdensity ($\sim 2.8$ kpc; this value is derived assuming a tidal radius of 16.1 arcmin and a distance of 16 kpc, and adopting the framework of \citet{2008MNRAS.387.1248K} for a cluster embedded in a point-mass Galactic potential) lies farther from the cluster center than the region where we detect these fluctuations. Although our analysis uses only data above the 50\% detection completeness limit and shows no evidence for spatially periodic completeness variations, the low-mass nature of the stellar sample makes it susceptible to overcorrections for completeness and to spatially varying contamination from unresolved background galaxies. We therefore caution that the apparent periodic structures may, at least in part, be artificial, and further investigation is required to robustly establish their physical origin.

\section{Conclusions}\label{sec:conclusions}
We have carried out deep imaging observations of NGC~5466 using Subaru/HSC with the {\it NB395} filter. The observations covered the stellar stream out to a projected distance of 400 pc from the cluster center and reached depths 1-2 magnitudes fainter than previous {\it CaHK} surveys. By applying a kNN algorithm optimized for our CCMD, we have selected member stars and successfully detected both the stellar stream and its associated gap structure.

Using {\it NB395}-selected stars, photometric metallicities estimation has been performed based on the {\it NB395} color index. Estimated photometric metallicities are consistent with previous reported values, indicating our calibration is well-performed. From this estimation, metallicity distributions have been constructed, and this result suggests a trend toward higher metallicities in the outer regions of NGC~5466. Although this would nominally indicate the presence of a metallicity gradient in NGC~5466, the observed trend is likely dominated by contamination effects rather than representing an intrinsic property of the cluster, or the presence of MPs may have caused a systematic shift in the flux covered by the {\it NB395} filter in the outer regions.

To investigate the structure of the stellar stream, we have constructed the radial surface density profiles by combining our data with Gaia measurements. The main body of GC is well described by a King profile with a core radius of $r_c = 1.75$ pc and a tidal radius of $r_t = 16.1$ pc, while the stream region is well reproduced by a power-law profile with a slope of $\alpha=-4.53$. These results are in good agreement with previous studies (\cite{2012MNRAS.419...14C}) and enable us to identify individual faint stars belonging to the NGC 5466 stream. Moreover, we have detected power-law components even in the direction perpendicular to the prominent stream, suggesting that NGC~5466 has experienced multiple passages through apogalacticon.

By comparing the observed gap structure with the N-body simulation project eTidal, we have found that a gap is also present in the simulation data at approximately the same location as detected in this study. Because the eTidal simulations do not include encounters with dark matter subhaloes, this agreement suggests that the gap can be reproduced by the orbital evolution of NGC~5466 alone. In particular, given that NGC 5466 is thought to have experienced a recent pericentric passage and a crossing of the Galactic disc, one or both of these events may have contributed to the formation of the observed gap.

To investigate the dynamical evolution of NGC~5466, we have constructed the mass profile of low-mass MS stars. We have found that the slope of the MF becomes progressively steeper with increasing radius, indicating a clear radial evolution of the MF. This behavior suggests that NGC~5466 is a dynamically evolved system in which two-body relaxation has played a significant role (e.g., \cite{2017MNRAS.471.3668S}). Furthermore, the MF profile measured along the stream direction (major-axis) exhibits a flatter slope than that measured perpendicular to the stream (minor axis) in regions beyond the tidal radius. This anisotropy implies that low-mass stars, having migrated to larger radii through internal relaxation and mass segregation, have subsequently been preferentially removed by tidal stripping. Such a scenario is consistent with a recent pericentric passage of NGC~5466, during which tidal effects are expected to be strongest.

In this study, using Subaru/HSC and {\it NB395} deep photometry, we have found new structural features traced by low-mass MS stars in addition to the previously reported results. Furthermore, this work has demonstrated the feasibility of calibration and photometric metallicity estimation using the {\it NB395} filter, so we conclude that HSC/{\it NB395} performs comparably to established metallicity-sensitive systems such as Pristine/{\it CaHK}, MAGIC/{\it CaHK}, and the SkyMapper $\nu$-band. This fact highlights the suitability of HSC/{\it NB395} for future wide-field surveys targeting low-metallicity populations and/or identifying metal-poor stars in the Galactic stellar halo, such as the ongoing Zero Enrichment Rare Objects (ZERO) survey (PI: M. Chiba).

\begin{ack}
Data analysis was in part carried out on the Multi-wavelength Data Analysis System (MDAS) operated by the Astronomy Data Center (NAOJ/ADC) and the Large-scale data analysis system (LSC) co-operated by the ADC and Subaru Telescope, NAOJ.
Based in part on data collected at Subaru Telescope and obtained from the Subaru-Mitaka-Okayama-Kiso Archive System (SMOKA), which is operated by the NAOJ/ADC. 
This research is based on data collected at the Subaru Telescope, which is operated by the National Astronomical Observatory of Japan. We are honored and grateful for the opportunity of observing the Universe from Maunakea, which has the cultural, historical, and natural significance in Hawaii.
This work was supported by the NAOJ Research Coordination Committee, NINS (NAOJ-RCC-2402-0401).
This work was supported by JSPS KAKENHI grant Nos. JP25K01046, JP24K00669, JP25H00394, JP23KF0290, JP22K14076, JP21H04499, and JP20H05855. 
We especially thank JP17H01101 for funding {\it NB395}.
This work was supported by JSPS Core-to-Core Program (grant number: JPJSCCA20210003), the Overseas Travel Fund for Students (2024) of the Astronomical Science Program, the Graduate University for Advanced Studies, SOKENDAI and the NAOJ Research Coordination Committee, NINS (NAOJ-RCC-2402-0401).  AMNF is supported by UK Research and Innovation (UKRI) under the UK government's Horizon Europe funding guarantee [grant number EP/Z534353/1] and by the Science and Technology Facilities
Council [grant number ST/Y001281/1].
The Pan-STARRS1 Surveys (PS1) and the PS1 public science archive have been made possible through contributions by the Institute for Astronomy, the University of Hawaii, the Pan-STARRS Project Office, the Max-Planck Society and its participating institutes, the Max Planck Institute for Astronomy, Heidelberg and the Max Planck Institute for Extraterrestrial Physics, Garching, The Johns Hopkins University, Durham University, the University of Edinburgh, the Queen's University Belfast, the Harvard-Smithsonian Center for Astrophysics, the Las Cumbres Observatory Global Telescope Network Incorporated, the National Central University of Taiwan, the Space Telescope Science Institute, the National Aeronautics and Space Administration under Grant No. NNX08AR22G issued through the Planetary Science Division of the NASA Science Mission Directorate, the National Science Foundation Grant No. AST–1238877, the University of Maryland, Eotvos Lorand University (ELTE), the Los Alamos National Laboratory, and the Gordon and Betty Moore Foundation. 
\end{ack}



\appendix 
\section{Photometric transformation between Subaru/HSC and CFHT/MegaCam systems}\label{appendix:PhtometricTransformation}
Photometric transformations are required to compare our results with the Pristine dataset, which is carried out in different filter systems. In this section, color conversion formulae relevant to our study are summarized. The method to obtain the formulae is similar to those used in previous studies, such as \citet{2010AJ....140.1814Y} and \citet{1995PASP..107..945F}. We calculate the colors of stars using the Bruzual–Persson–Gunn–Stryker (BPGS) Atlas by convolving the transmission curve of a CFHT/MegaCam filter system with the stellar spectra. The transmission curves for HSC can be obtained from the Subaru Telescope HSC website.

\begin{figure*}
 \begin{center}
  \includegraphics[width=\linewidth]{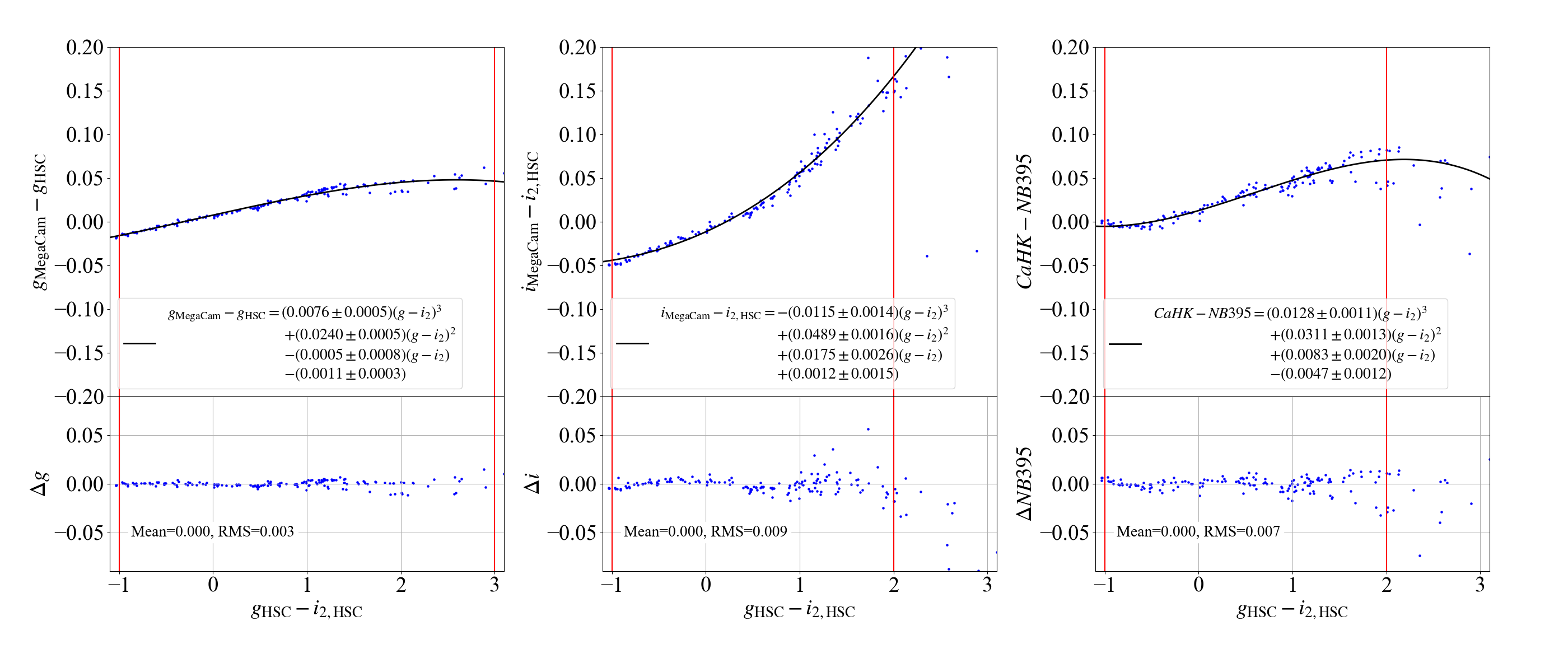}
 \end{center}
 \caption{Color conversion diagrams between Subaru/HSC and CFHT/MegaCam (left) ${\it g_{\rm HSC} - i_{2,\rm HSC}}$ and ${\it g_{\rm MegaCam} - g_{\rm HSC}}$, (middle) ${\it g_{\rm HSC} - i_{2,\rm HSC}}$ and ${\it i_{\rm MegaCam} - i_{2,\rm HSC}}$, and (right) ${\it g_{\rm HSC} - i_{2,\rm HSC}}$ and ${\it CaHK_{\rm MegaCam} - NB395_{\rm HSC}}$. For each of the panels: (top) the calculated colors for BPGS stars are plotted against $(g_{\rm HSC}-i_{2,\rm HSC}$. The fitted 3rd-order polynomial function is plotted as a black solid line, and the color range used for the fitting is plotted as vertical red solid lines, respectively. (Bottom) The residuals of the calculated color from the fitted functions.
  {Alt text: This figure shows color–color conversion diagrams between Subaru/HSC and CFHT/MegaCam for three color combinations. In each panel, the upper plot displays the calculated colors of BPGS stars as a function of $g_{\rm HSC} - i_{2,{\rm HSC}}$, with the third-order polynomial fit shown as a black solid line and the fitting range marked by vertical red lines. The lower plot presents the residuals between the calculated colors and the fitted functions.} }\label{fig:ColorConversion_HSC}
\end{figure*}

\begin{figure*}
 \begin{center}
  \includegraphics[width=\linewidth]{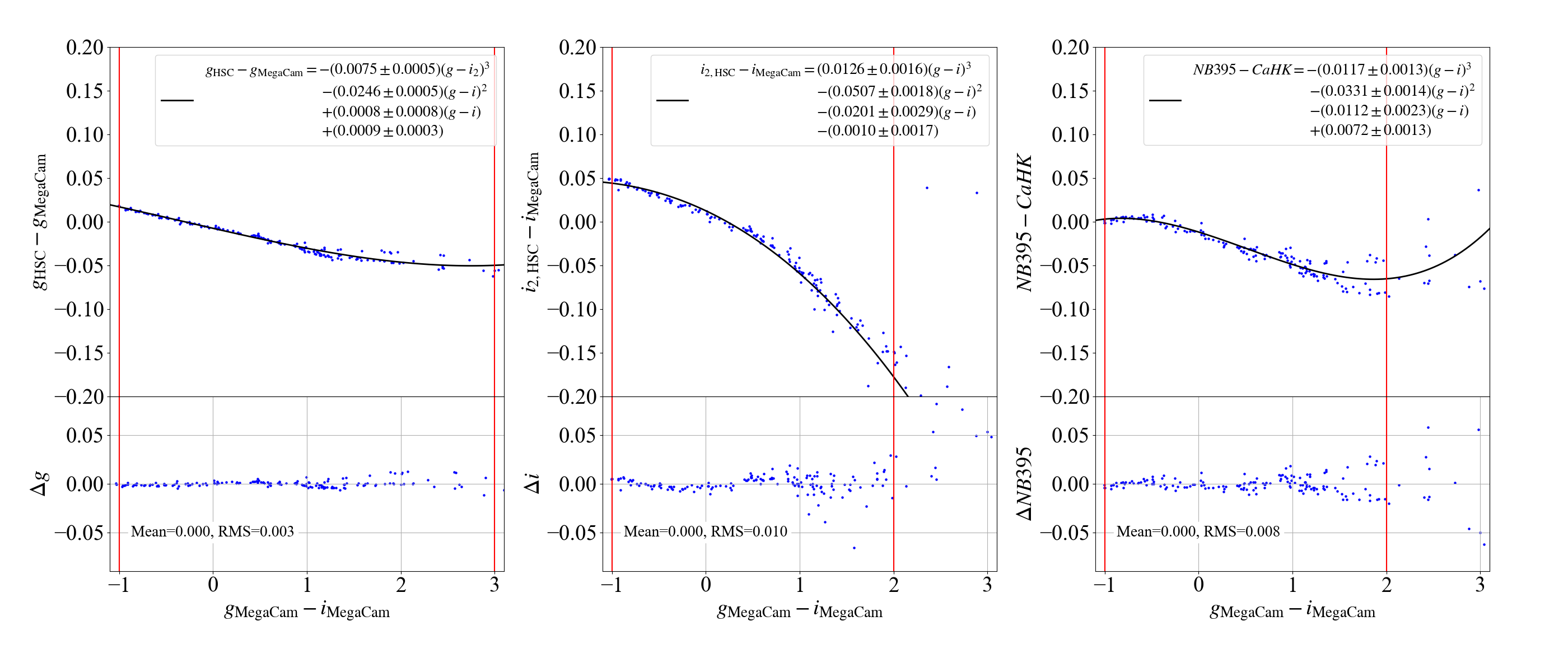}
 \end{center}
 \caption{Color conversion diagrams between Subaru/HSC and CFHT/MegaCam (left) ${\it g_{\rm MegaCam} - i_{\rm MegaCam}}$ and ${\it g_{\rm HSC} - g_{\rm MegaCam}}$, (middle) ${\it g_{\rm MegaCam} - i_{\rm MegaCam}}$ and ${\it i_{2,\rm HSC} - i_{\rm MegaCam}}$, and (right) ${\it g_{\rm MegaCam} - i_{\rm MegaCam}}$  and ${\it NB395_{\rm HSC} - CaHK_{\rm MegaCam}}$. For each of the panels: (top) the calculated colors for BPGS stars are plotted against $(g_{\rm MegaCam}-i_{\rm MegaCam}$. The fitted 3rd-order polynomial function is plotted as a black solid line, and the color range used for the fitting is plotted as vertical red solid lines, respectively. (Bottom) The residuals of the calculated color from the fitted functions.
 {Alt text: The same figure as figure \ref{fig:ColorConversion_HSC}, but the vertical axes correspond to ${\it g_{\rm HSC} - g_{\rm MegaCam}}$, ${\it g_{\rm MegaCam} - i_{\rm MegaCam}}$ and ${\it i_{2,\rm HSC} - i_{\rm MegaCam}}$ and ${\it g_{\rm MegaCam} - i_{\rm MegaCam}}$, while the horizontal axis has been replaced with ${\it g_{\rm MegaCam} - i_{\rm MegaCam}}$.} }\label{fig:ColorConversion_MegaCam}
\end{figure*}

Figures \ref{fig:ColorConversion_HSC} and \ref{fig:ColorConversion_MegaCam} show our calculation. The top panel of each of these two figures shows the calculated colors ($g_{\rm MegaCam} - g_{\rm HSC}$, $i_{\rm MegaCam} - i_{2,{\rm HSC}}$, ${\it CaHK} - {\it NB395}$ in figure \ref{fig:ColorConversion_HSC}, and $g_{\rm HSC}-g_{\rm MegaCam}$, $i_{2,{\rm HSC}}-i_{\rm MegaCam}$, ${\it NB395}-{\it CaHK}$ in figure \ref{fig:ColorConversion_MegaCam}) for BPGS stars plotted against
$g_{\rm MegaCam} - i_{\rm MegaCam}$ and $g_{\rm HSC} - i_{2,{\rm HSC}}$, respectively. The black solid curves are third-order polynomial functions fitted to BPGS stars (blue dots). To mitigate the influence of outliers, the polynomial fitting is performed only on the stars located between the red vertical lines in each panel. In the bottom panels of figures \ref{fig:ColorConversion_HSC} and \ref{fig:ColorConversion_MegaCam}, the residuals from the polynomial functions are plotted against $g_{\rm MegaCam} - i_{\rm MegaCam}$ and $g_{\rm HSC} - i_{2,{\rm HSC}}$. All fitting results yield root mean square (RMS) values below 0.01 within the range indicated by the red lines, demonstrating that these fits are performed very well in this interval. The color conversion formulae derived from our fitting procedure are described below:
\begin{align}
g_{\rm MegaCam} - g_{\rm HSC} = & (0.0076 \pm 0.0005)(g_{\rm HSC}-i_{2,{\rm HSC}})^3 \notag \\
& + (0.0240 \pm 0.0005)(g_{\rm HSC}-i_{2,{\rm HSC}})^2 \notag \\
& - (0.0005 \pm 0.0008)(g_{\rm HSC}-i_{2,{\rm HSC}}) \notag \\
& - (0.0011 \pm 0.0003)
\label{eq:gM-gH}
\end{align}
for $-1<g_{\rm HSC}-i_{2,\rm HSC} < 3$.
\begin{align}
i_{\rm MegaCam} - i_{2,{\rm HSC}} = & -(0.0115 \pm 0.0014)(g_{\rm HSC}-i_{2,{\rm HSC}})^3 \notag \\
& + (0.0489 \pm 0.0016)(g_{\rm HSC}-i_{2,{\rm HSC}})^2 \notag \\
& + (0.0175 \pm 0.0026)(g_{\rm HSC}-i_{2,{\rm HSC}}) \notag \\
& + (0.0012 \pm 0.0015)
\label{eq:iM-iH}
\end{align}
for $-1<g_{\rm HSC}-i_{2,\rm HSC} < 2$,
\begin{align}
{\it CaHK} - {\it NB395} = & (0.0128 \pm 0.0011)(g_{\rm HSC}-i_{2,{\rm HSC}})^3 \notag \\
& + (0.0311 \pm 0.0013)(g_{\rm HSC}-i_{2,{\rm HSC}})^2 \notag \\
& + (0.0083 \pm 0.0020)(g_{\rm HSC}-i_{2,{\rm HSC}}) \notag \\
& - (0.0047 \pm 0.0012)
\label{eq:CM-NH}
\end{align}
for $-1<g_{\rm HSC}-i_{2,\rm HSC} < 2$,
\begin{align}
g_{\rm HSC} - g_{\rm MegaCam} = & -(0.0075 \pm 0.0005)(g_{\rm MegaCam}-i_{\rm MegaCam})^3 \notag \\ 
& - (0.0246 \pm 0.0005)(g_{\rm MegaCam}-i_{\rm MegaCam})^2 \notag \\
& + (0.0008 \pm 0.0008)(g_{\rm MegaCam}-i_{\rm MegaCam}) \notag \\
& + (0.0009 \pm 0.0003)
\label{eq:gH-gM}
\end{align}
for $-1<g_{\rm MegaCam}-i_{\rm MegaCam} < 3$,
\begin{align}
i_{2, {\rm HSC}} - i_{\rm MegaCam} = & (0.0126 \pm 0.0016)(g_{\rm MegaCam}-i_{\rm MegaCam})^3 \notag \\ 
& - (0.0507 \pm 0.0018)(g_{\rm MegaCam}-i_{\rm MegaCam})^2 \notag \\ 
& - (0.0201 \pm 0.0029)(g_{\rm MegaCam}-i_{\rm MegaCam}) \notag \\
& - (0.0010 \pm 0.0017)
\label{eq:iH-iM}
\end{align}
for $-1<g_{\rm MegaCam}-i_{\rm MegaCam} < 2$,
\begin{align}
NB395 - {\it CaHK} = & -(0.0117 \pm 0.0013)(g_{\rm MegaCam}-i_{\rm MegaCam})^3 \notag \\ 
& - (0.0331 \pm 0.0014)(g_{\rm MegaCam}-i_{\rm MegaCam})^2 \notag \\ 
& - (0.0112 \pm 0.0023)(g_{\rm MegaCam}-i_{\rm MegaCam}) \notag \\
& + (0.0072 \pm 0.0013)
\label{eq:NH-CM}
\end{align}
for $-1<g_{\rm MegaCam}-i_{\rm MegaCam} < 2$.

\section{MCMC results}\label{appendix:MCMC_Results}
\begin{figure*}
 \begin{center}
  \includegraphics[width=0.5\linewidth]{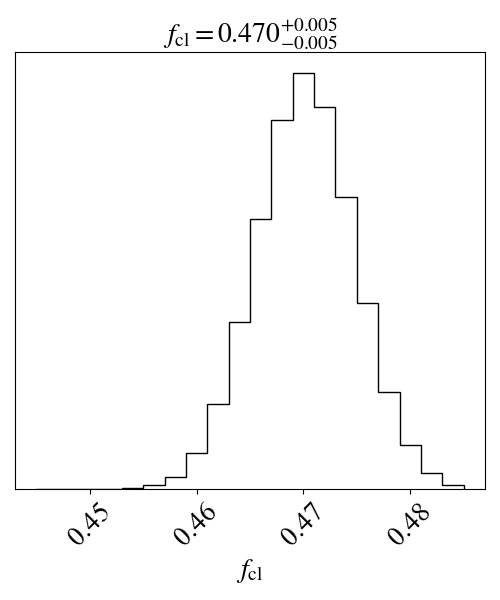}
 \end{center}
 \caption{The posterior distribution of the relative fraction of clusters, $f_{\rm cl}$. {Alt text: MCMC result of the relative fraction of clusters, $f_{\rm cl}$.} }\label{fig:MCMC_Pmem}
\end{figure*}

\begin{figure*}
 \begin{center}
  \includegraphics[width=\linewidth]{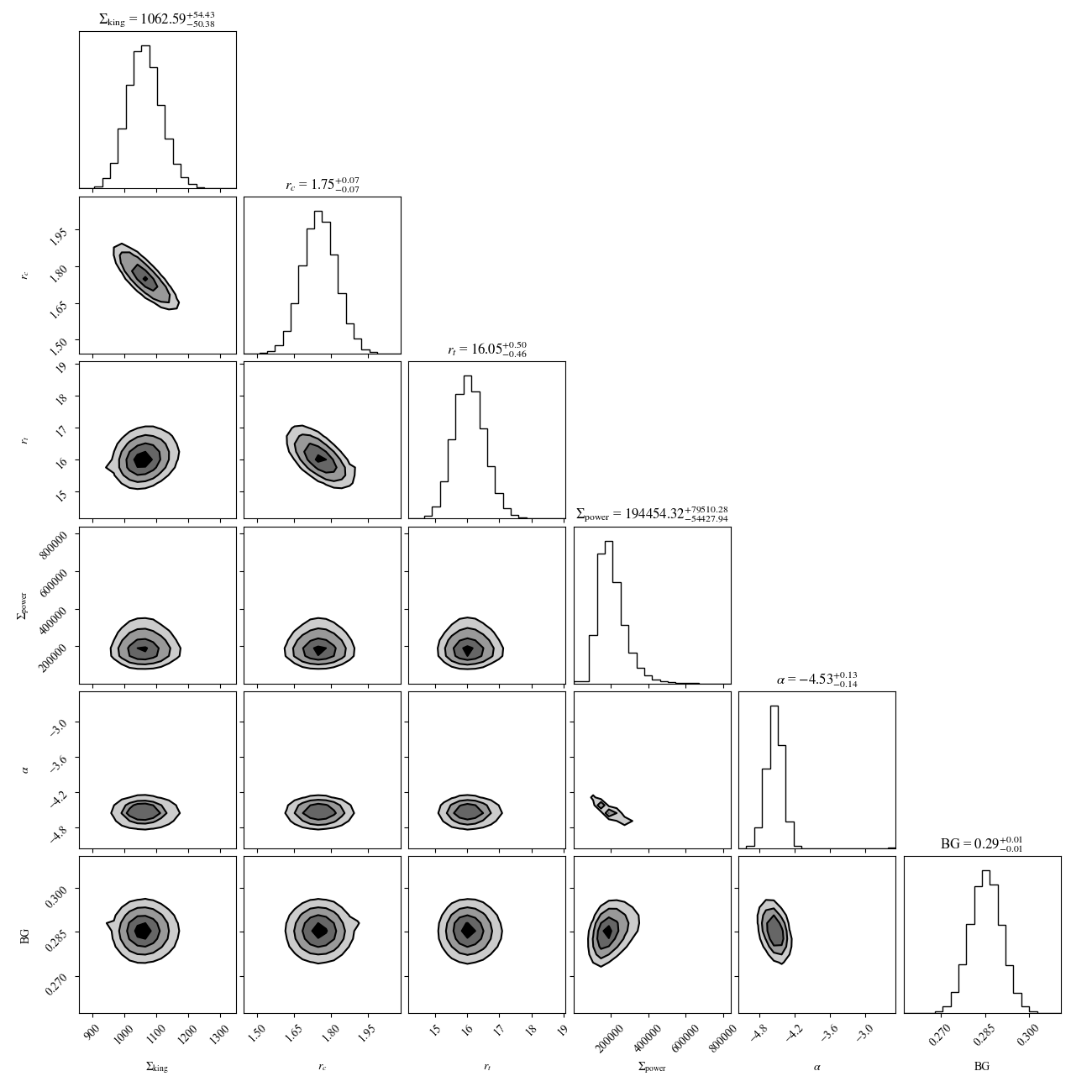}
 \end{center}
 \caption{The posterior distributions and marginalized distributions in the radial density profile. Each panel shows the central surface density scale ($\Sigma_{\rm king}$), core radius ($r_c$), tidal radius ($r_t$), scaledisk scale radius, surface density scale of the halo, projected power-law index, and contamination scale, from left (top) to right (bottom). {Alt text: MCMC results of the radial density profile for full data. Two-dimensional posterior contours illustrate parameter covariances, while the diagonal panels show the marginalized distributions with their peak likelihood values.} }\label{fig:MCMC_RadialProfile}
\end{figure*}

\begin{figure*}
 \begin{center}
  \includegraphics[width=\linewidth]{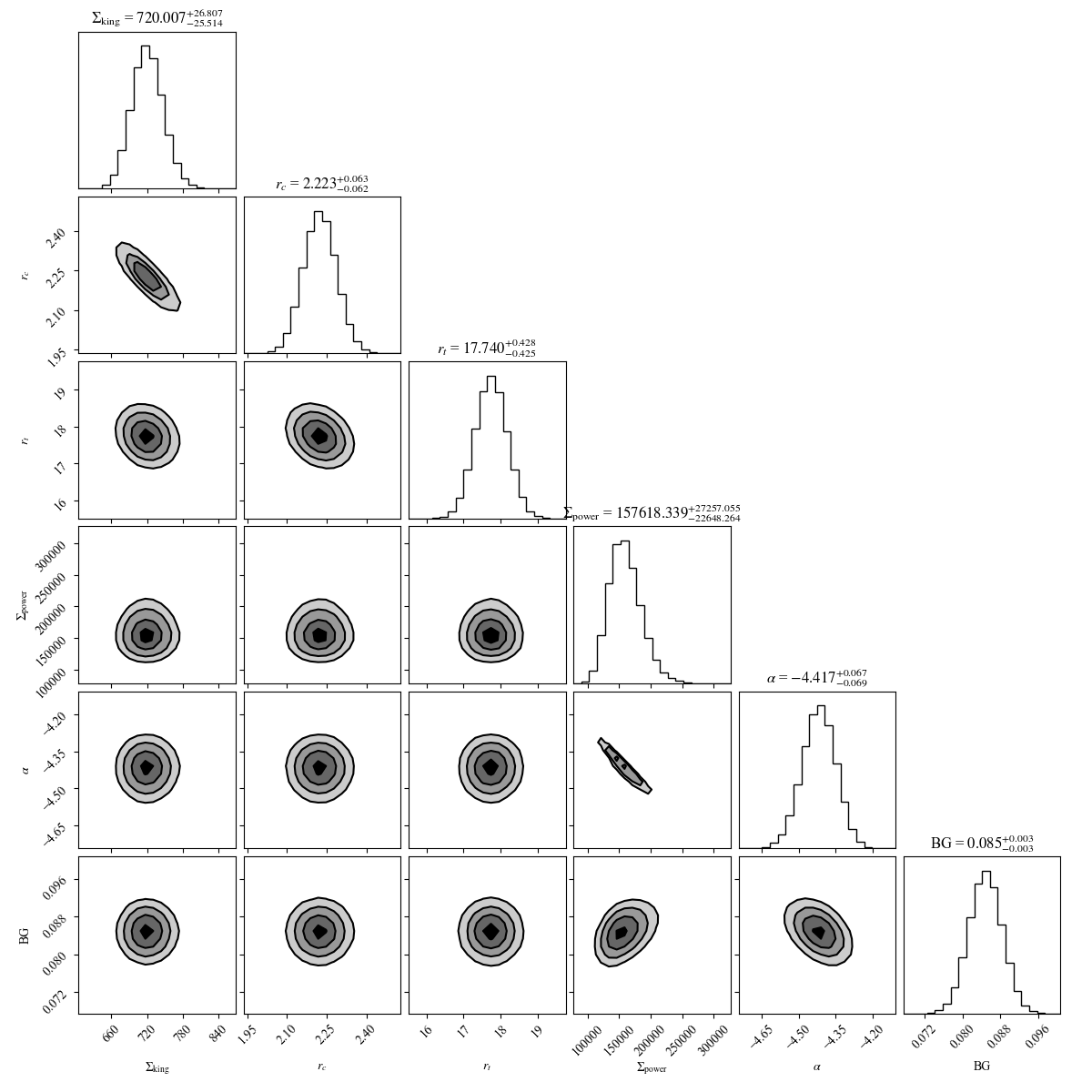}
 \end{center}
 \caption{The same as in figure \ref{fig:MCMC_RadialProfile}, but for major-axis data. {Alt text: MCMC results of the radial density profile for major-axis data.} }\label{fig:MCMC_RadialProfile_MajorAxis}
\end{figure*}

\begin{figure*}
 \begin{center}
  \includegraphics[width=\linewidth]{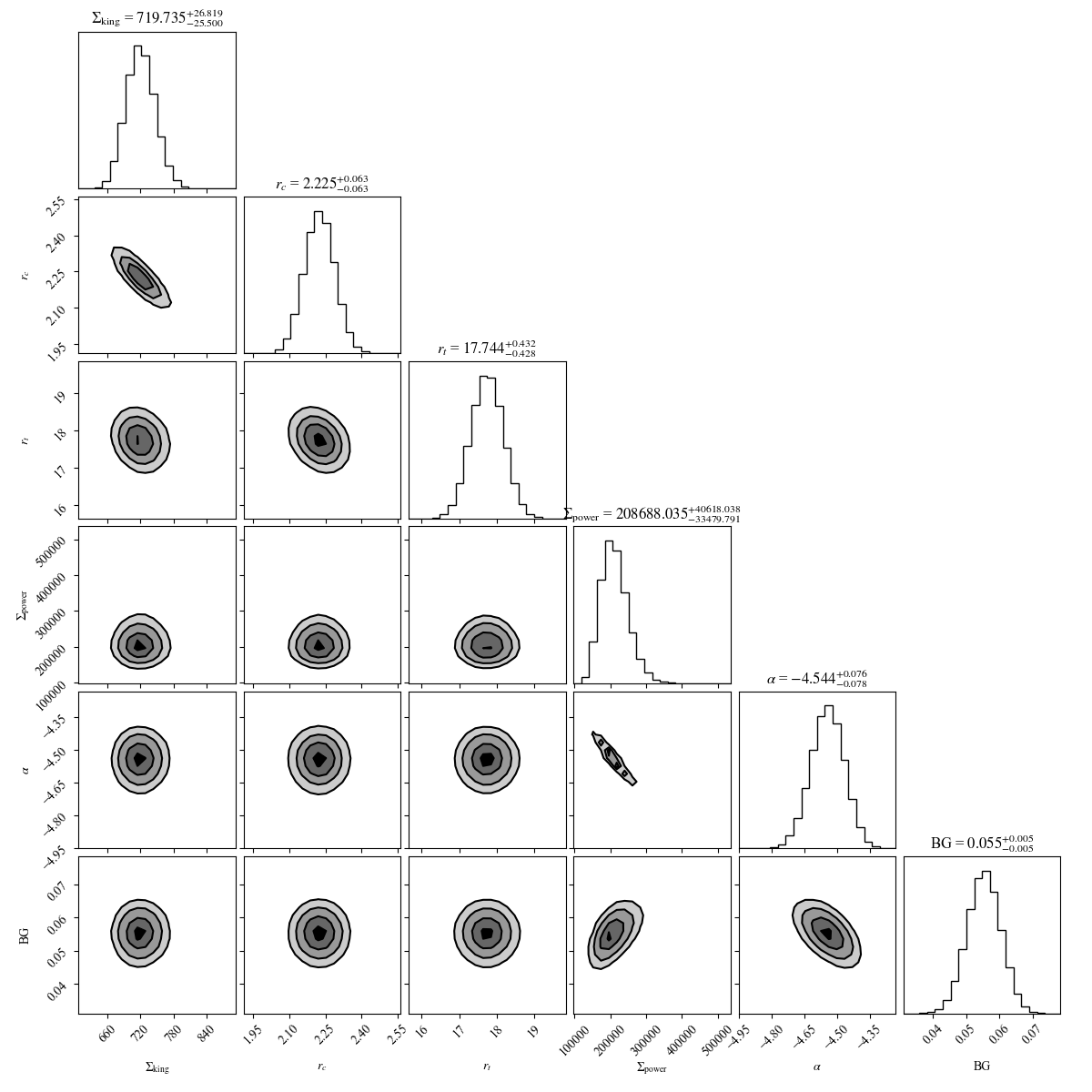}
 \end{center}
 \caption{The same as in figure \ref{fig:MCMC_RadialProfile}, but for minor-axis data. {Alt text: MCMC results of the radial density profile for minor-axis data.} }\label{fig:MCMC_RadialProfile_MinorAxis}
\end{figure*}

\begin{figure*}
 \begin{center}
  \includegraphics[width=\linewidth]{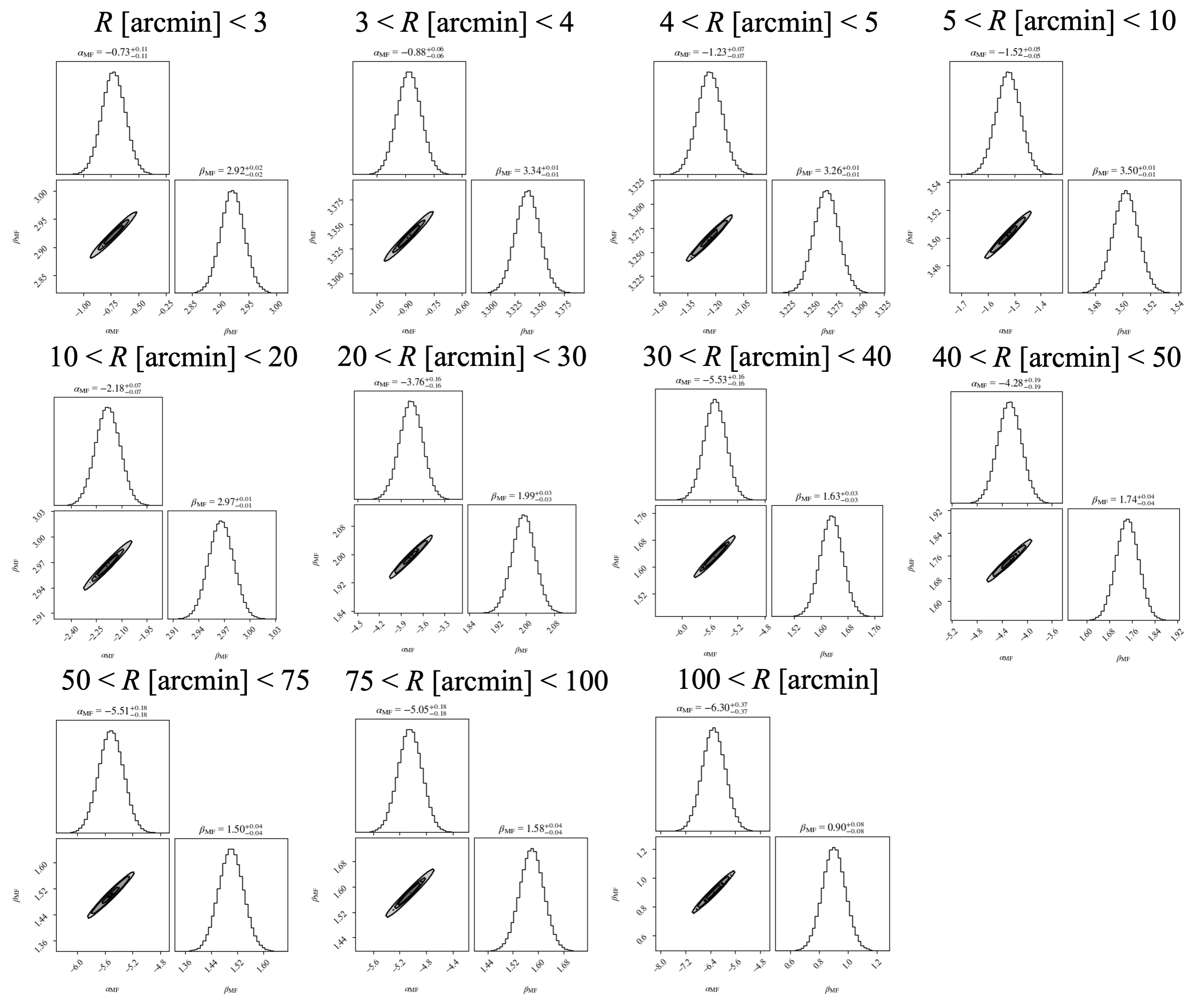}
 \end{center}
 \caption{The posterior distributions and marginalized distributions in the MF of ``full data''. Each panel shows the slope of MF, $\alpha_{\rm MF}$, and intercept, $\beta_{\rm MF}$, from left (top) to right (bottom).
  {Alt text: Same layout as figure \ref{fig:MCMC_RadialProfile}, but for the MF of ``full data''.} }\label{fig:MCMC_MF_AllData}
\end{figure*}

\begin{figure*}
 \begin{center}
  \includegraphics[width=\linewidth]{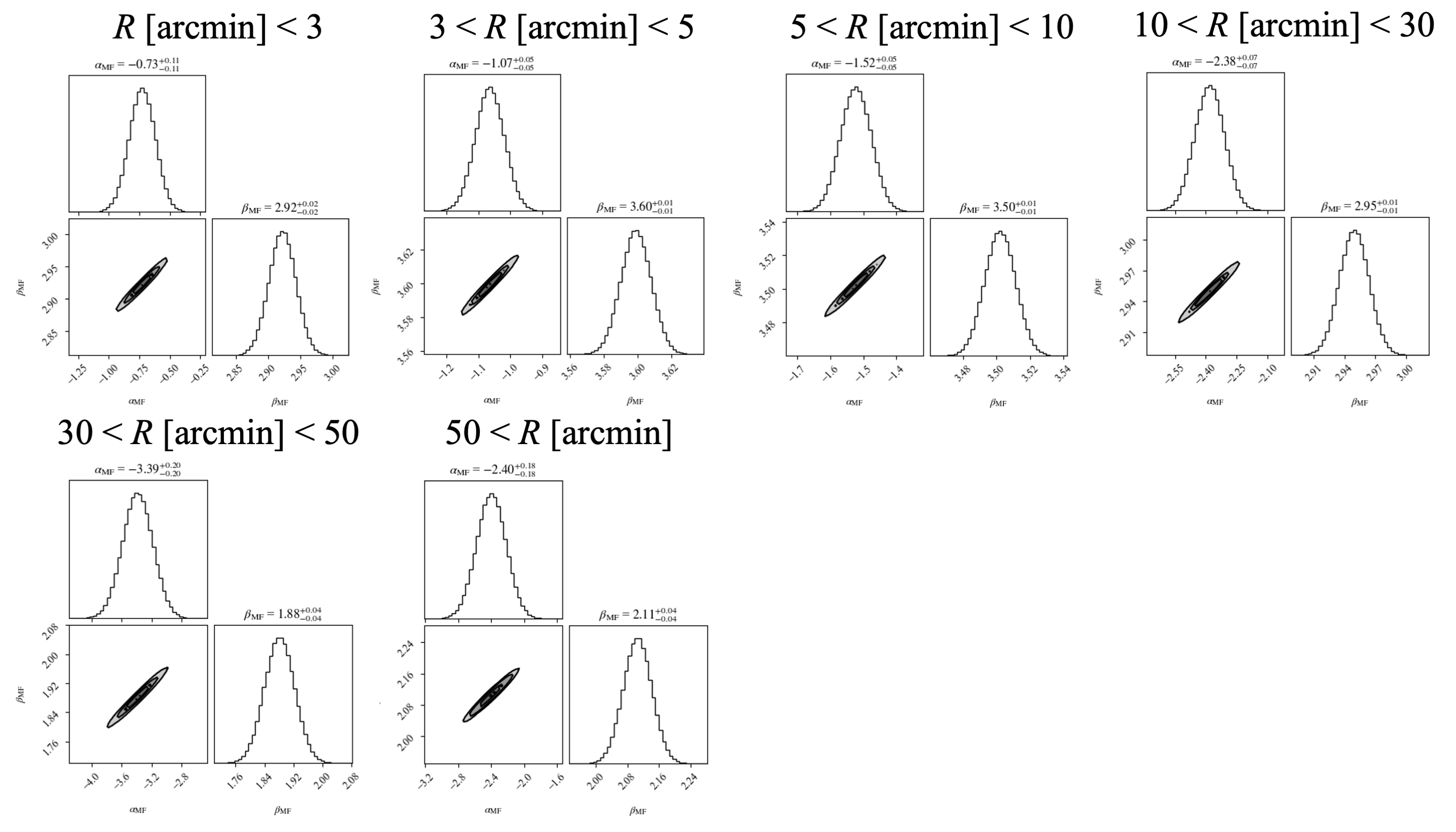}
 \end{center}
 \caption{The same as in figure \ref{fig:MCMC_MF_AllData} but for the case of the ``major-axis data''.
  {Alt text: Same plot as figure \ref{fig:MCMC_MF_AllData}, but for the ``major-axis data''.} }\label{fig:MCMC_MF_MajorAxis}
\end{figure*}

\begin{figure*}
 \begin{center}
  \includegraphics[width=\linewidth]{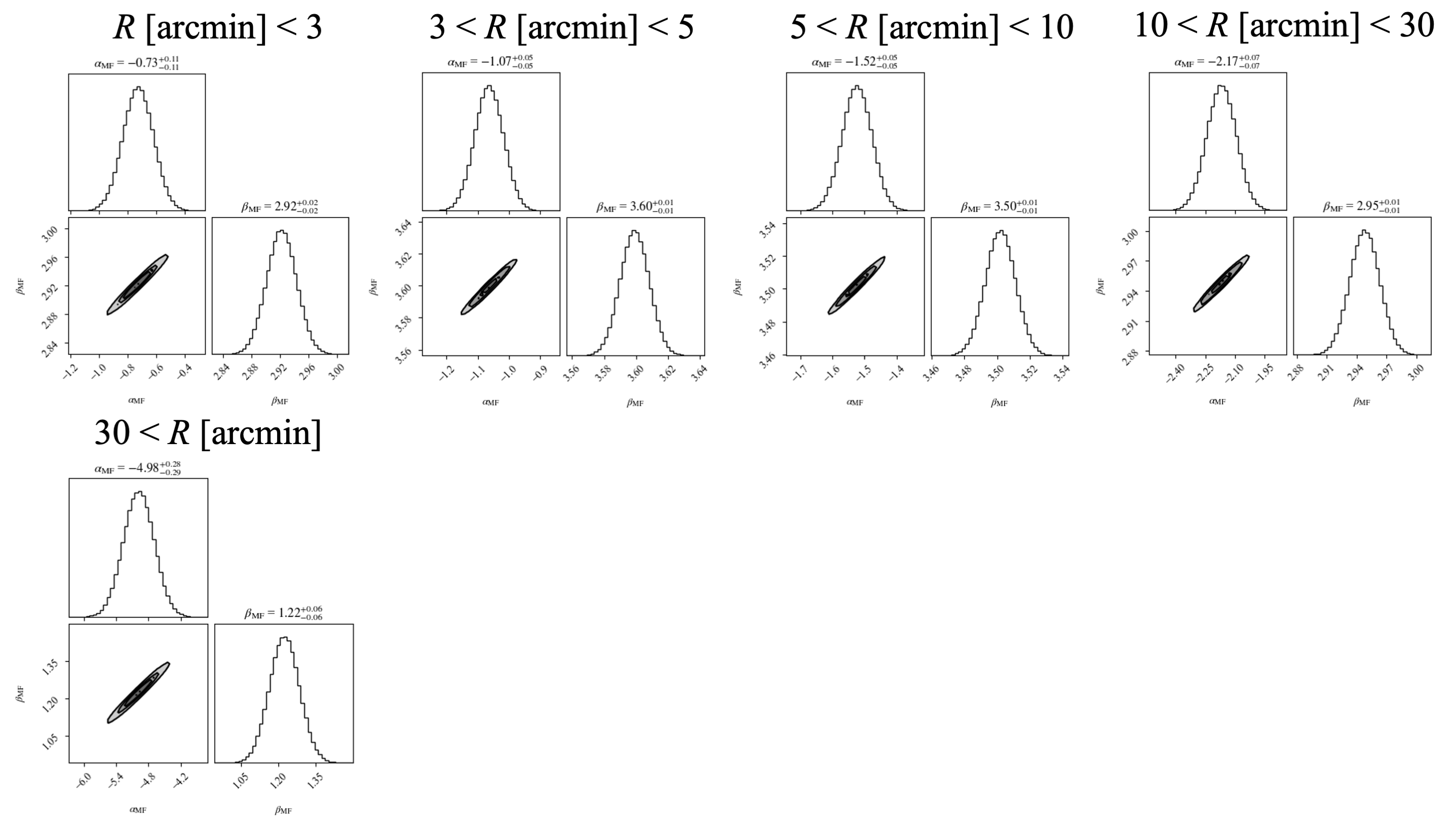}
 \end{center}
 \caption{The same as in figure \ref{fig:MCMC_MF_AllData} but for the case of the ``minor-axis data''.
  {Alt text: Same plot as figure \ref{fig:MCMC_MF_AllData}, but for the ``minor-axis data''.} }\label{fig:MCMC_MF_MinorAxis}
\end{figure*}

\begin{figure*}
 \begin{center}
  \includegraphics[width=\linewidth]{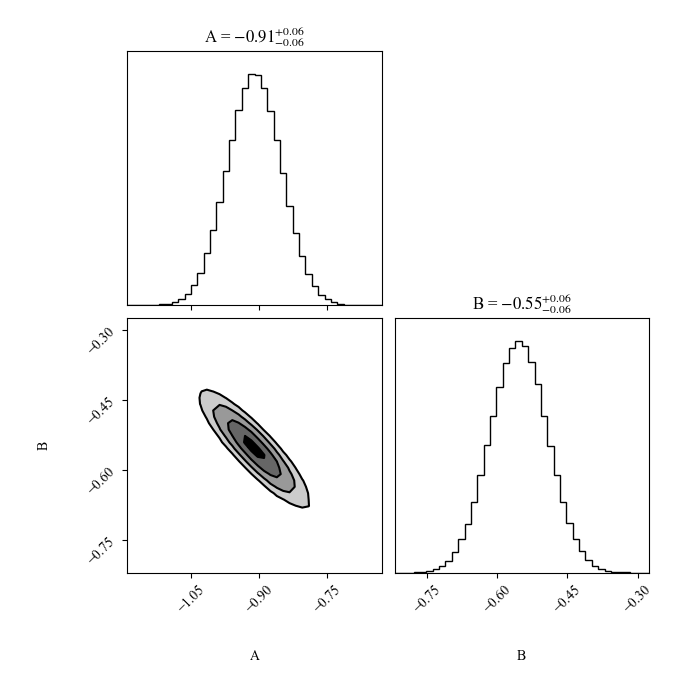}
 \end{center}
 \caption{The posterior distributions and marginalized distributions for the linear-fitting of the radial variation of slopes in MFs. Each panel shows the slope, $A$, and intercept, $B$, from left (top) to right (bottom).
  {Alt text: Same layout as figure \ref{fig:MCMC_RadialProfile}, but for the linear-fitting of radial variation of MF slopes.} }\label{fig:MCMC_MF_MFvariation}
\end{figure*}

Here, we show the MCMC results as corner plots (posterior distribution and marginal distributions for all estimated parameters) in the case of the membership estimation (figure \ref{fig:MCMC_Pmem}), the radial density profiles (figures \ref{fig:MCMC_RadialProfile}-\ref{fig:MCMC_RadialProfile_MinorAxis}, MFs (figure \ref{fig:MCMC_MF_AllData}-\ref{fig:MCMC_MF_MinorAxis}), and linear-fitting to the radial variation of MF slopes (figure \ref{fig:MCMC_MF_MFvariation}).

\bibliography{NGC5466_bibtex}{}

@ARTICLE{2025ApJ...993...77B,
       author = {{Barbosa}, F.~O. and {Chiti}, A. and {Limberg}, G. and {Pace}, A.~B. and {Cerny}, W. and {Rossi}, S. and {Carlin}, J.~L. and {Stringfellow}, G.~S. and {Placco}, V.~M. and {Atzberger}, K. and {Carballo-Bello}, J.~A. and {Chaturvedi}, A. and {Choi}, Y. and {Crnojevi{\'c}}, D. and {Drlica-Wagner}, A. and {Ji}, A.~P. and {Kallivayalil}, N. and {Mart{\'\i}nez-V{\'a}zquez}, C.~E. and {Medina}, G.~E. and {No{\"e}l}, N.~E.~D. and {Riley}, A.~H. and {Sand}, D.~J. and {Vivas}, A.~K. and {Bom}, C.~R. and {Ferguson}, P.~S. and {Mutlu-Pakdil}, B. and {Navabi}, M. and {Sakowska}, J.~D. and {Zenteno}, A. and {MAGIC Collaboration} and {Delve Collaboration}},
        title = "{The DECam MAGIC Survey: A Wide-field Photometric Metallicity Study of the Sculptor Dwarf Spheroidal Galaxy}",
      journal = {\apj},
         year = 2025,
        month = nov,
       volume = {993},
       number = {1},
          eid = {77},
        pages = {77},
          doi = {10.3847/1538-4357/ae0039},
archivePrefix = {arXiv},
       eprint = {2504.03593},
 primaryClass = {astro-ph.GA},
       adsurl = {https://ui.adsabs.harvard.edu/abs/2025ApJ...993...77B}
}

@ARTICLE{2025AJ....170..157K,
       author = {{Kuzma}, Pete B. and {Ishigaki}, Miho N. and {Kirihara}, Takanobu and {Ogami}, Itsuki},
        title = "{Constructing a Pristine View of Extended Globular Cluster Structure}",
      journal = {\aj},
         year = 2025,
        month = sep,
       volume = {170},
       number = {3},
          eid = {157},
        pages = {157},
          doi = {10.3847/1538-3881/aded8e},
archivePrefix = {arXiv},
       eprint = {2507.05590},
 primaryClass = {astro-ph.GA},
       adsurl = {https://ui.adsabs.harvard.edu/abs/2025AJ....170..157K}
}

@ARTICLE{2025MNRAS.537.2752K,
       author = {{Kuzma}, P.~B. and {Ishigaki}, M.~N.},
        title = "{A pristine view of galactic globular clusters and their peripheries: Omega Centauri}",
      journal = {\mnras},
         year = 2025,
        month = mar,
       volume = {537},
       number = {3},
        pages = {2752-2762},
          doi = {10.1093/mnras/staf212},
archivePrefix = {arXiv},
       eprint = {2502.01135},
 primaryClass = {astro-ph.GA},
       adsurl = {https://ui.adsabs.harvard.edu/abs/2025MNRAS.537.2752K}
}

@software{2022zndo....596036B,
       author = {{Bradley}, Larry and {Sip{\H{o}}cz}, Brigitta and {Robitaille}, Thomas and {Tollerud}, Erik and {Vin{\'\i}cius}, Z{\'e} and {Deil}, Christoph and {Barbary}, Kyle and {Wilson}, Tom J and {Busko}, Ivo and {Donath}, Axel and {G{\"u}nther}, Hans Moritz and {Cara}, Mihai and {Lim}, P.~L. and {Me{\ss}linger}, Sebastian and {Burnett}, Zach and {Conseil}, Simon and {Droettboom}, Michael and {Bostroem}, Azalee and {Bray}, E.~M. and {Andersen Bratholm}, Lars and {Jamieson}, William and {Ginsburg}, Adam and {Barentsen}, Geert and {Craig}, Matt and {Pascual}, Sergio and {Rathi}, Shivangee and {Perrin}, Marshall and {Morris}, Brett M.},
        title = "{astropy/photutils: 2.2.0}",
         year = 2025,
        month = feb,
          eid = {10.5281/zenodo.596036},
          doi = {10.5281/zenodo.596036},
      version = {2.2.0},
    publisher = {Zenodo},
       adsurl = {https://ui.adsabs.harvard.edu/abs/2022zndo....596036B}
}

@ARTICLE{2025MNRAS.536..530O,
       author = {{Ogami}, Itsuki and {Tanaka}, Mikito and {Komiyama}, Yutaka and {Chiba}, Masashi and {Guhathakurta}, Puragra and {Kirby}, Evan N. and {Wyse}, Rosemary F.~G. and {Filion}, Carrie and {Gilbert}, Karoline M. and {Escala}, Ivanna and {Mori}, Masao and {Kirihara}, Takanobu and {Tanaka}, Masayuki and {Ishigaki}, Miho N. and {Hayashi}, Kohei and {Lee}, Myun Gyoon and {Sharma}, Sanjib and {Kalirai}, Jason S. and {Lupton}, Robert H.},
        title = "{The structure of the stellar halo of the Andromeda galaxy explored with the NB515 for Subaru/HSC - I. New insights on the stellar halo up to 120 kpc}",
      journal = {\mnras},
         year = 2025,
        month = jan,
       volume = {536},
       number = {1},
        pages = {530-553},
          doi = {10.1093/mnras/stae2527},
archivePrefix = {arXiv},
       eprint = {2401.00668},
 primaryClass = {astro-ph.GA},
       adsurl = {https://ui.adsabs.harvard.edu/abs/2025MNRAS.536..530O}
}

@ARTICLE{2024A&A...692A.115M,
       author = {{Martin}, Nicolas F. and {Starkenburg}, Else and {Yuan}, Zhen and {Fouesneau}, Morgan and {Ardern-Arentsen}, Anke and {De Angeli}, Francesca and {Gran}, Felipe and {Montelius}, Martin and {Rusterucci}, Samuel and {Andrae}, Ren{\'e} and {Bellazzini}, Michele and {Montegriffo}, Paolo and {Esselink}, Anna F. and {Zhang}, Hanyuan and {Venn}, Kim A. and {Viswanathan}, Akshara and {Aguado}, David S. and {Battaglia}, Giuseppina and {Bayer}, Manuel and {Bonifacio}, Piercarlo and {Caffau}, Elisabetta and {C{\^o}t{\'e}}, Patrick and {Carlberg}, Raymond and {Fabbro}, S{\'e}bastien and {Fern{\'a}ndez-Alvar}, Emma and {Gonz{\'a}lez Hern{\'a}ndez}, Jonay I. and {Gonz{\'a}lez Rivera de La Vernhe}, Isaure and {Hill}, Vanessa and {Ibata}, Rodrigo A. and {Jablonka}, Pascale and {Kordopatis}, Georges and {Lardo}, Carmela and {McConnachie}, Alan W. and {Navarrete}, Camila and {Navarro}, Julio and {Recio-Blanco}, Alejandra and {S{\'a}nchez-Janssen}, Rub{\'e}n and {Sestito}, Federico and {Thomas}, Guillaume F. and {Vitali}, Sara and {Youakim}, Kristopher},
        title = "{The Pristine survey: XXIII. Data Release 1 and an all-sky metallicity catalogue based on Gaia DR3 BP/RP spectro-photometry}",
      journal = {\aap},
         year = 2024,
        month = dec,
       volume = {692},
          eid = {A115},
        pages = {A115},
          doi = {10.1051/0004-6361/202347633},
archivePrefix = {arXiv},
       eprint = {2308.01344},
 primaryClass = {astro-ph.GA},
       adsurl = {https://ui.adsabs.harvard.edu/abs/2024A&A...692A.115M}
}

@ARTICLE{2024ApJ...971..107O,
       author = {{Ogami}, Itsuki and {Komiyama}, Yutaka and {Chiba}, Masashi and {Tanaka}, Mikito and {Guhathakurta}, Puragra and {Kirby}, Evan N. and {Wyse}, Rosemary F.~G. and {Filion}, Carrie and {Kirihara}, Takanobu and {Ishigaki}, Miho N. and {Hayashi}, Kohei},
        title = "{Detection of a Spatially Extended Stellar Population in M33: A Shallow Stellar Halo?}",
      journal = {\apj},
         year = 2024,
        month = aug,
       volume = {971},
       number = {1},
          eid = {107},
        pages = {107},
          doi = {10.3847/1538-4357/ad5445},
archivePrefix = {arXiv},
       eprint = {2403.14234},
 primaryClass = {astro-ph.GA},
       adsurl = {https://ui.adsabs.harvard.edu/abs/2024ApJ...971..107O}
}

@ARTICLE{2024A&A...688A.197M,
       author = {{M{\'e}sz{\'a}ros}, Szabolcs and {Bohlin}, Ralph and {Allende Prieto}, Carlos and {Cseh}, Borb{\'a}la and {Kov{\'a}cs}, J{\'o}zsef and {Fleming}, Scott W. and {Dencs}, Zolt{\'a}n and {Deustua}, Susana and {Gordon}, Karl D. and {Hubeny}, Ivan and {Mez{\H{o}}}, Gy{\"o}rgy and {Truszek}, M{\'a}rton},
        title = "{The updated BOSZ synthetic stellar spectral library}",
      journal = {\aap},
         year = 2024,
        month = aug,
       volume = {688},
          eid = {A197},
        pages = {A197},
          doi = {10.1051/0004-6361/202449306},
archivePrefix = {arXiv},
       eprint = {2407.10872},
 primaryClass = {astro-ph.SR},
       adsurl = {https://ui.adsabs.harvard.edu/abs/2024A&A...688A.197M}
}

@ARTICLE{2024RAA....24f5014G,
       author = {{Gontcharov}, G.~A. and {Savchenko}, S.~S. and {Marchuk}, A.~A. and {Bonatto}, C.~J. and {Ryutina}, O.~S. and {Khovritchev}, M. Yu. and {Il'in}, V.~B. and {Mosenkov}, A.~V. and {Poliakov}, D.~M. and {Smirnov}, A.~A.},
        title = "{Isochrone Fitting of Galactic Globular Clusters{\textemdash}VI. High-latitude Clusters NGC 5024 (M53), NGC 5053, NGC 5272 (M3), NGC 5466, and NGC 7099 (M30)}",
      journal = {Research in Astronomy and Astrophysics},
         year = 2024,
        month = jun,
       volume = {24},
       number = {6},
          eid = {065014},
        pages = {065014},
          doi = {10.1088/1674-4527/ad420f},
archivePrefix = {arXiv},
       eprint = {2404.14797},
 primaryClass = {astro-ph.GA},
       adsurl = {https://ui.adsabs.harvard.edu/abs/2024RAA....24f5014G}
}

@ARTICLE{2024MNRAS.527.2065P,
       author = {{Pietrinferni}, Adriano and {Salaris}, Maurizio and {Cassisi}, Santi and {Savino}, Alessandro and {Mucciarelli}, Alessio and {Hyder}, David and {Hidalgo}, Sebastian},
        title = "{The updated BaSTI stellar evolution models and isochrones - IV. {\ensuremath{\alpha}}-Depleted calculations}",
      journal = {\mnras},
         year = 2024,
        month = jan,
       volume = {527},
       number = {2},
        pages = {2065-2070},
          doi = {10.1093/mnras/stad3267},
archivePrefix = {arXiv},
       eprint = {2311.05985},
 primaryClass = {astro-ph.SR},
       adsurl = {https://ui.adsabs.harvard.edu/abs/2024MNRAS.527.2065P}
}

@ARTICLE{2023A&A...674A...3M,
       author = {{Montegriffo}, P. and {De Angeli}, F. and {Andrae}, R. and {Riello}, M. and {Pancino}, E. and {Sanna}, N. and {Bellazzini}, M. and {Evans}, D.~W. and {Carrasco}, J.~M. and {Sordo}, R. and {Busso}, G. and {Cacciari}, C. and {Jordi}, C. and {van Leeuwen}, F. and {Vallenari}, A. and {Altavilla}, G. and {Barstow}, M.~A. and {Brown}, A.~G.~A. and {Burgess}, P.~W. and {Castellani}, M. and {Cowell}, S. and {Davidson}, M. and {De Luise}, F. and {Delchambre}, L. and {Diener}, C. and {Fabricius}, C. and {Fr{\'e}mat}, Y. and {Fouesneau}, M. and {Gilmore}, G. and {Giuffrida}, G. and {Hambly}, N.~C. and {Harrison}, D.~L. and {Hidalgo}, S. and {Hodgkin}, S.~T. and {Holland}, G. and {Marinoni}, S. and {Osborne}, P.~J. and {Pagani}, C. and {Palaversa}, L. and {Piersimoni}, A.~M. and {Pulone}, L. and {Ragaini}, S. and {Rainer}, M. and {Richards}, P.~J. and {Rowell}, N. and {Ruz-Mieres}, D. and {Sarro}, L.~M. and {Walton}, N.~A. and {Yoldas}, A.},
        title = "{Gaia Data Release 3. External calibration of BP/RP low-resolution spectroscopic data}",
      journal = {\aap},
         year = 2023,
        month = jun,
       volume = {674},
          eid = {A3},
        pages = {A3},
          doi = {10.1051/0004-6361/202243880},
archivePrefix = {arXiv},
       eprint = {2206.06205},
 primaryClass = {astro-ph.IM},
       adsurl = {https://ui.adsabs.harvard.edu/abs/2023A&A...674A...3M}
}

@ARTICLE{2023A&A...674A...2D,
       author = {{De Angeli}, F. and {Weiler}, M. and {Montegriffo}, P. and {Evans}, D.~W. and {Riello}, M. and {Andrae}, R. and {Carrasco}, J.~M. and {Busso}, G. and {Burgess}, P.~W. and {Cacciari}, C. and {Davidson}, M. and {Harrison}, D.~L. and {Hodgkin}, S.~T. and {Jordi}, C. and {Osborne}, P.~J. and {Pancino}, E. and {Altavilla}, G. and {Barstow}, M.~A. and {Bailer-Jones}, C.~A.~L. and {Bellazzini}, M. and {Brown}, A.~G.~A. and {Castellani}, M. and {Cowell}, S. and {Delchambre}, L. and {De Luise}, F. and {Diener}, C. and {Fabricius}, C. and {Fouesneau}, M. and {Fr{\'e}mat}, Y. and {Gilmore}, G. and {Giuffrida}, G. and {Hambly}, N.~C. and {Hidalgo}, S. and {Holland}, G. and {Kostrzewa-Rutkowska}, Z. and {van Leeuwen}, F. and {Lobel}, A. and {Marinoni}, S. and {Miller}, N. and {Pagani}, C. and {Palaversa}, L. and {Piersimoni}, A.~M. and {Pulone}, L. and {Ragaini}, S. and {Rainer}, M. and {Richards}, P.~J. and {Rixon}, G.~T. and {Ruz-Mieres}, D. and {Sanna}, N. and {Sarro}, L.~M. and {Rowell}, N. and {Sordo}, R. and {Walton}, N.~A. and {Yoldas}, A.},
        title = "{Gaia Data Release 3. Processing and validation of BP/RP low-resolution spectral data}",
      journal = {\aap},
         year = 2023,
        month = jun,
       volume = {674},
          eid = {A2},
        pages = {A2},
          doi = {10.1051/0004-6361/202243680},
archivePrefix = {arXiv},
       eprint = {2206.06143},
 primaryClass = {astro-ph.IM},
       adsurl = {https://ui.adsabs.harvard.edu/abs/2023A&A...674A...2D}
}

@ARTICLE{2023A&A...674A...1G,
       author = {{Gaia Collaboration} and {Vallenari}, A. and {Brown}, A.~G.~A. and {Prusti}, T. and {de Bruijne}, J.~H.~J. and {Arenou}, F. and {Babusiaux}, C. and {Biermann}, M. and {Creevey}, O.~L. and {Ducourant}, C. and {Evans}, D.~W. and {Eyer}, L. and {Guerra}, R. and {Hutton}, A. and {Jordi}, C. and {Klioner}, S.~A. and {Lammers}, U.~L. and {Lindegren}, L. and {Luri}, X. and {Mignard}, F. and {Panem}, C. and {Pourbaix}, D. and {Randich}, S. and {Sartoretti}, P. and {Soubiran}, C. and {Tanga}, P. and {Walton}, N.~A. and {Bailer-Jones}, C.~A.~L. and {Bastian}, U. and {Drimmel}, R. and {Jansen}, F. and {Katz}, D. and {Lattanzi}, M.~G. and {van Leeuwen}, F. and {Bakker}, J. and {Cacciari}, C. and {Casta{\~n}eda}, J. and {De Angeli}, F. and {Fabricius}, C. and {Fouesneau}, M. and {Fr{\'e}mat}, Y. and {Galluccio}, L. and {Guerrier}, A. and {Heiter}, U. and {Masana}, E. and {Messineo}, R. and {Mowlavi}, N. and {Nicolas}, C. and {Nienartowicz}, K. and {Pailler}, F. and {Panuzzo}, P. and {Riclet}, F. and {Roux}, W. and {Seabroke}, G.~M. and {Sordo}, R. and {Th{\'e}venin}, F. and {Gracia-Abril}, G. and {Portell}, J. and {Teyssier}, D. and {Altmann}, M. and {Andrae}, R. and {Audard}, M. and {Bellas-Velidis}, I. and {Benson}, K. and {Berthier}, J. and {Blomme}, R. and {Burgess}, P.~W. and {Busonero}, D. and {Busso}, G. and {C{\'a}novas}, H. and {Carry}, B. and {Cellino}, A. and {Cheek}, N. and {Clementini}, G. and {Damerdji}, Y. and {Davidson}, M. and {de Teodoro}, P. and {Nu{\~n}ez Campos}, M. and {Delchambre}, L. and {Dell'Oro}, A. and {Esquej}, P. and {Fern{\'a}ndez-Hern{\'a}ndez}, J. and {Fraile}, E. and {Garabato}, D. and {Garc{\'\i}a-Lario}, P. and {Gosset}, E. and {Haigron}, R. and {Halbwachs}, J.-L. and {Hambly}, N.~C. and {Harrison}, D.~L. and {Hern{\'a}ndez}, J. and {Hestroffer}, D. and {Hodgkin}, S.~T. and {Holl}, B. and {Jan{\ss}en}, K. and {Jevardat de Fombelle}, G. and {Jordan}, S. and {Krone-Martins}, A. and {Lanzafame}, A.~C. and {L{\"o}ffler}, W. and {Marchal}, O. and {Marrese}, P.~M. and {Moitinho}, A. and {Muinonen}, K. and {Osborne}, P. and {Pancino}, E. and {Pauwels}, T. and {Recio-Blanco}, A. and {Reyl{\'e}}, C. and {Riello}, M. and {Rimoldini}, L. and {Roegiers}, T. and {Rybizki}, J. and {Sarro}, L.~M. and {Siopis}, C. and {Smith}, M. and {Sozzetti}, A. and {Utrilla}, E. and {van Leeuwen}, M. and {Abbas}, U. and {{\'A}brah{\'a}m}, P. and {Abreu Aramburu}, A. and {Aerts}, C. and {Aguado}, J.~J. and {Ajaj}, M. and {Aldea-Montero}, F. and {Altavilla}, G. and {{\'A}lvarez}, M.~A. and {Alves}, J. and {Anders}, F. and {Anderson}, R.~I. and {Anglada Varela}, E. and {Antoja}, T. and {Baines}, D. and {Baker}, S.~G. and {Balaguer-N{\'u}{\~n}ez}, L. and {Balbinot}, E. and {Balog}, Z. and {Barache}, C. and {Barbato}, D. and {Barros}, M. and {Barstow}, M.~A. and {Bartolom{\'e}}, S. and {Bassilana}, J.-L. and {Bauchet}, N. and {Becciani}, U. and {Bellazzini}, M. and {Berihuete}, A. and {Bernet}, M. and {Bertone}, S. and {Bianchi}, L. and {Binnenfeld}, A. and {Blanco-Cuaresma}, S. and {Blazere}, A. and {Boch}, T. and {Bombrun}, A. and {Bossini}, D. and {Bouquillon}, S. and {Bragaglia}, A. and {Bramante}, L. and {Breedt}, E. and {Bressan}, A. and {Brouillet}, N. and {Brugaletta}, E. and {Bucciarelli}, B. and {Burlacu}, A. and {Butkevich}, A.~G. and {Buzzi}, R. and {Caffau}, E. and {Cancelliere}, R. and {Cantat-Gaudin}, T. and {Carballo}, R. and {Carlucci}, T. and {Carnerero}, M.~I. and {Carrasco}, J.~M. and {Casamiquela}, L. and {Castellani}, M. and {Castro-Ginard}, A. and {Chaoul}, L. and {Charlot}, P. and {Chemin}, L. and {Chiaramida}, V. and {Chiavassa}, A. and {Chornay}, N. and {Comoretto}, G. and {Contursi}, G. and {Cooper}, W.~J. and {Cornez}, T. and {Cowell}, S. and {Crifo}, F. and {Cropper}, M. and {Crosta}, M. and {Crowley}, C. and {Dafonte}, C. and {Dapergolas}, A. and {David}, M. and {David}, P. and {de Laverny}, P. and {De Luise}, F. and {De March}, R.},
        title = "{Gaia Data Release 3. Summary of the content and survey properties}",
      journal = {\aap},
         year = 2023,
        month = jun,
       volume = {674},
          eid = {A1},
        pages = {A1},
          doi = {10.1051/0004-6361/202243940},
archivePrefix = {arXiv},
       eprint = {2208.00211},
 primaryClass = {astro-ph.GA},
       adsurl = {https://ui.adsabs.harvard.edu/abs/2023A&A...674A...1G}
}

@ARTICLE{2023A&A...673A..44F,
       author = {{Ferrone}, Salvatore and {Di Matteo}, Paola and {Mastrobuono-Battisti}, Alessandra and {Haywood}, Misha and {Snaith}, Owain N. and {Montuori}, Marco and {Khoperskov}, Sergey and {Valls-Gabaud}, David},
        title = "{The e-TidalGCs project. Modeling the extra-tidal features generated by Galactic globular clusters}",
      journal = {\aap},
         year = 2023,
        month = may,
       volume = {673},
          eid = {A44},
        pages = {A44},
          doi = {10.1051/0004-6361/202244141},
archivePrefix = {arXiv},
       eprint = {2301.05166},
 primaryClass = {astro-ph.GA},
       adsurl = {https://ui.adsabs.harvard.edu/abs/2023A&A...673A..44F}
}

@ARTICLE{2023MNRAS.520.5225M,
       author = {{Mateu}, Cecilia},
        title = "{galstreams: A library of Milky Way stellar stream footprints and tracks}",
      journal = {\mnras},
         year = 2023,
        month = apr,
       volume = {520},
       number = {4},
        pages = {5225-5258},
          doi = {10.1093/mnras/stad321},
archivePrefix = {arXiv},
       eprint = {2204.10326},
 primaryClass = {astro-ph.GA},
       adsurl = {https://ui.adsabs.harvard.edu/abs/2023MNRAS.520.5225M}
}

@ARTICLE{2022ApJS..261...38D,
       author = {{Drlica-Wagner}, A. and {Ferguson}, P.~S. and {Adam{\'o}w}, M. and {Aguena}, M. and {Allam}, S. and {Andrade-Oliveira}, F. and {Bacon}, D. and {Bechtol}, K. and {Bell}, E.~F. and {Bertin}, E. and {Bilaji}, P. and {Bocquet}, S. and {Bom}, C.~R. and {Brooks}, D. and {Burke}, D.~L. and {Carballo-Bello}, J.~A. and {Carlin}, J.~L. and {Carnero Rosell}, A. and {Carrasco Kind}, M. and {Carretero}, J. and {Castander}, F.~J. and {Cerny}, W. and {Chang}, C. and {Choi}, Y. and {Conselice}, C. and {Costanzi}, M. and {Crnojevi{\'c}}, D. and {da Costa}, L.~N. and {de Vicente}, J. and {Desai}, S. and {Esteves}, J. and {Everett}, S. and {Ferrero}, I. and {Fitzpatrick}, M. and {Flaugher}, B. and {Friedel}, D. and {Frieman}, J. and {Garc{\'\i}a-Bellido}, J. and {Gatti}, M. and {Gaztanaga}, E. and {Gerdes}, D.~W. and {Gruen}, D. and {Gruendl}, R.~A. and {Gschwend}, J. and {Hartley}, W.~G. and {Hernandez-Lang}, D. and {Hinton}, S.~R. and {Hollowood}, D.~L. and {Honscheid}, K. and {Hughes}, A.~K. and {Jacques}, A. and {James}, D.~J. and {Johnson}, M.~D. and {Kuehn}, K. and {Kuropatkin}, N. and {Lahav}, O. and {Li}, T.~S. and {Lidman}, C. and {Lin}, H. and {March}, M. and {Marshall}, J.~L. and {Mart{\'\i}nez-Delgado}, D. and {Mart{\'\i}nez-V{\'a}zquez}, C.~E. and {Massana}, P. and {Mau}, S. and {McNanna}, M. and {Melchior}, P. and {Menanteau}, F. and {Miller}, A.~E. and {Miquel}, R. and {Mohr}, J.~J. and {Morgan}, R. and {Mutlu-Pakdil}, B. and {Mu{\~n}oz}, R.~R. and {Neilsen}, E.~H. and {Nidever}, D.~L. and {Nikutta}, R. and {Nilo Castellon}, J.~L. and {No{\"e}l}, N.~E.~D. and {Ogando}, R.~L.~C. and {Olsen}, K.~A.~G. and {Pace}, A.~B. and {Palmese}, A. and {Paz-Chinch{\'o}n}, F. and {Pereira}, M.~E.~S. and {Pieres}, A. and {Plazas Malag{\'o}n}, A.~A. and {Prat}, J. and {Riley}, A.~H. and {Rodriguez-Monroy}, M. and {Romer}, A.~K. and {Roodman}, A. and {Sako}, M. and {Sakowska}, J.~D. and {Sanchez}, E. and {S{\'a}nchez}, F.~J. and {Sand}, D.~J. and {Santana-Silva}, L. and {Santiago}, B. and {Schubnell}, M. and {Serrano}, S. and {Sevilla-Noarbe}, I. and {Simon}, J.~D. and {Smith}, M. and {Soares-Santos}, M. and {Stringfellow}, G.~S. and {Suchyta}, E. and {Suson}, D.~J. and {Tan}, C.~Y. and {Tarle}, G. and {Tavangar}, K. and {Thomas}, D. and {To}, C. and {Tollerud}, E.~J. and {Troxel}, M.~A. and {Tucker}, D.~L. and {Varga}, T.~N. and {Vivas}, A.~K. and {Walker}, A.~R. and {Weller}, J. and {Wilkinson}, R.~D. and {Wu}, J.~F. and {Yanny}, B. and {Zaborowski}, E. and {Zenteno}, A. and {Delve Collaboration} and {Des Collaboration} and {Astro Data Lab}},
        title = "{The DECam Local Volume Exploration Survey Data Release 2}",
      journal = {\apjs},
         year = 2022,
        month = aug,
       volume = {261},
       number = {2},
          eid = {38},
        pages = {38},
          doi = {10.3847/1538-4365/ac78eb},
archivePrefix = {arXiv},
       eprint = {2203.16565},
 primaryClass = {astro-ph.IM},
       adsurl = {https://ui.adsabs.harvard.edu/abs/2022ApJS..261...38D}
}

@ARTICLE{2022MNRAS.513..853Y,
       author = {{Yang}, Yong and {Zhao}, Jing-Kun and {Ishigaki}, Miho N. and {Zhou}, Jian-Zhao and {Yang}, Cheng-Qun and {Xue}, Xiang-Xiang and {Ye}, Xian-Hao and {Zhao}, Gang},
        title = "{Revisit NGC 5466 tidal stream with Gaia, SDSS/SEGUE, and LAMOST}",
      journal = {\mnras},
         year = 2022,
        month = jun,
       volume = {513},
       number = {1},
        pages = {853-863},
          doi = {10.1093/mnras/stac860},
archivePrefix = {arXiv},
       eprint = {2203.13414},
 primaryClass = {astro-ph.GA},
       adsurl = {https://ui.adsabs.harvard.edu/abs/2022MNRAS.513..853Y}
}

@ARTICLE{2022PASJ...74..247A,
       author = {{Aihara}, Hiroaki and {AlSayyad}, Yusra and {Ando}, Makoto and {Armstrong}, Robert and {Bosch}, James and {Egami}, Eiichi and {Furusawa}, Hisanori and {Furusawa}, Junko and {Harasawa}, Sumiko and {Harikane}, Yuichi and {Hsieh}, Bau-Ching and {Ikeda}, Hiroyuki and {Ito}, Kei and {Iwata}, Ikuru and {Kodama}, Tadayuki and {Koike}, Michitaro and {Kokubo}, Mitsuru and {Komiyama}, Yutaka and {Li}, Xiangchong and {Liang}, Yongming and {Lin}, Yen-Ting and {Lupton}, Robert H. and {Lust}, Nate B. and {MacArthur}, Lauren A. and {Mawatari}, Ken and {Mineo}, Sogo and {Miyatake}, Hironao and {Miyazaki}, Satoshi and {More}, Surhud and {Morishima}, Takahiro and {Murayama}, Hitoshi and {Nakajima}, Kimihiko and {Nakata}, Fumiaki and {Nishizawa}, Atsushi J. and {Oguri}, Masamune and {Okabe}, Nobuhiro and {Okura}, Yuki and {Ono}, Yoshiaki and {Osato}, Ken and {Ouchi}, Masami and {Pan}, Yen-Chen and {Plazas Malag{\'o}n}, Andr{\'e}s A. and {Price}, Paul A. and {Reed}, Sophie L. and {Rykoff}, Eli S. and {Shibuya}, Takatoshi and {Simunovic}, Mirko and {Strauss}, Michael A. and {Sugimori}, Kanako and {Suto}, Yasushi and {Suzuki}, Nao and {Takada}, Masahiro and {Takagi}, Yuhei and {Takata}, Tadafumi and {Takita}, Satoshi and {Tanaka}, Masayuki and {Tang}, Shenli and {Taranu}, Dan S. and {Terai}, Tsuyoshi and {Toba}, Yoshiki and {Turner}, Edwin L. and {Uchiyama}, Hisakazu and {Vijarnwannaluk}, Bovornpratch and {Waters}, Christopher Z. and {Yamada}, Yoshihiko and {Yamamoto}, Naoaki and {Yamashita}, Takuji},
        title = "{Third data release of the Hyper Suprime-Cam Subaru Strategic Program}",
      journal = {\pasj},
         year = 2022,
        month = apr,
       volume = {74},
       number = {2},
        pages = {247-272},
          doi = {10.1093/pasj/psab122},
archivePrefix = {arXiv},
       eprint = {2108.13045},
 primaryClass = {astro-ph.IM},
       adsurl = {https://ui.adsabs.harvard.edu/abs/2022PASJ...74..247A}
}

@ARTICLE{2022MNRAS.509.5197S,
       author = {{Salaris}, Maurizio and {Cassisi}, Santi and {Pietrinferni}, Adriano and {Hidalgo}, Sebastian},
        title = "{The updated BASTI stellar evolution models and isochrones - III. White dwarfs}",
      journal = {\mnras},
         year = 2022,
        month = feb,
       volume = {509},
       number = {4},
        pages = {5197-5208},
          doi = {10.1093/mnras/stab3359},
archivePrefix = {arXiv},
       eprint = {2111.09285},
 primaryClass = {astro-ph.SR},
       adsurl = {https://ui.adsabs.harvard.edu/abs/2022MNRAS.509.5197S}
}

@ARTICLE{2022ApJ...925....6F,
       author = {{Fu}, Sal Wanying and {Weisz}, Daniel R. and {Starkenburg}, Else and {Martin}, Nicolas and {Ji}, Alexander P. and {Patel}, Ekta and {Boylan-Kolchin}, Michael and {C{\^o}t{\'e}}, Patrick and {Dolphin}, Andrew E. and {Longeard}, Nicolas and {Mateo}, Mario L. and {Sandford}, Nathan R.},
        title = "{Metallicity Distribution Function of the Eridanus II Ultra-faint Dwarf Galaxy from Hubble Space Telescope Narrowband Imaging}",
      journal = {\apj},
         year = 2022,
        month = jan,
       volume = {925},
       number = {1},
          eid = {6},
        pages = {6},
          doi = {10.3847/1538-4357/ac3665},
archivePrefix = {arXiv},
       eprint = {2111.00045},
 primaryClass = {astro-ph.GA},
       adsurl = {https://ui.adsabs.harvard.edu/abs/2022ApJ...925....6F}
}

@ARTICLE{2021MNRAS.507.1923J,
       author = {{Jensen}, Jaclyn and {Thomas}, Guillaume and {McConnachie}, Alan W. and {Starkenburg}, Else and {Malhan}, Khyati and {Navarro}, Julio and {Martin}, Nicolas and {Famaey}, Benoit and {Ibata}, Rodrigo and {Chapman}, Scott and {Cuillandre}, Jean-Charles and {Gwyn}, Stephen},
        title = "{Uncovering fossils of the distant Milky Way with UNIONS: NGC 5466 and its stellar stream}",
      journal = {\mnras},
         year = 2021,
        month = oct,
       volume = {507},
       number = {2},
        pages = {1923-1936},
          doi = {10.1093/mnras/stab2325},
archivePrefix = {arXiv},
       eprint = {2108.04340},
 primaryClass = {astro-ph.GA},
       adsurl = {https://ui.adsabs.harvard.edu/abs/2021MNRAS.507.1923J}
}

@ARTICLE{2021ApJS..256....2D,
       author = {{Drlica-Wagner}, A. and {Carlin}, J.~L. and {Nidever}, D.~L. and {Ferguson}, P.~S. and {Kuropatkin}, N. and {Adam{\'o}w}, M. and {Cerny}, W. and {Choi}, Y. and {Esteves}, J.~H. and {Mart{\'\i}nez-V{\'a}zquez}, C.~E. and {Mau}, S. and {Miller}, A.~E. and {Mutlu-Pakdil}, B. and {Neilsen}, E.~H. and {Olsen}, K.~A.~G. and {Pace}, A.~B. and {Riley}, A.~H. and {Sakowska}, J.~D. and {Sand}, D.~J. and {Santana-Silva}, L. and {Tollerud}, E.~J. and {Tucker}, D.~L. and {Vivas}, A.~K. and {Zaborowski}, E. and {Zenteno}, A. and {Abbott}, T.~M.~C. and {Allam}, S. and {Bechtol}, K. and {Bell}, C.~P.~M. and {Bell}, E.~F. and {Bilaji}, P. and {Bom}, C.~R. and {Carballo-Bello}, J.~A. and {Crnojevi{\'c}}, D. and {Cioni}, M.-R.~L. and {Diaz-Ocampo}, A. and {de Boer}, T.~J.~L. and {Erkal}, D. and {Gruendl}, R.~A. and {Hernandez-Lang}, D. and {Hughes}, A.~K. and {James}, D.~J. and {Johnson}, L.~C. and {Li}, T.~S. and {Mao}, Y.-Y. and {Mart{\'\i}nez-Delgado}, D. and {Massana}, P. and {McNanna}, M. and {Morgan}, R. and {Nadler}, E.~O. and {No{\"e}l}, N.~E.~D. and {Palmese}, A. and {Peter}, A.~H.~G. and {Rykoff}, E.~S. and {S{\'a}nchez}, J. and {Shipp}, N. and {Simon}, J.~D. and {Smercina}, A. and {Soares-Santos}, M. and {Stringfellow}, G.~S. and {Tavangar}, K. and {van der Marel}, R.~P. and {Walker}, A.~R. and {Wechsler}, R.~H. and {Wu}, J.~F. and {Yanny}, B. and {Fitzpatrick}, M. and {Huang}, L. and {Jacques}, A. and {Nikutta}, R. and {Scott}, A. and {Astro Data Lab}},
        title = "{The DECam Local Volume Exploration Survey: Overview and First Data Release}",
      journal = {\apjs},
         year = 2021,
        month = sep,
       volume = {256},
       number = {1},
          eid = {2},
        pages = {2},
          doi = {10.3847/1538-4365/ac079d},
archivePrefix = {arXiv},
       eprint = {2103.07476},
 primaryClass = {astro-ph.GA},
       adsurl = {https://ui.adsabs.harvard.edu/abs/2021ApJS..256....2D}
}

@ARTICLE{2021MNRAS.505.5957B,
       author = {{Baumgardt}, H. and {Vasiliev}, E.},
        title = "{Accurate distances to Galactic globular clusters through a combination of Gaia EDR3, HST, and literature data}",
      journal = {\mnras},
         year = 2021,
        month = aug,
       volume = {505},
       number = {4},
        pages = {5957-5977},
          doi = {10.1093/mnras/stab1474},
archivePrefix = {arXiv},
       eprint = {2105.09526},
 primaryClass = {astro-ph.GA},
       adsurl = {https://ui.adsabs.harvard.edu/abs/2021MNRAS.505.5957B}
}

@ARTICLE{2021MNRAS.505.5978V,
       author = {{Vasiliev}, Eugene and {Baumgardt}, Holger},
        title = "{Gaia EDR3 view on galactic globular clusters}",
      journal = {\mnras},
         year = 2021,
        month = aug,
       volume = {505},
       number = {4},
        pages = {5978-6002},
          doi = {10.1093/mnras/stab1475},
archivePrefix = {arXiv},
       eprint = {2102.09568},
 primaryClass = {astro-ph.GA},
       adsurl = {https://ui.adsabs.harvard.edu/abs/2021MNRAS.505.5978V}
}

@ARTICLE{2021A&A...649A...1G,
       author = {{Gaia Collaboration} and {Brown}, A.~G.~A. and {Vallenari}, A. and {Prusti}, T. and {de Bruijne}, J.~H.~J. and {Babusiaux}, C. and {Biermann}, M. and {Creevey}, O.~L. and {Evans}, D.~W. and {Eyer}, L. and {Hutton}, A. and {Jansen}, F. and {Jordi}, C. and {Klioner}, S.~A. and {Lammers}, U. and {Lindegren}, L. and {Luri}, X. and {Mignard}, F. and {Panem}, C. and {Pourbaix}, D. and {Randich}, S. and {Sartoretti}, P. and {Soubiran}, C. and {Walton}, N.~A. and {Arenou}, F. and {Bailer-Jones}, C.~A.~L. and {Bastian}, U. and {Cropper}, M. and {Drimmel}, R. and {Katz}, D. and {Lattanzi}, M.~G. and {van Leeuwen}, F. and {Bakker}, J. and {Cacciari}, C. and {Casta{\~n}eda}, J. and {De Angeli}, F. and {Ducourant}, C. and {Fabricius}, C. and {Fouesneau}, M. and {Fr{\'e}mat}, Y. and {Guerra}, R. and {Guerrier}, A. and {Guiraud}, J. and {Jean-Antoine Piccolo}, A. and {Masana}, E. and {Messineo}, R. and {Mowlavi}, N. and {Nicolas}, C. and {Nienartowicz}, K. and {Pailler}, F. and {Panuzzo}, P. and {Riclet}, F. and {Roux}, W. and {Seabroke}, G.~M. and {Sordo}, R. and {Tanga}, P. and {Th{\'e}venin}, F. and {Gracia-Abril}, G. and {Portell}, J. and {Teyssier}, D. and {Altmann}, M. and {Andrae}, R. and {Bellas-Velidis}, I. and {Benson}, K. and {Berthier}, J. and {Blomme}, R. and {Brugaletta}, E. and {Burgess}, P.~W. and {Busso}, G. and {Carry}, B. and {Cellino}, A. and {Cheek}, N. and {Clementini}, G. and {Damerdji}, Y. and {Davidson}, M. and {Delchambre}, L. and {Dell'Oro}, A. and {Fern{\'a}ndez-Hern{\'a}ndez}, J. and {Galluccio}, L. and {Garc{\'\i}a-Lario}, P. and {Garcia-Reinaldos}, M. and {Gonz{\'a}lez-N{\'u}{\~n}ez}, J. and {Gosset}, E. and {Haigron}, R. and {Halbwachs}, J.-L. and {Hambly}, N.~C. and {Harrison}, D.~L. and {Hatzidimitriou}, D. and {Heiter}, U. and {Hern{\'a}ndez}, J. and {Hestroffer}, D. and {Hodgkin}, S.~T. and {Holl}, B. and {Jan{\ss}en}, K. and {Jevardat de Fombelle}, G. and {Jordan}, S. and {Krone-Martins}, A. and {Lanzafame}, A.~C. and {L{\"o}ffler}, W. and {Lorca}, A. and {Manteiga}, M. and {Marchal}, O. and {Marrese}, P.~M. and {Moitinho}, A. and {Mora}, A. and {Muinonen}, K. and {Osborne}, P. and {Pancino}, E. and {Pauwels}, T. and {Petit}, J.-M. and {Recio-Blanco}, A. and {Richards}, P.~J. and {Riello}, M. and {Rimoldini}, L. and {Robin}, A.~C. and {Roegiers}, T. and {Rybizki}, J. and {Sarro}, L.~M. and {Siopis}, C. and {Smith}, M. and {Sozzetti}, A. and {Ulla}, A. and {Utrilla}, E. and {van Leeuwen}, M. and {van Reeven}, W. and {Abbas}, U. and {Abreu Aramburu}, A. and {Accart}, S. and {Aerts}, C. and {Aguado}, J.~J. and {Ajaj}, M. and {Altavilla}, G. and {{\'A}lvarez}, M.~A. and {{\'A}lvarez Cid-Fuentes}, J. and {Alves}, J. and {Anderson}, R.~I. and {Anglada Varela}, E. and {Antoja}, T. and {Audard}, M. and {Baines}, D. and {Baker}, S.~G. and {Balaguer-N{\'u}{\~n}ez}, L. and {Balbinot}, E. and {Balog}, Z. and {Barache}, C. and {Barbato}, D. and {Barros}, M. and {Barstow}, M.~A. and {Bartolom{\'e}}, S. and {Bassilana}, J.-L. and {Bauchet}, N. and {Baudesson-Stella}, A. and {Becciani}, U. and {Bellazzini}, M. and {Bernet}, M. and {Bertone}, S. and {Bianchi}, L. and {Blanco-Cuaresma}, S. and {Boch}, T. and {Bombrun}, A. and {Bossini}, D. and {Bouquillon}, S. and {Bragaglia}, A. and {Bramante}, L. and {Breedt}, E. and {Bressan}, A. and {Brouillet}, N. and {Bucciarelli}, B. and {Burlacu}, A. and {Busonero}, D. and {Butkevich}, A.~G. and {Buzzi}, R. and {Caffau}, E. and {Cancelliere}, R. and {C{\'a}novas}, H. and {Cantat-Gaudin}, T. and {Carballo}, R. and {Carlucci}, T. and {Carnerero}, M.~I. and {Carrasco}, J.~M. and {Casamiquela}, L. and {Castellani}, M. and {Castro-Ginard}, A. and {Castro Sampol}, P. and {Chaoul}, L. and {Charlot}, P. and {Chemin}, L. and {Chiavassa}, A. and {Cioni}, M.-R.~L. and {Comoretto}, G. and {Cooper}, W.~J. and {Cornez}, T. and {Cowell}, S. and {Crifo}, F. and {Crosta}, M. and {Crowley}, C. and {Dafonte}, C. and {Dapergolas}, A. and {David}, M. and {David}, P.},
        title = "{Gaia Early Data Release 3. Summary of the contents and survey properties}",
      journal = {\aap},
         year = 2021,
        month = may,
       volume = {649},
          eid = {A1},
        pages = {A1},
          doi = {10.1051/0004-6361/202039657},
archivePrefix = {arXiv},
       eprint = {2012.01533},
 primaryClass = {astro-ph.GA},
       adsurl = {https://ui.adsabs.harvard.edu/abs/2021A&A...649A...1G}
}

@ARTICLE{2021ApJ...908..102P,
       author = {{Pietrinferni}, Adriano and {Hidalgo}, Sebastian and {Cassisi}, Santi and {Salaris}, Maurizio and {Savino}, Alessandro and {Mucciarelli}, Alessio and {Verma}, Kuldeep and {Silva Aguirre}, Victor and {Aparicio}, Antonio and {Ferguson}, Jason W.},
        title = "{Updated BaSTI Stellar Evolution Models and Isochrones. II. {\ensuremath{\alpha}}-enhanced Calculations}",
      journal = {\apj},
         year = 2021,
        month = feb,
       volume = {908},
       number = {1},
          eid = {102},
        pages = {102},
          doi = {10.3847/1538-4357/abd4d5},
archivePrefix = {arXiv},
       eprint = {2012.10085},
 primaryClass = {astro-ph.SR},
       adsurl = {https://ui.adsabs.harvard.edu/abs/2021ApJ...908..102P}
}

@ARTICLE{2020ApJS..251....7F,
       author = {{Flewelling}, H.~A. and {Magnier}, E.~A. and {Chambers}, K.~C. and {Heasley}, J.~N. and {Holmberg}, C. and {Huber}, M.~E. and {Sweeney}, W. and {Waters}, C.~Z. and {Calamida}, A. and {Casertano}, S. and {Chen}, X. and {Farrow}, D. and {Hasinger}, G. and {Henderson}, R. and {Long}, K.~S. and {Metcalfe}, N. and {Narayan}, G. and {Nieto-Santisteban}, M.~A. and {Norberg}, P. and {Rest}, A. and {Saglia}, R.~P. and {Szalay}, A. and {Thakar}, A.~R. and {Tonry}, J.~L. and {Valenti}, J. and {Werner}, S. and {White}, R. and {Denneau}, L. and {Draper}, P.~W. and {Hodapp}, K.~W. and {Jedicke}, R. and {Kaiser}, N. and {Kudritzki}, R.~P. and {Price}, P.~A. and {Wainscoat}, R.~J. and {Chastel}, S. and {McLean}, B. and {Postman}, M. and {Shiao}, B.},
        title = "{The Pan-STARRS1 Database and Data Products}",
      journal = {\apjs},
         year = 2020,
        month = nov,
       volume = {251},
       number = {1},
          eid = {7},
        pages = {7},
          doi = {10.3847/1538-4365/abb82d},
archivePrefix = {arXiv},
       eprint = {1612.05243},
 primaryClass = {astro-ph.IM},
       adsurl = {https://ui.adsabs.harvard.edu/abs/2020ApJS..251....7F}
}

@ARTICLE{2020ApJ...891..161I,
       author = {{Ibata}, Rodrigo and {Thomas}, Guillaume and {Famaey}, Benoit and {Malhan}, Khyati and {Martin}, Nicolas and {Monari}, Giacomo},
        title = "{Detection of Strong Epicyclic Density Spikes in the GD-1 Stellar Stream: An Absence of Evidence for the Influence of Dark Matter Subhalos?}",
      journal = {\apj},
         year = 2020,
        month = mar,
       volume = {891},
       number = {2},
          eid = {161},
        pages = {161},
          doi = {10.3847/1538-4357/ab7303},
archivePrefix = {arXiv},
       eprint = {2002.01488},
 primaryClass = {astro-ph.GA},
       adsurl = {https://ui.adsabs.harvard.edu/abs/2020ApJ...891..161I}
}

@INPROCEEDINGS{2020IAUS..351..516S,
       author = {{Sollima}, A. and {Baumgardt}, H. and {Hilker}, M.},
        title = "{The eye of Gaia on globular cluster kinematics: Internal rotation}",
    booktitle = {Star Clusters: From the Milky Way to the Early Universe},
         year = 2020,
       editor = {{Bragaglia}, Angela and {Davies}, Melvyn and {Sills}, Alison and {Vesperini}, Enrico},
       series = {IAU Symposium},
       volume = {351},
        month = jan,
        pages = {516-519},
          doi = {10.1017/S1743921319007099},
       adsurl = {https://ui.adsabs.harvard.edu/abs/2020IAUS..351..516S}
}

@ARTICLE{2019ApJ...887L..12B,
       author = {{Bianchini}, P. and {Ibata}, R. and {Famaey}, B.},
        title = "{Exploring the Outskirts of Globular Clusters: The Peculiar Kinematics of NGC 3201}",
      journal = {\apjl},
         year = 2019,
        month = dec,
       volume = {887},
       number = {1},
          eid = {L12},
        pages = {L12},
          doi = {10.3847/2041-8213/ab58d1},
archivePrefix = {arXiv},
       eprint = {1912.02195},
 primaryClass = {astro-ph.GA},
       adsurl = {https://ui.adsabs.harvard.edu/abs/2019ApJ...887L..12B}
}

@ARTICLE{2019PASJ...71..114A,
       author = {{Aihara}, Hiroaki and {AlSayyad}, Yusra and {Ando}, Makoto and {Armstrong}, Robert and {Bosch}, James and {Egami}, Eiichi and {Furusawa}, Hisanori and {Furusawa}, Junko and {Goulding}, Andy and {Harikane}, Yuichi and {Hikage}, Chiaki and {Ho}, Paul T.~P. and {Hsieh}, Bau-Ching and {Huang}, Song and {Ikeda}, Hiroyuki and {Imanishi}, Masatoshi and {Ito}, Kei and {Iwata}, Ikuru and {Jaelani}, Anton T. and {Kakuma}, Ryota and {Kawana}, Kojiro and {Kikuta}, Satoshi and {Kobayashi}, Umi and {Koike}, Michitaro and {Komiyama}, Yutaka and {Li}, Xiangchong and {Liang}, Yongming and {Lin}, Yen-Ting and {Luo}, Wentao and {Lupton}, Robert and {Lust}, Nate B. and {MacArthur}, Lauren A. and {Matsuoka}, Yoshiki and {Mineo}, Sogo and {Miyatake}, Hironao and {Miyazaki}, Satoshi and {More}, Surhud and {Murata}, Ryoma and {Namiki}, Shigeru V. and {Nishizawa}, Atsushi J. and {Oguri}, Masamune and {Okabe}, Nobuhiro and {Okamoto}, Sakurako and {Okura}, Yuki and {Ono}, Yoshiaki and {Onodera}, Masato and {Onoue}, Masafusa and {Osato}, Ken and {Ouchi}, Masami and {Shibuya}, Takatoshi and {Strauss}, Michael A. and {Sugiyama}, Naoshi and {Suto}, Yasushi and {Takada}, Masahiro and {Takagi}, Yuhei and {Takata}, Tadafumi and {Takita}, Satoshi and {Tanaka}, Masayuki and {Terai}, Tsuyoshi and {Toba}, Yoshiki and {Uchiyama}, Hisakazu and {Utsumi}, Yousuke and {Wang}, Shiang-Yu and {Wang}, Wenting and {Yamada}, Yoshihiko},
        title = "{Second data release of the Hyper Suprime-Cam Subaru Strategic Program}",
      journal = {\pasj},
         year = 2019,
        month = dec,
       volume = {71},
       number = {6},
          eid = {114},
        pages = {114},
          doi = {10.1093/pasj/psz103},
archivePrefix = {arXiv},
       eprint = {1905.12221},
 primaryClass = {astro-ph.IM},
       adsurl = {https://ui.adsabs.harvard.edu/abs/2019PASJ...71..114A}
}

@ARTICLE{2019MNRAS.487.3815M,
       author = {{Marino}, A.~F. and {Milone}, A.~P. and {Renzini}, A. and {D'Antona}, F. and {Anderson}, J. and {Bedin}, L.~R. and {Bellini}, A. and {Cordoni}, G. and {Lagioia}, E.~P. and {Piotto}, G. and {Tailo}, M.},
        title = "{The Hubble Space Telescope UV Legacy Survey of Galactic Globular Clusters - XIX. A chemical tagging of the multiple stellar populations over the chromosome maps}",
      journal = {\mnras},
         year = 2019,
        month = aug,
       volume = {487},
       number = {3},
        pages = {3815-3844},
          doi = {10.1093/mnras/stz1415},
archivePrefix = {arXiv},
       eprint = {1904.05180},
 primaryClass = {astro-ph.SR},
       adsurl = {https://ui.adsabs.harvard.edu/abs/2019MNRAS.487.3815M}
}

@ARTICLE{2019ApJ...873..111I,
       author = {{Ivezi{\'c}}, {\v{Z}}eljko and {Kahn}, Steven M. and {Tyson}, J. Anthony and {Abel}, Bob and {Acosta}, Emily and {Allsman}, Robyn and {Alonso}, David and {AlSayyad}, Yusra and {Anderson}, Scott F. and {Andrew}, John and {Angel}, James Roger P. and {Angeli}, George Z. and {Ansari}, Reza and {Antilogus}, Pierre and {Araujo}, Constanza and {Armstrong}, Robert and {Arndt}, Kirk T. and {Astier}, Pierre and {Aubourg}, {\'E}ric and {Auza}, Nicole and {Axelrod}, Tim S. and {Bard}, Deborah J. and {Barr}, Jeff D. and {Barrau}, Aurelian and {Bartlett}, James G. and {Bauer}, Amanda E. and {Bauman}, Brian J. and {Baumont}, Sylvain and {Bechtol}, Ellen and {Bechtol}, Keith and {Becker}, Andrew C. and {Becla}, Jacek and {Beldica}, Cristina and {Bellavia}, Steve and {Bianco}, Federica B. and {Biswas}, Rahul and {Blanc}, Guillaume and {Blazek}, Jonathan and {Blandford}, Roger D. and {Bloom}, Josh S. and {Bogart}, Joanne and {Bond}, Tim W. and {Booth}, Michael T. and {Borgland}, Anders W. and {Borne}, Kirk and {Bosch}, James F. and {Boutigny}, Dominique and {Brackett}, Craig A. and {Bradshaw}, Andrew and {Brandt}, William Nielsen and {Brown}, Michael E. and {Bullock}, James S. and {Burchat}, Patricia and {Burke}, David L. and {Cagnoli}, Gianpietro and {Calabrese}, Daniel and {Callahan}, Shawn and {Callen}, Alice L. and {Carlin}, Jeffrey L. and {Carlson}, Erin L. and {Chandrasekharan}, Srinivasan and {Charles-Emerson}, Glenaver and {Chesley}, Steve and {Cheu}, Elliott C. and {Chiang}, Hsin-Fang and {Chiang}, James and {Chirino}, Carol and {Chow}, Derek and {Ciardi}, David R. and {Claver}, Charles F. and {Cohen-Tanugi}, Johann and {Cockrum}, Joseph J. and {Coles}, Rebecca and {Connolly}, Andrew J. and {Cook}, Kem H. and {Cooray}, Asantha and {Covey}, Kevin R. and {Cribbs}, Chris and {Cui}, Wei and {Cutri}, Roc and {Daly}, Philip N. and {Daniel}, Scott F. and {Daruich}, Felipe and {Daubard}, Guillaume and {Daues}, Greg and {Dawson}, William and {Delgado}, Francisco and {Dellapenna}, Alfred and {de Peyster}, Robert and {de Val-Borro}, Miguel and {Digel}, Seth W. and {Doherty}, Peter and {Dubois}, Richard and {Dubois-Felsmann}, Gregory P. and {Durech}, Josef and {Economou}, Frossie and {Eifler}, Tim and {Eracleous}, Michael and {Emmons}, Benjamin L. and {Fausti Neto}, Angelo and {Ferguson}, Henry and {Figueroa}, Enrique and {Fisher-Levine}, Merlin and {Focke}, Warren and {Foss}, Michael D. and {Frank}, James and {Freemon}, Michael D. and {Gangler}, Emmanuel and {Gawiser}, Eric and {Geary}, John C. and {Gee}, Perry and {Geha}, Marla and {Gessner}, Charles J.~B. and {Gibson}, Robert R. and {Gilmore}, D. Kirk and {Glanzman}, Thomas and {Glick}, William and {Goldina}, Tatiana and {Goldstein}, Daniel A. and {Goodenow}, Iain and {Graham}, Melissa L. and {Gressler}, William J. and {Gris}, Philippe and {Guy}, Leanne P. and {Guyonnet}, Augustin and {Haller}, Gunther and {Harris}, Ron and {Hascall}, Patrick A. and {Haupt}, Justine and {Hernandez}, Fabio and {Herrmann}, Sven and {Hileman}, Edward and {Hoblitt}, Joshua and {Hodgson}, John A. and {Hogan}, Craig and {Howard}, James D. and {Huang}, Dajun and {Huffer}, Michael E. and {Ingraham}, Patrick and {Innes}, Walter R. and {Jacoby}, Suzanne H. and {Jain}, Bhuvnesh and {Jammes}, Fabrice and {Jee}, M. James and {Jenness}, Tim and {Jernigan}, Garrett and {Jevremovi{\'c}}, Darko and {Johns}, Kenneth and {Johnson}, Anthony S. and {Johnson}, Margaret W.~G. and {Jones}, R. Lynne and {Juramy-Gilles}, Claire and {Juri{\'c}}, Mario and {Kalirai}, Jason S. and {Kallivayalil}, Nitya J. and {Kalmbach}, Bryce and {Kantor}, Jeffrey P. and {Karst}, Pierre and {Kasliwal}, Mansi M. and {Kelly}, Heather and {Kessler}, Richard and {Kinnison}, Veronica and {Kirkby}, David and {Knox}, Lloyd and {Kotov}, Ivan V. and {Krabbendam}, Victor L. and {Krughoff}, K. Simon and {Kub{\'a}nek}, Petr and {Kuczewski}, John and {Kulkarni}, Shri and {Ku}, John and {Kurita}, Nadine R. and {Lage}, Craig S. and {Lambert}, Ron and {Lange}, Travis and {Langton}, J. Brian and {Le Guillou}, Laurent and {Levine}, Deborah and {Liang}, Ming and {Lim}, Kian-Tat and {Lintott}, Chris J. and {Long}, Kevin E. and {Lopez}, Margaux and {Lotz}, Paul J. and {Lupton}, Robert H. and {Lust}, Nate B. and {MacArthur}, Lauren A. and {Mahabal}, Ashish and {Mandelbaum}, Rachel and {Markiewicz}, Thomas W. and {Marsh}, Darren S. and {Marshall}, Philip J. and {Marshall}, Stuart and {May}, Morgan and {McKercher}, Robert and {McQueen}, Michelle and {Meyers}, Joshua and {Migliore}, Myriam and {Miller}, Michelle and {Mills}, David J.},
        title = "{LSST: From Science Drivers to Reference Design and Anticipated Data Products}",
      journal = {\apj},
         year = 2019,
        month = mar,
       volume = {873},
       number = {2},
          eid = {111},
        pages = {111},
          doi = {10.3847/1538-4357/ab042c},
archivePrefix = {arXiv},
       eprint = {0805.2366},
 primaryClass = {astro-ph},
       adsurl = {https://ui.adsabs.harvard.edu/abs/2019ApJ...873..111I}
}

@ARTICLE{2018PASJ...70...66K,
       author = {{Kawanomoto}, Satoshi and {Uraguchi}, Fumihiro and {Komiyama}, Yutaka and {Miyazaki}, Satoshi and {Furusawa}, Hisanori and {Finet}, Fran{\c{c}}ois and {Hattori}, Takashi and {Wang}, Shiang-Yu and {Yasuda}, Naoki and {Suzuki}, Naotaka},
        title = "{Hyper Suprime-Cam: Filters}",
      journal = {\pasj},
         year = 2018,
        month = aug,
       volume = {70},
       number = {4},
          eid = {66},
        pages = {66},
          doi = {10.1093/pasj/psy056},
       adsurl = {https://ui.adsabs.harvard.edu/abs/2018PASJ...70...66K}
}

@ARTICLE{2018JOSS....3..695G,
       author = {{Green}, Gregory M.},
        title = "{dustmaps: A Python interface for maps of interstellar dust}",
      journal = {The Journal of Open Source Software},
         year = 2018,
        month = jun,
       volume = {3},
       number = {26},
        pages = {695},
          doi = {10.21105/joss.00695},
       adsurl = {https://ui.adsabs.harvard.edu/abs/2018JOSS....3..695G}
}

@ARTICLE{2018ApJ...856..125H,
       author = {{Hidalgo}, Sebastian L. and {Pietrinferni}, Adriano and {Cassisi}, Santi and {Salaris}, Maurizio and {Mucciarelli}, Alessio and {Savino}, Alessandro and {Aparicio}, Antonio and {Silva Aguirre}, Victor and {Verma}, Kuldeep},
        title = "{The Updated BaSTI Stellar Evolution Models and Isochrones. I. Solar-scaled Calculations}",
      journal = {\apj},
         year = 2018,
        month = apr,
       volume = {856},
       number = {2},
          eid = {125},
        pages = {125},
          doi = {10.3847/1538-4357/aab158},
archivePrefix = {arXiv},
       eprint = {1802.07319},
 primaryClass = {astro-ph.GA},
       adsurl = {https://ui.adsabs.harvard.edu/abs/2018ApJ...856..125H}
}

@ARTICLE{2018MNRAS.474..683C,
       author = {{Carballo-Bello}, Julio A. and {Mart{\'\i}nez-Delgado}, David and {Navarrete}, Camila and {Catelan}, M{\'a}rcio and {Mu{\~n}oz}, Ricardo R. and {Antoja}, Teresa and {Sollima}, Antonio},
        title = "{Tails and streams around the Galactic globular clusters NGC 1851, NGC 1904, NGC 2298 and NGC 2808}",
      journal = {\mnras},
         year = 2018,
        month = feb,
       volume = {474},
       number = {1},
        pages = {683-695},
          doi = {10.1093/mnras/stx2767},
archivePrefix = {arXiv},
       eprint = {1710.08927},
 primaryClass = {astro-ph.GA},
       adsurl = {https://ui.adsabs.harvard.edu/abs/2018MNRAS.474..683C}
}

@ARTICLE{2018PASJ...70S...1M,
       author = {{Miyazaki}, Satoshi and {Komiyama}, Yutaka and {Kawanomoto}, Satoshi and {Doi}, Yoshiyuki and {Furusawa}, Hisanori and {Hamana}, Takashi and {Hayashi}, Yusuke and {Ikeda}, Hiroyuki and {Kamata}, Yukiko and {Karoji}, Hiroshi and {Koike}, Michitaro and {Kurakami}, Tomio and {Miyama}, Shoken and {Morokuma}, Tomoki and {Nakata}, Fumiaki and {Namikawa}, Kazuhito and {Nakaya}, Hidehiko and {Nariai}, Kyoji and {Obuchi}, Yoshiyuki and {Oishi}, Yukie and {Okada}, Norio and {Okura}, Yuki and {Tait}, Philip and {Takata}, Tadafumi and {Tanaka}, Yoko and {Tanaka}, Masayuki and {Terai}, Tsuyoshi and {Tomono}, Daigo and {Uraguchi}, Fumihiro and {Usuda}, Tomonori and {Utsumi}, Yousuke and {Yamada}, Yoshihiko and {Yamanoi}, Hitomi and {Aihara}, Hiroaki and {Fujimori}, Hiroki and {Mineo}, Sogo and {Miyatake}, Hironao and {Oguri}, Masamune and {Uchida}, Tomohisa and {Tanaka}, Manobu M. and {Yasuda}, Naoki and {Takada}, Masahiro and {Murayama}, Hitoshi and {Nishizawa}, Atsushi J. and {Sugiyama}, Naoshi and {Chiba}, Masashi and {Futamase}, Toshifumi and {Wang}, Shiang-Yu and {Chen}, Hsin-Yo and {Ho}, Paul T.~P. and {Liaw}, Eric J.~Y. and {Chiu}, Chi-Fang and {Ho}, Cheng-Lin and {Lai}, Tsang-Chih and {Lee}, Yao-Cheng and {Jeng}, Dun-Zen and {Iwamura}, Satoru and {Armstrong}, Robert and {Bickerton}, Steve and {Bosch}, James and {Gunn}, James E. and {Lupton}, Robert H. and {Loomis}, Craig and {Price}, Paul and {Smith}, Steward and {Strauss}, Michael A. and {Turner}, Edwin L. and {Suzuki}, Hisanori and {Miyazaki}, Yasuhito and {Muramatsu}, Masaharu and {Yamamoto}, Koei and {Endo}, Makoto and {Ezaki}, Yutaka and {Ito}, Noboru and {Kawaguchi}, Noboru and {Sofuku}, Satoshi and {Taniike}, Tomoaki and {Akutsu}, Kotaro and {Dojo}, Naoto and {Kasumi}, Kazuyuki and {Matsuda}, Toru and {Imoto}, Kohei and {Miwa}, Yoshinori and {Suzuki}, Masayuki and {Takeshi}, Kunio and {Yokota}, Hideo},
        title = "{Hyper Suprime-Cam: System design and verification of image quality}",
      journal = {\pasj},
         year = 2018,
        month = jan,
       volume = {70},
          eid = {S1},
        pages = {S1},
          doi = {10.1093/pasj/psx063},
       adsurl = {https://ui.adsabs.harvard.edu/abs/2018PASJ...70S...1M}
}

@ARTICLE{2018PASJ...70S...5B,
       author = {{Bosch}, James and {Armstrong}, Robert and {Bickerton}, Steven and {Furusawa}, Hisanori and {Ikeda}, Hiroyuki and {Koike}, Michitaro and {Lupton}, Robert and {Mineo}, Sogo and {Price}, Paul and {Takata}, Tadafumi and {Tanaka}, Masayuki and {Yasuda}, Naoki and {AlSayyad}, Yusra and {Becker}, Andrew C. and {Coulton}, William and {Coupon}, Jean and {Garmilla}, Jose and {Huang}, Song and {Krughoff}, K. Simon and {Lang}, Dustin and {Leauthaud}, Alexie and {Lim}, Kian-Tat and {Lust}, Nate B. and {MacArthur}, Lauren A. and {Mandelbaum}, Rachel and {Miyatake}, Hironao and {Miyazaki}, Satoshi and {Murata}, Ryoma and {More}, Surhud and {Okura}, Yuki and {Owen}, Russell and {Swinbank}, John D. and {Strauss}, Michael A. and {Yamada}, Yoshihiko and {Yamanoi}, Hitomi},
        title = "{The Hyper Suprime-Cam software pipeline}",
      journal = {\pasj},
         year = 2018,
        month = jan,
       volume = {70},
          eid = {S5},
        pages = {S5},
          doi = {10.1093/pasj/psx080},
archivePrefix = {arXiv},
       eprint = {1705.06766},
 primaryClass = {astro-ph.IM},
       adsurl = {https://ui.adsabs.harvard.edu/abs/2018PASJ...70S...5B}
}

@ARTICLE{2018PASJ...70S...2K,
       author = {{Komiyama}, Yutaka and {Obuchi}, Yoshiyuki and {Nakaya}, Hidehiko and {Kamata}, Yukiko and {Kawanomoto}, Satoshi and {Utsumi}, Yousuke and {Miyazaki}, Satoshi and {Uraguchi}, Fumihiro and {Furusawa}, Hisanori and {Morokuma}, Tomoki and {Uchida}, Tomohisa and {Miyatake}, Hironao and {Mineo}, Sogo and {Fujimori}, Hiroki and {Aihara}, Hiroaki and {Karoji}, Hiroshi and {Gunn}, James E. and {Wang}, Shiang-Yu},
        title = "{Hyper Suprime-Cam: Camera dewar design}",
      journal = {\pasj},
         year = 2018,
        month = jan,
       volume = {70},
          eid = {S2},
        pages = {S2},
          doi = {10.1093/pasj/psx069},
       adsurl = {https://ui.adsabs.harvard.edu/abs/2018PASJ...70S...2K}
}

@ARTICLE{2018PASJ...70S...3F,
       author = {{Furusawa}, Hisanori and {Koike}, Michitaro and {Takata}, Tadafumi and {Okura}, Yuki and {Miyatake}, Hironao and {Lupton}, Robert H. and {Bickerton}, Steven and {Price}, Paul A. and {Bosch}, James and {Yasuda}, Naoki and {Mineo}, Sogo and {Yamada}, Yoshihiko and {Miyazaki}, Satoshi and {Nakata}, Fumiaki and {Koshida}, Shintaro and {Komiyama}, Yutaka and {Utsumi}, Yousuke and {Kawanomoto}, Satoshi and {Jeschke}, Eric and {Noumaru}, Junichi and {Schubert}, Kiaina and {Iwata}, Ikuru and {Finet}, Francois and {Fujiyoshi}, Takuya and {Tajitsu}, Akito and {Terai}, Tsuyoshi and {Lee}, Chien-Hsiu},
        title = "{The on-site quality-assurance system for Hyper Suprime-Cam: OSQAH}",
      journal = {\pasj},
         year = 2018,
        month = jan,
       volume = {70},
          eid = {S3},
        pages = {S3},
          doi = {10.1093/pasj/psx079},
       adsurl = {https://ui.adsabs.harvard.edu/abs/2018PASJ...70S...3F}
}

@ARTICLE{2018PASJ...70S...8A,
       author = {{Aihara}, Hiroaki and {Armstrong}, Robert and {Bickerton}, Steven and {Bosch}, James and {Coupon}, Jean and {Furusawa}, Hisanori and {Hayashi}, Yusuke and {Ikeda}, Hiroyuki and {Kamata}, Yukiko and {Karoji}, Hiroshi and {Kawanomoto}, Satoshi and {Koike}, Michitaro and {Komiyama}, Yutaka and {Lang}, Dustin and {Lupton}, Robert H. and {Mineo}, Sogo and {Miyatake}, Hironao and {Miyazaki}, Satoshi and {Morokuma}, Tomoki and {Obuchi}, Yoshiyuki and {Oishi}, Yukie and {Okura}, Yuki and {Price}, Paul A. and {Takata}, Tadafumi and {Tanaka}, Manobu M. and {Tanaka}, Masayuki and {Tanaka}, Yoko and {Uchida}, Tomohisa and {Uraguchi}, Fumihiro and {Utsumi}, Yousuke and {Wang}, Shiang-Yu and {Yamada}, Yoshihiko and {Yamanoi}, Hitomi and {Yasuda}, Naoki and {Arimoto}, Nobuo and {Chiba}, Masashi and {Finet}, Francois and {Fujimori}, Hiroki and {Fujimoto}, Seiji and {Furusawa}, Junko and {Goto}, Tomotsugu and {Goulding}, Andy and {Gunn}, James E. and {Harikane}, Yuichi and {Hattori}, Takashi and {Hayashi}, Masao and {He{\l}miniak}, Krzysztof G. and {Higuchi}, Ryo and {Hikage}, Chiaki and {Ho}, Paul T.~P. and {Hsieh}, Bau-Ching and {Huang}, Kuiyun and {Huang}, Song and {Imanishi}, Masatoshi and {Iwata}, Ikuru and {Jaelani}, Anton T. and {Jian}, Hung-Yu and {Kashikawa}, Nobunari and {Katayama}, Nobuhiko and {Kojima}, Takashi and {Konno}, Akira and {Koshida}, Shintaro and {Kusakabe}, Haruka and {Leauthaud}, Alexie and {Lee}, Chien-Hsiu and {Lin}, Lihwai and {Lin}, Yen-Ting and {Mandelbaum}, Rachel and {Matsuoka}, Yoshiki and {Medezinski}, Elinor and {Miyama}, Shoken and {Momose}, Rieko and {More}, Anupreeta and {More}, Surhud and {Mukae}, Shiro and {Murata}, Ryoma and {Murayama}, Hitoshi and {Nagao}, Tohru and {Nakata}, Fumiaki and {Niida}, Mana and {Niikura}, Hiroko and {Nishizawa}, Atsushi J. and {Oguri}, Masamune and {Okabe}, Nobuhiro and {Ono}, Yoshiaki and {Onodera}, Masato and {Onoue}, Masafusa and {Ouchi}, Masami and {Pyo}, Tae-Soo and {Shibuya}, Takatoshi and {Shimasaku}, Kazuhiro and {Simet}, Melanie and {Speagle}, Joshua and {Spergel}, David N. and {Strauss}, Michael A. and {Sugahara}, Yuma and {Sugiyama}, Naoshi and {Suto}, Yasushi and {Suzuki}, Nao and {Tait}, Philip J. and {Takada}, Masahiro and {Terai}, Tsuyoshi and {Toba}, Yoshiki and {Turner}, Edwin L. and {Uchiyama}, Hisakazu and {Umetsu}, Keiichi and {Urata}, Yuji and {Usuda}, Tomonori and {Yeh}, Sherry and {Yuma}, Suraphong},
        title = "{First data release of the Hyper Suprime-Cam Subaru Strategic Program}",
      journal = {\pasj},
         year = 2018,
        month = jan,
       volume = {70},
          eid = {S8},
        pages = {S8},
          doi = {10.1093/pasj/psx081},
archivePrefix = {arXiv},
       eprint = {1702.08449},
 primaryClass = {astro-ph.IM},
       adsurl = {https://ui.adsabs.harvard.edu/abs/2018PASJ...70S...8A}
}

@INPROCEEDINGS{2017ASPC..512..279J,
       author = {{Juri{\'c}}, M. and {Kantor}, J. and {Lim}, K.-T. and {Lupton}, R.~H. and {Dubois-Felsmann}, G. and {Jenness}, T. and {Axelrod}, T.~S. and {Aleksi{\'c}}, J. and {Allsman}, R.~A. and {AlSayyad}, Y. and {Alt}, J. and {Armstrong}, R. and {Basney}, J. and {Becker}, A.~C. and {Becla}, J. and {Biswas}, R. and {Bosch}, J. and {Boutigny}, D. and {Kind}, M.~C. and {Ciardi}, D.~R. and {Connolly}, A.~J. and {Daniel}, S.~F. and {Daues}, G.~E. and {Economou}, F. and {Chiang}, H.-F. and {Fausti}, A. and {Fisher-Levine}, M. and {Freemon}, D.~M. and {Gris}, P. and {Hernandez}, F. and {Hoblitt}, J. and {Ivezi{\'c}}, Z. and {Jammes}, F. and {Jevremovi{\'c}}, D. and {Jones}, R.~L. and {Kalmbach}, J.~B. and {Kasliwal}, V.~P. and {Krughoff}, K.~S. and {Lurie}, J. and {Lust}, N.~B. and {MacArthur}, L.~A. and {Melchior}, P. and {Moeyens}, J. and {Nidever}, D.~L. and {Owen}, R. and {Parejko}, J.~K. and {Peterson}, J.~M. and {Petravick}, D. and {Pietrowicz}, S.~R. and {Price}, P.~A. and {Reiss}, D.~J. and {Shaw}, R.~A. and {Sick}, J. and {Slater}, C.~T. and {Strauss}, M.~A. and {Sullivan}, I.~S. and {Swinbank}, J.~D. and {Van Dyk}, S. and {Vuj{\v{c}}i{\'c}}, V. and {Withers}, A. and {Yoachim}, P.},
        title = "{The LSST Data Management System}",
    booktitle = {Astronomical Data Analysis Software and Systems XXV},
         year = 2017,
       editor = {{Lorente}, N.~P.~F. and {Shortridge}, K. and {Wayth}, R.},
       series = {Astronomical Society of the Pacific Conference Series},
       volume = {512},
        month = dec,
        pages = {279},
          doi = {10.48550/arXiv.1512.07914},
archivePrefix = {arXiv},
       eprint = {1512.07914},
 primaryClass = {astro-ph.IM},
       adsurl = {https://ui.adsabs.harvard.edu/abs/2017ASPC..512..279J}
}

@ARTICLE{2017MNRAS.471.3668S,
       author = {{Sollima}, A. and {Baumgardt}, H.},
        title = "{The global mass functions of 35 Galactic globular clusters: I. Observational data and correlations with cluster parameters}",
      journal = {\mnras},
         year = 2017,
        month = nov,
       volume = {471},
       number = {3},
        pages = {3668-3679},
          doi = {10.1093/mnras/stx1856},
archivePrefix = {arXiv},
       eprint = {1708.09529},
 primaryClass = {astro-ph.GA},
       adsurl = {https://ui.adsabs.harvard.edu/abs/2017MNRAS.471.3668S}
}

@ARTICLE{2017MNRAS.471.2587S,
       author = {{Starkenburg}, Else and {Martin}, Nicolas and {Youakim}, Kris and {Aguado}, David S. and {Allende Prieto}, Carlos and {Arentsen}, Anke and {Bernard}, Edouard J. and {Bonifacio}, Piercarlo and {Caffau}, Elisabetta and {Carlberg}, Raymond G. and {C{\^o}t{\'e}}, Patrick and {Fouesneau}, Morgan and {Fran{\c{c}}ois}, Patrick and {Franke}, Oliver and {Gonz{\'a}lez Hern{\'a}ndez}, Jonay I. and {Gwyn}, Stephen D.~J. and {Hill}, Vanessa and {Ibata}, Rodrigo A. and {Jablonka}, Pascale and {Longeard}, Nicolas and {McConnachie}, Alan W. and {Navarro}, Julio F. and {S{\'a}nchez-Janssen}, Rub{\'e}n and {Tolstoy}, Eline and {Venn}, Kim A.},
        title = "{The Pristine survey - I. Mining the Galaxy for the most metal-poor stars}",
      journal = {\mnras},
         year = 2017,
        month = nov,
       volume = {471},
       number = {3},
        pages = {2587-2604},
          doi = {10.1093/mnras/stx1068},
archivePrefix = {arXiv},
       eprint = {1705.01113},
 primaryClass = {astro-ph.GA},
       adsurl = {https://ui.adsabs.harvard.edu/abs/2017MNRAS.471.2587S}
}

@ARTICLE{2017MNRAS.470...60E,
       author = {{Erkal}, Denis and {Koposov}, Sergey E. and {Belokurov}, Vasily},
        title = "{A sharper view of Pal 5's tails: discovery of stream perturbations with a novel non-parametric technique}",
      journal = {\mnras},
         year = 2017,
        month = sep,
       volume = {470},
       number = {1},
        pages = {60-84},
          doi = {10.1093/mnras/stx1208},
archivePrefix = {arXiv},
       eprint = {1609.01282},
 primaryClass = {astro-ph.GA},
       adsurl = {https://ui.adsabs.harvard.edu/abs/2017MNRAS.470...60E}
}

@ARTICLE{2017AJ....153..234B,
       author = {{Bohlin}, Ralph C. and {M{\'e}sz{\'a}ros}, Szabolcs and {Fleming}, Scott W. and {Gordon}, Karl D. and {Koekemoer}, Anton M. and {Kov{\'a}cs}, J{\'o}zsef},
        title = "{A New Stellar Atmosphere Grid and Comparisons with HST/STIS CALSPEC Flux Distributions}",
      journal = {\aj},
         year = 2017,
        month = may,
       volume = {153},
       number = {5},
          eid = {234},
        pages = {234},
          doi = {10.3847/1538-3881/aa6ba9},
archivePrefix = {arXiv},
       eprint = {1704.00653},
 primaryClass = {astro-ph.SR},
       adsurl = {https://ui.adsabs.harvard.edu/abs/2017AJ....153..234B}
}

@ARTICLE{2017MNRAS.464.3636M,
       author = {{Milone}, A.~P. and {Piotto}, G. and {Renzini}, A. and {Marino}, A.~F. and {Bedin}, L.~R. and {Vesperini}, E. and {D'Antona}, F. and {Nardiello}, D. and {Anderson}, J. and {King}, I.~R. and {Yong}, D. and {Bellini}, A. and {Aparicio}, A. and {Barbuy}, B. and {Brown}, T.~M. and {Cassisi}, S. and {Ortolani}, S. and {Salaris}, M. and {Sarajedini}, A. and {van der Marel}, R.~P.},
        title = "{The Hubble Space Telescope UV Legacy Survey of Galactic globular clusters - IX. The Atlas of multiple stellar populations}",
      journal = {\mnras},
         year = 2017,
        month = jan,
       volume = {464},
       number = {3},
        pages = {3636-3656},
          doi = {10.1093/mnras/stw2531},
archivePrefix = {arXiv},
       eprint = {1610.00451},
 primaryClass = {astro-ph.SR},
       adsurl = {https://ui.adsabs.harvard.edu/abs/2017MNRAS.464.3636M}
}

@ARTICLE{2016MNRAS.463.2383W,
       author = {{Webb}, Jeremy J. and {Vesperini}, Enrico},
        title = "{Radial variation in the stellar mass functions of star clusters}",
      journal = {\mnras},
         year = 2016,
        month = dec,
       volume = {463},
       number = {3},
        pages = {2383-2393},
          doi = {10.1093/mnras/stw2186},
archivePrefix = {arXiv},
       eprint = {1608.07293},
 primaryClass = {astro-ph.GA},
       adsurl = {https://ui.adsabs.harvard.edu/abs/2016MNRAS.463.2383W}
}

@ARTICLE{2016ApJ...833..167M,
       author = {{Martin}, Nicolas F. and {Ibata}, Rodrigo A. and {Lewis}, Geraint F. and {McConnachie}, Alan and {Babul}, Arif and {Bate}, Nicholas F. and {Bernard}, Edouard and {Chapman}, Scott C. and {Collins}, Michelle M.~L. and {Conn}, Anthony R. and {Crnojevi{\'c}}, Denija and {Fardal}, Mark A. and {Ferguson}, Annette M.~N. and {Irwin}, Michael and {Mackey}, A. Dougal and {McMonigal}, Brendan and {Navarro}, Julio F. and {Rich}, R. Michael},
        title = "{The PAndAS View of the Andromeda Satellite System. II. Detailed Properties of 23 M31 Dwarf Spheroidal Galaxies}",
      journal = {\apj},
         year = 2016,
        month = dec,
       volume = {833},
       number = {2},
          eid = {167},
        pages = {167},
          doi = {10.3847/1538-4357/833/2/167},
archivePrefix = {arXiv},
       eprint = {1610.01158},
 primaryClass = {astro-ph.GA},
       adsurl = {https://ui.adsabs.harvard.edu/abs/2016ApJ...833..167M}
}

@ARTICLE{2016A&A...595A...1G,
       author = {{Gaia Collaboration} and {Prusti}, T. and {de Bruijne}, J.~H.~J. and {Brown}, A.~G.~A. and {Vallenari}, A. and {Babusiaux}, C. and {Bailer-Jones}, C.~A.~L. and {Bastian}, U. and {Biermann}, M. and {Evans}, D.~W. and {Eyer}, L. and {Jansen}, F. and {Jordi}, C. and {Klioner}, S.~A. and {Lammers}, U. and {Lindegren}, L. and {Luri}, X. and {Mignard}, F. and {Milligan}, D.~J. and {Panem}, C. and {Poinsignon}, V. and {Pourbaix}, D. and {Randich}, S. and {Sarri}, G. and {Sartoretti}, P. and {Siddiqui}, H.~I. and {Soubiran}, C. and {Valette}, V. and {van Leeuwen}, F. and {Walton}, N.~A. and {Aerts}, C. and {Arenou}, F. and {Cropper}, M. and {Drimmel}, R. and {H{\o}g}, E. and {Katz}, D. and {Lattanzi}, M.~G. and {O'Mullane}, W. and {Grebel}, E.~K. and {Holland}, A.~D. and {Huc}, C. and {Passot}, X. and {Bramante}, L. and {Cacciari}, C. and {Casta{\~n}eda}, J. and {Chaoul}, L. and {Cheek}, N. and {De Angeli}, F. and {Fabricius}, C. and {Guerra}, R. and {Hern{\'a}ndez}, J. and {Jean-Antoine-Piccolo}, A. and {Masana}, E. and {Messineo}, R. and {Mowlavi}, N. and {Nienartowicz}, K. and {Ord{\'o}{\~n}ez-Blanco}, D. and {Panuzzo}, P. and {Portell}, J. and {Richards}, P.~J. and {Riello}, M. and {Seabroke}, G.~M. and {Tanga}, P. and {Th{\'e}venin}, F. and {Torra}, J. and {Els}, S.~G. and {Gracia-Abril}, G. and {Comoretto}, G. and {Garcia-Reinaldos}, M. and {Lock}, T. and {Mercier}, E. and {Altmann}, M. and {Andrae}, R. and {Astraatmadja}, T.~L. and {Bellas-Velidis}, I. and {Benson}, K. and {Berthier}, J. and {Blomme}, R. and {Busso}, G. and {Carry}, B. and {Cellino}, A. and {Clementini}, G. and {Cowell}, S. and {Creevey}, O. and {Cuypers}, J. and {Davidson}, M. and {De Ridder}, J. and {de Torres}, A. and {Delchambre}, L. and {Dell'Oro}, A. and {Ducourant}, C. and {Fr{\'e}mat}, Y. and {Garc{\'\i}a-Torres}, M. and {Gosset}, E. and {Halbwachs}, J.-L. and {Hambly}, N.~C. and {Harrison}, D.~L. and {Hauser}, M. and {Hestroffer}, D. and {Hodgkin}, S.~T. and {Huckle}, H.~E. and {Hutton}, A. and {Jasniewicz}, G. and {Jordan}, S. and {Kontizas}, M. and {Korn}, A.~J. and {Lanzafame}, A.~C. and {Manteiga}, M. and {Moitinho}, A. and {Muinonen}, K. and {Osinde}, J. and {Pancino}, E. and {Pauwels}, T. and {Petit}, J.-M. and {Recio-Blanco}, A. and {Robin}, A.~C. and {Sarro}, L.~M. and {Siopis}, C. and {Smith}, M. and {Smith}, K.~W. and {Sozzetti}, A. and {Thuillot}, W. and {van Reeven}, W. and {Viala}, Y. and {Abbas}, U. and {Abreu Aramburu}, A. and {Accart}, S. and {Aguado}, J.~J. and {Allan}, P.~M. and {Allasia}, W. and {Altavilla}, G. and {{\'A}lvarez}, M.~A. and {Alves}, J. and {Anderson}, R.~I. and {Andrei}, A.~H. and {Anglada Varela}, E. and {Antiche}, E. and {Antoja}, T. and {Ant{\'o}n}, S. and {Arcay}, B. and {Atzei}, A. and {Ayache}, L. and {Bach}, N. and {Baker}, S.~G. and {Balaguer-N{\'u}{\~n}ez}, L. and {Barache}, C. and {Barata}, C. and {Barbier}, A. and {Barblan}, F. and {Baroni}, M. and {Barrado y Navascu{\'e}s}, D. and {Barros}, M. and {Barstow}, M.~A. and {Becciani}, U. and {Bellazzini}, M. and {Bellei}, G. and {Bello Garc{\'\i}a}, A. and {Belokurov}, V. and {Bendjoya}, P. and {Berihuete}, A. and {Bianchi}, L. and {Bienaym{\'e}}, O. and {Billebaud}, F. and {Blagorodnova}, N. and {Blanco-Cuaresma}, S. and {Boch}, T. and {Bombrun}, A. and {Borrachero}, R. and {Bouquillon}, S. and {Bourda}, G. and {Bouy}, H. and {Bragaglia}, A. and {Breddels}, M.~A. and {Brouillet}, N. and {Br{\"u}semeister}, T. and {Bucciarelli}, B. and {Budnik}, F. and {Burgess}, P. and {Burgon}, R. and {Burlacu}, A. and {Busonero}, D. and {Buzzi}, R. and {Caffau}, E. and {Cambras}, J. and {Campbell}, H. and {Cancelliere}, R. and {Cantat-Gaudin}, T. and {Carlucci}, T. and {Carrasco}, J.~M. and {Castellani}, M. and {Charlot}, P. and {Charnas}, J. and {Charvet}, P. and {Chassat}, F. and {Chiavassa}, A. and {Clotet}, M. and {Cocozza}, G. and {Collins}, R.~S. and {Collins}, P. and {Costigan}, G.},
        title = "{The Gaia mission}",
      journal = {\aap},
         year = 2016,
        month = nov,
       volume = {595},
          eid = {A1},
        pages = {A1},
          doi = {10.1051/0004-6361/201629272},
archivePrefix = {arXiv},
       eprint = {1609.04153},
 primaryClass = {astro-ph.IM},
       adsurl = {https://ui.adsabs.harvard.edu/abs/2016A&A...595A...1G}
}

@ARTICLE{2016MNRAS.461.3639K,
       author = {{Kuzma}, P.~B. and {Da Costa}, G.~S. and {Mackey}, A.~D. and {Roderick}, T.~A.},
        title = "{The outer envelopes of globular clusters - I. NGC 7089 (M2)}",
      journal = {\mnras},
         year = 2016,
        month = oct,
       volume = {461},
       number = {4},
        pages = {3639-3652},
          doi = {10.1093/mnras/stw1561},
archivePrefix = {arXiv},
       eprint = {1606.05949},
 primaryClass = {astro-ph.GA},
       adsurl = {https://ui.adsabs.harvard.edu/abs/2016MNRAS.461.3639K}
}

@ARTICLE{2015MNRAS.454.3542E,
       author = {{Erkal}, Denis and {Belokurov}, Vasily},
        title = "{Properties of dark subhaloes from gaps in tidal streams}",
      journal = {\mnras},
         year = 2015,
        month = dec,
       volume = {454},
       number = {4},
        pages = {3542-3558},
          doi = {10.1093/mnras/stv2122},
archivePrefix = {arXiv},
       eprint = {1507.05625},
 primaryClass = {astro-ph.GA},
       adsurl = {https://ui.adsabs.harvard.edu/abs/2015MNRAS.454.3542E}
}

@ARTICLE{2015ApJ...814..144B,
       author = {{Beccari}, G. and {Dalessandro}, E. and {Lanzoni}, B. and {Ferraro}, F.~R. and {Bellazzini}, M. and {Sollima}, A.},
        title = "{Deep Multi-telescope Photometry of NGC 5466. II. The Radial Behavior of the Mass Function Slope}",
      journal = {\apj},
         year = 2015,
        month = dec,
       volume = {814},
       number = {2},
          eid = {144},
        pages = {144},
          doi = {10.1088/0004-637X/814/2/144},
archivePrefix = {arXiv},
       eprint = {1511.00829},
 primaryClass = {astro-ph.SR},
       adsurl = {https://ui.adsabs.harvard.edu/abs/2015ApJ...814..144B}
}

@ARTICLE{2015ApJ...811..123F,
       author = {{Fritz}, T.~K. and {Kallivayalil}, N.},
        title = "{The Proper Motion of Palomar 5}",
      journal = {\apj},
         year = 2015,
        month = oct,
       volume = {811},
       number = {2},
          eid = {123},
        pages = {123},
          doi = {10.1088/0004-637X/811/2/123},
archivePrefix = {arXiv},
       eprint = {1508.06647},
 primaryClass = {astro-ph.GA},
       adsurl = {https://ui.adsabs.harvard.edu/abs/2015ApJ...811..123F}
}

@ARTICLE{2015MNRAS.450..575A,
       author = {{Amorisco}, N.~C.},
        title = "{On feathers, bifurcations and shells: the dynamics of tidal streams across the mass scale}",
      journal = {\mnras},
         year = 2015,
        month = jun,
       volume = {450},
       number = {1},
        pages = {575-591},
          doi = {10.1093/mnras/stv648},
archivePrefix = {arXiv},
       eprint = {1410.0360},
 primaryClass = {astro-ph.GA},
       adsurl = {https://ui.adsabs.harvard.edu/abs/2015MNRAS.450..575A}
}

@ARTICLE{2015MNRAS.450.1136E,
       author = {{Erkal}, Denis and {Belokurov}, Vasily},
        title = "{Forensics of subhalo-stream encounters: the three phases of gap growth}",
      journal = {\mnras},
         year = 2015,
        month = jun,
       volume = {450},
       number = {1},
        pages = {1136-1149},
          doi = {10.1093/mnras/stv655},
archivePrefix = {arXiv},
       eprint = {1412.6035},
 primaryClass = {astro-ph.GA},
       adsurl = {https://ui.adsabs.harvard.edu/abs/2015MNRAS.450.1136E}
}

@ARTICLE{2015MNRAS.448...42L,
       author = {{Lamb}, M.~P. and {Venn}, K.~A. and {Shetrone}, M.~D. and {Sakari}, C.~M. and {Pritzl}, B.~J.},
        title = "{Chemical abundances in the globular clusters NGC 5024 and NGC 5466 from optical and infrared spectroscopy}",
      journal = {\mnras},
         year = 2015,
        month = mar,
       volume = {448},
       number = {1},
        pages = {42-58},
          doi = {10.1093/mnras/stu2674},
archivePrefix = {arXiv},
       eprint = {1412.5210},
 primaryClass = {astro-ph.SR},
       adsurl = {https://ui.adsabs.harvard.edu/abs/2015MNRAS.448...42L}
}

@ARTICLE{2015MNRAS.446.3100H,
       author = {{Hozumi}, Shunsuke and {Burkert}, Andreas},
        title = "{Development of multiple tidal tails around globular clusters and dwarf satellite galaxies}",
      journal = {\mnras},
         year = 2015,
        month = jan,
       volume = {446},
       number = {3},
        pages = {3100-3109},
          doi = {10.1093/mnras/stu2318},
archivePrefix = {arXiv},
       eprint = {1409.2157},
 primaryClass = {astro-ph.GA},
       adsurl = {https://ui.adsabs.harvard.edu/abs/2015MNRAS.446.3100H}
}

@ARTICLE{2014MNRAS.445.1048W,
       author = {{Webb}, Jeremy J. and {Sills}, Alison and {Harris}, William E. and {Hurley}, Jarrod R.},
        title = "{The effects of orbital inclination on the scale size and evolution of tidally filling star clusters}",
      journal = {\mnras},
         year = 2014,
        month = nov,
       volume = {445},
       number = {1},
        pages = {1048-1055},
          doi = {10.1093/mnras/stu1763},
archivePrefix = {arXiv},
       eprint = {1409.0879},
 primaryClass = {astro-ph.GA},
       adsurl = {https://ui.adsabs.harvard.edu/abs/2014MNRAS.445.1048W}
}

@ARTICLE{2014ApJ...793..110M,
       author = {{Moreno}, Edmundo and {Pichardo}, B{\'a}rbara and {Vel{\'a}zquez}, H{\'e}ctor},
        title = "{Tidal Radii and Destruction Rates of Globular Clusters in the Milky Way due to Bulge-Bar and Disk Shocking}",
      journal = {\apj},
         year = 2014,
        month = oct,
       volume = {793},
       number = {2},
          eid = {110},
        pages = {110},
          doi = {10.1088/0004-637X/793/2/110},
archivePrefix = {arXiv},
       eprint = {1408.0457},
 primaryClass = {astro-ph.GA},
       adsurl = {https://ui.adsabs.harvard.edu/abs/2014ApJ...793..110M}
}

@ARTICLE{2014ApJ...788..181N,
       author = {{Ngan}, W.~H.~W. and {Carlberg}, R.~G.},
        title = "{Using Gaps in N-body Tidal Streams to Probe Missing Satellites}",
      journal = {\apj},
         year = 2014,
        month = jun,
       volume = {788},
       number = {2},
          eid = {181},
        pages = {181},
          doi = {10.1088/0004-637X/788/2/181},
archivePrefix = {arXiv},
       eprint = {1311.1710},
 primaryClass = {astro-ph.CO},
       adsurl = {https://ui.adsabs.harvard.edu/abs/2014ApJ...788..181N}
}

@ARTICLE{2014PASJ...66R...1T,
       author = {{Takada}, Masahiro and {Ellis}, Richard S. and {Chiba}, Masashi and {Greene}, Jenny E. and {Aihara}, Hiroaki and {Arimoto}, Nobuo and {Bundy}, Kevin and {Cohen}, Judith and {Dor{\'e}}, Olivier and {Graves}, Genevieve and {Gunn}, James E. and {Heckman}, Timothy and {Hirata}, Christopher M. and {Ho}, Paul and {Kneib}, Jean-Paul and {Le F{\`e}vre}, Olivier and {Lin}, Lihwai and {More}, Surhud and {Murayama}, Hitoshi and {Nagao}, Tohru and {Ouchi}, Masami and {Seiffert}, Michael and {Silverman}, John D. and {Sodr{\'e}}, Laerte and {Spergel}, David N. and {Strauss}, Michael A. and {Sugai}, Hajime and {Suto}, Yasushi and {Takami}, Hideki and {Wyse}, Rosemary},
        title = "{Extragalactic science, cosmology, and Galactic archaeology with the Subaru Prime Focus Spectrograph}",
      journal = {\pasj},
         year = 2014,
        month = feb,
       volume = {66},
       number = {1},
          eid = {R1},
        pages = {R1},
          doi = {10.1093/pasj/pst019},
archivePrefix = {arXiv},
       eprint = {1206.0737},
 primaryClass = {astro-ph.CO},
       adsurl = {https://ui.adsabs.harvard.edu/abs/2014PASJ...66R...1T}
}

@ARTICLE{2013ApJ...775..134V,
       author = {{VandenBerg}, Don A. and {Brogaard}, K. and {Leaman}, R. and {Casagrande}, L.},
        title = "{The Ages of 55 Globular Clusters as Determined Using an Improved \textbackslashDelta V\^HB\_TO Method along with Color-Magnitude Diagram Constraints, and Their Implications for Broader Issues}",
      journal = {\apj},
         year = 2013,
        month = oct,
       volume = {775},
       number = {2},
          eid = {134},
        pages = {134},
          doi = {10.1088/0004-637X/775/2/134},
archivePrefix = {arXiv},
       eprint = {1308.2257},
 primaryClass = {astro-ph.GA},
       adsurl = {https://ui.adsabs.harvard.edu/abs/2013ApJ...775..134V}
}

@ARTICLE{2013ApJ...776...60B,
       author = {{Beccari}, G. and {Dalessandro}, E. and {Lanzoni}, B. and {Ferraro}, F.~R. and {Sollima}, A. and {Bellazzini}, M. and {Miocchi}, P.},
        title = "{Deep Multi-telescope Photometry of NGC 5466. I. Blue Stragglers and Binary Systems}",
      journal = {\apj},
         year = 2013,
        month = oct,
       volume = {776},
       number = {1},
          eid = {60},
        pages = {60},
          doi = {10.1088/0004-637X/776/1/60},
archivePrefix = {arXiv},
       eprint = {1308.5810},
 primaryClass = {astro-ph.SR},
       adsurl = {https://ui.adsabs.harvard.edu/abs/2013ApJ...776...60B}
}

@ARTICLE{2013MNRAS.433.1378L,
       author = {{Lamers}, Henny J.~G.~L.~M. and {Baumgardt}, Holger and {Gieles}, Mark},
        title = "{The evolution of the global stellar mass function of star clusters: an analytic description}",
      journal = {\mnras},
         year = 2013,
        month = aug,
       volume = {433},
       number = {2},
        pages = {1378-1388},
          doi = {10.1093/mnras/stt808},
archivePrefix = {arXiv},
       eprint = {1305.2652},
 primaryClass = {astro-ph.GA},
       adsurl = {https://ui.adsabs.harvard.edu/abs/2013MNRAS.433.1378L}
}

@ARTICLE{2013ApJ...768..171C,
       author = {{Carlberg}, R.~G. and {Grillmair}, C.~J.},
        title = "{Gaps in the GD-1 Star Stream}",
      journal = {\apj},
         year = 2013,
        month = may,
       volume = {768},
       number = {2},
          eid = {171},
        pages = {171},
          doi = {10.1088/0004-637X/768/2/171},
archivePrefix = {arXiv},
       eprint = {1303.4342},
 primaryClass = {astro-ph.CO},
       adsurl = {https://ui.adsabs.harvard.edu/abs/2013ApJ...768..171C}
}

@ARTICLE{2013ApJS..205...20M,
       author = {{Magnier}, E.~A. and {Schlafly}, E. and {Finkbeiner}, D. and {Juric}, M. and {Tonry}, J.~L. and {Burgett}, W.~S. and {Chambers}, K.~C. and {Flewelling}, H.~A. and {Kaiser}, N. and {Kudritzki}, R.-P. and {Morgan}, J.~S. and {Price}, P.~A. and {Sweeney}, W.~E. and {Stubbs}, C.~W.},
        title = "{The Pan-STARRS 1 Photometric Reference Ladder, Release 12.01}",
      journal = {\apjs},
         year = 2013,
        month = apr,
       volume = {205},
       number = {2},
          eid = {20},
        pages = {20},
          doi = {10.1088/0067-0049/205/2/20},
archivePrefix = {arXiv},
       eprint = {1303.3634},
 primaryClass = {astro-ph.IM},
       adsurl = {https://ui.adsabs.harvard.edu/abs/2013ApJS..205...20M}
}

@ARTICLE{2013PASP..125..306F,
       author = {{Foreman-Mackey}, Daniel and {Hogg}, David W. and {Lang}, Dustin and {Goodman}, Jonathan},
        title = "{emcee: The MCMC Hammer}",
      journal = {\pasp},
         year = 2013,
        month = mar,
       volume = {125},
       number = {925},
        pages = {306},
          doi = {10.1086/670067},
archivePrefix = {arXiv},
       eprint = {1202.3665},
 primaryClass = {astro-ph.IM},
       adsurl = {https://ui.adsabs.harvard.edu/abs/2013PASP..125..306F}
}

@ARTICLE{2012ApJ...756..158S,
       author = {{Schlafly}, E.~F. and {Finkbeiner}, D.~P. and {Juri{\'c}}, M. and {Magnier}, E.~A. and {Burgett}, W.~S. and {Chambers}, K.~C. and {Grav}, T. and {Hodapp}, K.~W. and {Kaiser}, N. and {Kudritzki}, R.-P. and {Martin}, N.~F. and {Morgan}, J.~S. and {Price}, P.~A. and {Rix}, H.-W. and {Stubbs}, C.~W. and {Tonry}, J.~L. and {Wainscoat}, R.~J.},
        title = "{Photometric Calibration of the First 1.5 Years of the Pan-STARRS1 Survey}",
      journal = {\apj},
         year = 2012,
        month = sep,
       volume = {756},
       number = {2},
          eid = {158},
        pages = {158},
          doi = {10.1088/0004-637X/756/2/158},
archivePrefix = {arXiv},
       eprint = {1201.2208},
 primaryClass = {astro-ph.IM},
       adsurl = {https://ui.adsabs.harvard.edu/abs/2012ApJ...756..158S}
}

@INPROCEEDINGS{2012SPIE.8446E..0ZM,
       author = {{Miyazaki}, Satoshi and {Komiyama}, Yutaka and {Nakaya}, Hidehiko and {Kamata}, Yukiko and {Doi}, Yoshi and {Hamana}, Takashi and {Karoji}, Hiroshi and {Furusawa}, Hisanori and {Kawanomoto}, Satoshi and {Morokuma}, Tomoki and {Ishizuka}, Yuki and {Nariai}, Kyoji and {Tanaka}, Yoko and {Uraguchi}, Fumihiro and {Utsumi}, Yousuke and {Obuchi}, Yoshiyuki and {Okura}, Yuki and {Oguri}, Masamune and {Takata}, Tadafumi and {Tomono}, Daigo and {Kurakami}, Tomio and {Namikawa}, Kazuhito and {Usuda}, Tomonori and {Yamanoi}, Hitomi and {Terai}, Tsuyoshi and {Uekiyo}, Hatsue and {Yamada}, Yoshihiko and {Koike}, Michitaro and {Aihara}, Hiro and {Fujimori}, Yuki and {Mineo}, Sogo and {Miyatake}, Hironao and {Yasuda}, Naoki and {Nishizawa}, Jun and {Saito}, Tomoki and {Tanaka}, Manobu and {Uchida}, Tomohisa and {Katayama}, Nobu and {Wang}, Shiang-Yu and {Chen}, Hsin-Yo and {Lupton}, Robert and {Loomis}, Craig and {Bickerton}, Steve and {Price}, Paul and {Gunn}, Jim and {Suzuki}, Hisanori and {Miyazaki}, Yasuhito and {Muramatsu}, Masaharu and {Yamamoto}, Koei and {Endo}, Makoto and {Ezaki}, Yutaka and {Itoh}, Noboru and {Miwa}, Yoshinori and {Yokota}, Hideo and {Matsuda}, Toru and {Ebinuma}, Ryuichi and {Takeshi}, Kunio},
        title = "{Hyper Suprime-Cam}",
    booktitle = {Ground-based and Airborne Instrumentation for Astronomy IV},
         year = 2012,
       editor = {{McLean}, Ian S. and {Ramsay}, Suzanne K. and {Takami}, Hideki},
       series = {Society of Photo-Optical Instrumentation Engineers (SPIE) Conference Series},
       volume = {8446},
        month = sep,
          eid = {84460Z},
        pages = {84460Z},
          doi = {10.1117/12.926844},
       adsurl = {https://ui.adsabs.harvard.edu/abs/2012SPIE.8446E..0ZM}
}

@ARTICLE{2012MNRAS.423.2845L,
       author = {{Lane}, Richard R. and {K{\"u}pper}, Andreas H.~W. and {Heggie}, Douglas C.},
        title = "{The tidal tails of 47 Tucanae}",
      journal = {\mnras},
         year = 2012,
        month = jul,
       volume = {423},
       number = {3},
        pages = {2845-2853},
          doi = {10.1111/j.1365-2966.2012.21093.x},
archivePrefix = {arXiv},
       eprint = {1204.2549},
 primaryClass = {astro-ph.GA},
       adsurl = {https://ui.adsabs.harvard.edu/abs/2012MNRAS.423.2845L}
}

@ARTICLE{2012MNRAS.424L..16L,
       author = {{Lux}, H. and {Read}, J.~I. and {Lake}, G. and {Johnston}, K.~V.},
        title = "{NGC 5466: a unique probe of the Galactic halo shape}",
      journal = {\mnras},
         year = 2012,
        month = jul,
       volume = {424},
       number = {1},
        pages = {L16-L20},
          doi = {10.1111/j.1745-3933.2012.01276.x},
archivePrefix = {arXiv},
       eprint = {1204.5771},
 primaryClass = {astro-ph.GA},
       adsurl = {https://ui.adsabs.harvard.edu/abs/2012MNRAS.424L..16L}
}

@ARTICLE{2012ApJ...750...99T,
       author = {{Tonry}, J.~L. and {Stubbs}, C.~W. and {Lykke}, K.~R. and {Doherty}, P. and {Shivvers}, I.~S. and {Burgett}, W.~S. and {Chambers}, K.~C. and {Hodapp}, K.~W. and {Kaiser}, N. and {Kudritzki}, R.-P. and {Magnier}, E.~A. and {Morgan}, J.~S. and {Price}, P.~A. and {Wainscoat}, R.~J.},
        title = "{The Pan-STARRS1 Photometric System}",
      journal = {\apj},
         year = 2012,
        month = may,
       volume = {750},
       number = {2},
          eid = {99},
        pages = {99},
          doi = {10.1088/0004-637X/750/2/99},
archivePrefix = {arXiv},
       eprint = {1203.0297},
 primaryClass = {astro-ph.IM},
       adsurl = {https://ui.adsabs.harvard.edu/abs/2012ApJ...750...99T}
}

@ARTICLE{2012ApJ...748...20C,
       author = {{Carlberg}, R.~G.},
        title = "{Dark Matter Sub-halo Counts via Star Stream Crossings}",
      journal = {\apj},
         year = 2012,
        month = mar,
       volume = {748},
       number = {1},
          eid = {20},
        pages = {20},
          doi = {10.1088/0004-637X/748/1/20},
archivePrefix = {arXiv},
       eprint = {1109.6022},
 primaryClass = {astro-ph.CO},
       adsurl = {https://ui.adsabs.harvard.edu/abs/2012ApJ...748...20C}
}

@ARTICLE{2012ApJ...744...58M,
       author = {{Milone}, A.~P. and {Piotto}, G. and {Bedin}, L.~R. and {King}, I.~R. and {Anderson}, J. and {Marino}, A.~F. and {Bellini}, A. and {Gratton}, R. and {Renzini}, A. and {Stetson}, P.~B. and {Cassisi}, S. and {Aparicio}, A. and {Bragaglia}, A. and {Carretta}, E. and {D'Antona}, F. and {Di Criscienzo}, M. and {Lucatello}, S. and {Monelli}, M. and {Pietrinferni}, A.},
        title = "{Multiple Stellar Populations in 47 Tucanae}",
      journal = {\apj},
         year = 2012,
        month = jan,
       volume = {744},
       number = {1},
          eid = {58},
        pages = {58},
          doi = {10.1088/0004-637X/744/1/5810.1086/141918},
archivePrefix = {arXiv},
       eprint = {1109.0900},
 primaryClass = {astro-ph.SR},
       adsurl = {https://ui.adsabs.harvard.edu/abs/2012ApJ...744...58M}
}

@ARTICLE{2012MNRAS.419...14C,
       author = {{Carballo-Bello}, Julio A. and {Gieles}, Mark and {Sollima}, Antonio and {Koposov}, Sergey and {Mart{\'\i}nez-Delgado}, David and {Pe{\~n}arrubia}, Jorge},
        title = "{Outer density profiles of 19 Galactic globular clusters from deep and wide-field imaging}",
      journal = {\mnras},
         year = 2012,
        month = jan,
       volume = {419},
       number = {1},
        pages = {14-28},
          doi = {10.1111/j.1365-2966.2011.19663.x},
archivePrefix = {arXiv},
       eprint = {1108.4018},
 primaryClass = {astro-ph.GA},
       adsurl = {https://ui.adsabs.harvard.edu/abs/2012MNRAS.419...14C}
}

@ARTICLE{2011A&A...534A...9S,
       author = {{Sbordone}, L. and {Salaris}, M. and {Weiss}, A. and {Cassisi}, S.},
        title = "{Photometric signatures of multiple stellar populations in Galactic globular clusters}",
      journal = {\aap},
         year = 2011,
        month = oct,
       volume = {534},
          eid = {A9},
        pages = {A9},
          doi = {10.1051/0004-6361/201116714},
archivePrefix = {arXiv},
       eprint = {1103.5863},
 primaryClass = {astro-ph.SR},
       adsurl = {https://ui.adsabs.harvard.edu/abs/2011A&A...534A...9S}
}

@ARTICLE{2011A&A...525A.114L,
       author = {{Lardo}, C. and {Bellazzini}, M. and {Pancino}, E. and {Carretta}, E. and {Bragaglia}, A. and {Dalessandro}, E.},
        title = "{Mining SDSS in search of multiple populations in globular clusters}",
      journal = {\aap},
         year = 2011,
        month = jan,
       volume = {525},
          eid = {A114},
        pages = {A114},
          doi = {10.1051/0004-6361/201015662},
archivePrefix = {arXiv},
       eprint = {1010.4697},
 primaryClass = {astro-ph.GA},
       adsurl = {https://ui.adsabs.harvard.edu/abs/2011A&A...525A.114L}
}

@ARTICLE{2010AJ....140.1814Y,
       author = {{Yagi}, Masafumi and {Yoshida}, Michitoshi and {Komiyama}, Yutaka and {Kashikawa}, Nobunari and {Furusawa}, Hisanori and {Okamura}, Sadanori and {Graham}, Alister W. and {Miller}, Neal A. and {Carter}, David and {Mobasher}, Bahram and {Jogee}, Shardha},
        title = "{A Dozen New Galaxies Caught in the Act: Gas Stripping and Extended Emission Line Regions in the Coma Cluster}",
      journal = {\aj},
         year = 2010,
        month = dec,
       volume = {140},
       number = {6},
        pages = {1814-1829},
          doi = {10.1088/0004-6256/140/6/1814},
archivePrefix = {arXiv},
       eprint = {1005.3874},
 primaryClass = {astro-ph.CO},
       adsurl = {https://ui.adsabs.harvard.edu/abs/2010AJ....140.1814Y}
}

@ARTICLE{2010arXiv1012.3224H,
       author = {{Harris}, William E.},
        title = "{A New Catalog of Globular Clusters in the Milky Way}",
      journal = {arXiv e-prints},
         year = 2010,
        month = dec,
          eid = {arXiv:1012.3224},
        pages = {arXiv:1012.3224},
          doi = {10.48550/arXiv.1012.3224},
archivePrefix = {arXiv},
       eprint = {1012.3224},
 primaryClass = {astro-ph.GA},
       adsurl = {https://ui.adsabs.harvard.edu/abs/2010arXiv1012.3224H}
}

@INPROCEEDINGS{2010SPIE.7740E..15A,
       author = {{Axelrod}, T. and {Kantor}, J. and {Lupton}, R.~H. and {Pierfederici}, F.},
        title = "{An open source application framework for astronomical imaging pipelines}",
    booktitle = {Software and Cyberinfrastructure for Astronomy},
         year = 2010,
       editor = {{Radziwill}, Nicole M. and {Bridger}, Alan},
       series = {Society of Photo-Optical Instrumentation Engineers (SPIE) Conference Series},
       volume = {7740},
        month = jul,
          eid = {774015},
        pages = {774015},
          doi = {10.1117/12.857297},
       adsurl = {https://ui.adsabs.harvard.edu/abs/2010SPIE.7740E..15A}
}

@ARTICLE{2010MNRAS.401..105K,
       author = {{K{\"u}pper}, Andreas H.~W. and {Kroupa}, Pavel and {Baumgardt}, Holger and {Heggie}, Douglas C.},
        title = "{Tidal tails of star clusters}",
      journal = {\mnras},
         year = 2010,
        month = jan,
       volume = {401},
       number = {1},
        pages = {105-120},
          doi = {10.1111/j.1365-2966.2009.15690.x},
archivePrefix = {arXiv},
       eprint = {0909.2619},
 primaryClass = {astro-ph.SR},
       adsurl = {https://ui.adsabs.harvard.edu/abs/2010MNRAS.401..105K}
}

@ARTICLE{2008ApJ...689..936J,
       author = {{Johnston}, Kathryn V. and {Bullock}, James S. and {Sharma}, Sanjib and {Font}, Andreea and {Robertson}, Brant E. and {Leitner}, Samuel N.},
        title = "{Tracing Galaxy Formation with Stellar Halos. II. Relating Substructure in Phase and Abundance Space to Accretion Histories}",
      journal = {\apj},
         year = 2008,
        month = dec,
       volume = {689},
       number = {2},
        pages = {936-957},
          doi = {10.1086/592228},
archivePrefix = {arXiv},
       eprint = {0807.3911},
 primaryClass = {astro-ph},
       adsurl = {https://ui.adsabs.harvard.edu/abs/2008ApJ...689..936J}
}

@ARTICLE{2008MNRAS.387.1248K,
       author = {{K{\"u}pper}, Andreas H.~W. and {MacLeod}, Andrew and {Heggie}, Douglas C.},
        title = "{On the structure of tidal tails}",
      journal = {\mnras},
         year = 2008,
        month = jul,
       volume = {387},
       number = {3},
        pages = {1248-1252},
          doi = {10.1111/j.1365-2966.2008.13323.x},
archivePrefix = {arXiv},
       eprint = {0804.2476},
 primaryClass = {astro-ph},
       adsurl = {https://ui.adsabs.harvard.edu/abs/2008MNRAS.387.1248K}
}

@ARTICLE{2007MNRAS.380..749F,
       author = {{Fellhauer}, M. and {Evans}, N.~W. and {Belokurov}, V. and {Wilkinson}, M.~I. and {Gilmore}, G.},
        title = "{The tidal tails of NGC 5466}",
      journal = {\mnras},
         year = 2007,
        month = sep,
       volume = {380},
       number = {2},
        pages = {749-756},
          doi = {10.1111/j.1365-2966.2007.12111.x},
archivePrefix = {arXiv},
       eprint = {0706.2627},
 primaryClass = {astro-ph},
       adsurl = {https://ui.adsabs.harvard.edu/abs/2007MNRAS.380..749F}
}

@ARTICLE{2006ApJ...639L..17G,
       author = {{Grillmair}, C.~J. and {Johnson}, R.},
        title = "{The Detection of a 45{\textdegree} Tidal Stream Associated with the Globular Cluster NGC 5466}",
      journal = {\apjl},
         year = 2006,
        month = mar,
       volume = {639},
       number = {1},
        pages = {L17-L20},
          doi = {10.1086/501439},
archivePrefix = {arXiv},
       eprint = {astro-ph/0602602},
 primaryClass = {astro-ph},
       adsurl = {https://ui.adsabs.harvard.edu/abs/2006ApJ...639L..17G}
}

@ARTICLE{2006ApJ...637L..29B,
       author = {{Belokurov}, V. and {Evans}, N.~W. and {Irwin}, M.~J. and {Hewett}, P.~C. and {Wilkinson}, M.~I.},
        title = "{The Discovery of Tidal Tails around the Globular Cluster NGC 5466}",
      journal = {\apjl},
         year = 2006,
        month = jan,
       volume = {637},
       number = {1},
        pages = {L29-L32},
          doi = {10.1086/500362},
archivePrefix = {arXiv},
       eprint = {astro-ph/0511767},
 primaryClass = {astro-ph},
       adsurl = {https://ui.adsabs.harvard.edu/abs/2006ApJ...637L..29B}
}

@INPROCEEDINGS{2004ASPC..327..333K,
       author = {{Koch}, A. and {Odenkirchen}, M. and {Grebel}, E.~K. and {Mart{\'\i}nez-Delgado}, D. and {Caldwell}, J.~A.~R.},
        title = "{The Luminosity Function of Palomar 5 and Its Tidal Tails}",
    booktitle = {Satellites and Tidal Streams},
         year = 2004,
       editor = {{Prada}, F. and {Martinez Delgado}, D. and {Mahoney}, T.~J.},
       series = {Astronomical Society of the Pacific Conference Series},
       volume = {327},
        month = dec,
        pages = {333},
          doi = {10.48550/arXiv.astro-ph/0307496},
archivePrefix = {arXiv},
       eprint = {astro-ph/0307496},
 primaryClass = {astro-ph},
       adsurl = {https://ui.adsabs.harvard.edu/abs/2004ASPC..327..333K}
}

@INPROCEEDINGS{2004ASPC..327..284O,
       author = {{Odenkirchen}, M. and {Grebel}, E.~K.},
        title = "{The Tidal Perturbation of the Low Mass Globular Cluster NGC 5466}",
    booktitle = {Satellites and Tidal Streams},
         year = 2004,
       editor = {{Prada}, F. and {Martinez Delgado}, D. and {Mahoney}, T.~J.},
       series = {Astronomical Society of the Pacific Conference Series},
       volume = {327},
        month = dec,
        pages = {284},
          doi = {10.48550/arXiv.astro-ph/0307481},
archivePrefix = {arXiv},
       eprint = {astro-ph/0307481},
 primaryClass = {astro-ph},
       adsurl = {https://ui.adsabs.harvard.edu/abs/2004ASPC..327..284O}
}

@ARTICLE{2003AJ....126.2385O,
       author = {{Odenkirchen}, Michael and {Grebel}, Eva K. and {Dehnen}, Walter and {Rix}, Hans-Walter and {Yanny}, Brian and {Newberg}, Heidi Jo and {Rockosi}, Constance M. and {Mart{\'\i}nez-Delgado}, David and {Brinkmann}, Jon and {Pier}, Jeffrey R.},
        title = "{The Extended Tails of Palomar 5: A 10{\textdegree} Arc of Globular Cluster Tidal Debris}",
      journal = {\aj},
         year = 2003,
        month = nov,
       volume = {126},
       number = {5},
        pages = {2385-2407},
          doi = {10.1086/378601},
archivePrefix = {arXiv},
       eprint = {astro-ph/0307446},
 primaryClass = {astro-ph},
       adsurl = {https://ui.adsabs.harvard.edu/abs/2003AJ....126.2385O}
}

@ARTICLE{2003MNRAS.340..227B,
       author = {{Baumgardt}, Holger and {Makino}, Junichiro},
        title = "{Dynamical evolution of star clusters in tidal fields}",
      journal = {\mnras},
         year = 2003,
        month = mar,
       volume = {340},
       number = {1},
        pages = {227-246},
          doi = {10.1046/j.1365-8711.2003.06286.x},
archivePrefix = {arXiv},
       eprint = {astro-ph/0211471},
 primaryClass = {astro-ph},
       adsurl = {https://ui.adsabs.harvard.edu/abs/2003MNRAS.340..227B}
}

@ARTICLE{2001AJ....122.3231G,
       author = {{Grillmair}, Carl J. and {Smith}, Graeme H.},
        title = "{The Main-Sequence Luminosity Function of Palomar 5 from THE HUBBLE SPACE TELESCOPE}",
      journal = {\aj},
         year = 2001,
        month = dec,
       volume = {122},
       number = {6},
        pages = {3231-3238},
          doi = {10.1086/323916},
archivePrefix = {arXiv},
       eprint = {astro-ph/0110411},
 primaryClass = {astro-ph},
       adsurl = {https://ui.adsabs.harvard.edu/abs/2001AJ....122.3231G}
}

@ARTICLE{1999PASP..111...63F,
       author = {{Fitzpatrick}, Edward L.},
        title = "{Correcting for the Effects of Interstellar Extinction}",
      journal = {\pasp},
         year = 1999,
        month = jan,
       volume = {111},
       number = {755},
        pages = {63-75},
          doi = {10.1086/316293},
archivePrefix = {arXiv},
       eprint = {astro-ph/9809387},
 primaryClass = {astro-ph},
       adsurl = {https://ui.adsabs.harvard.edu/abs/1999PASP..111...63F}
}

@ARTICLE{1998ApJ...500..525S,
       author = {{Schlegel}, David J. and {Finkbeiner}, Douglas P. and {Davis}, Marc},
        title = "{Maps of Dust Infrared Emission for Use in Estimation of Reddening and Cosmic Microwave Background Radiation Foregrounds}",
      journal = {\apj},
         year = 1998,
        month = jun,
       volume = {500},
       number = {2},
        pages = {525-553},
          doi = {10.1086/305772},
archivePrefix = {arXiv},
       eprint = {astro-ph/9710327},
 primaryClass = {astro-ph},
       adsurl = {https://ui.adsabs.harvard.edu/abs/1998ApJ...500..525S}
}

@ARTICLE{1997MNRAS.287..915V,
       author = {{Vesperini}, E.},
        title = "{On the evolution of the Galactic globular cluster system}",
      journal = {\mnras},
         year = 1997,
        month = jun,
       volume = {287},
       number = {4},
        pages = {915-928},
          doi = {10.1093/mnras/287.4.915},
archivePrefix = {arXiv},
       eprint = {astro-ph/9702067},
 primaryClass = {astro-ph},
       adsurl = {https://ui.adsabs.harvard.edu/abs/1997MNRAS.287..915V}
}

@ARTICLE{1996AJ....112.1487H,
       author = {{Harris}, William E.},
        title = "{A Catalog of Parameters for Globular Clusters in the Milky Way}",
      journal = {\aj},
         year = 1996,
        month = oct,
       volume = {112},
        pages = {1487},
          doi = {10.1086/118116},
       adsurl = {https://ui.adsabs.harvard.edu/abs/1996AJ....112.1487H}
}

@ARTICLE{1995PASP..107..945F,
       author = {{Fukugita}, M. and {Shimasaku}, K. and {Ichikawa}, T.},
        title = "{Galaxy Colors in Various Photometric Band Systems}",
      journal = {\pasp},
         year = 1995,
        month = oct,
       volume = {107},
        pages = {945},
          doi = {10.1086/133643},
       adsurl = {https://ui.adsabs.harvard.edu/abs/1995PASP..107..945F}
}

@ARTICLE{1991ApJ...379...52W,
       author = {{White}, Simon D.~M. and {Frenk}, Carlos S.},
        title = "{Galaxy Formation through Hierarchical Clustering}",
      journal = {\apj},
         year = 1991,
        month = sep,
       volume = {379},
        pages = {52},
          doi = {10.1086/170483},
       adsurl = {https://ui.adsabs.harvard.edu/abs/1991ApJ...379...52W}
}

@ARTICLE{1967ITIT...13...21C,
       author = {{Cover}, T. and {Hart}, P.},
        title = "{Nearest neighbor pattern classification}",
      journal = {IEEE Transactions on Information Theory},
         year = 1967,
        month = jan,
       volume = {13},
       number = {1},
        pages = {21-27},
          doi = {10.1109/TIT.1967.1053964},
       adsurl = {https://ui.adsabs.harvard.edu/abs/1967ITIT...13...21C}
}

@ARTICLE{1962AJ.....67..471K,
       author = {{King}, Ivan},
        title = "{The structure of star clusters. I. an empirical density law}",
      journal = {\aj},
         year = 1962,
        month = oct,
       volume = {67},
        pages = {471},
          doi = {10.1086/108756},
       adsurl = {https://ui.adsabs.harvard.edu/abs/1962AJ.....67..471K}
}
\bibliographystyle{aasjournal}

\end{document}